\documentclass[format=acmsmall, review=false, screen=true]{acmart}
\usepackage{booktabs} 
\usepackage{makecell}
\usepackage{lipsum}
\usepackage[utf8]{inputenc}
\usepackage{caption}
\usepackage[labelformat=simple]{subcaption} 
\usepackage{array}
\usepackage{wrapfig}
\usepackage{multirow}
\usepackage{tabularx}
\usepackage{pifont}
\usepackage{lineno}
\usepackage{balance}
\usepackage{xcolor,pifont}

\usepackage{soul}

\usepackage[linesnumbered, ruled,vlined]{algorithm2e}
\RequirePackage{pdflscape}
\usepackage{flushend}
\usepackage{multicol}
\usepackage{placeins}
\usepackage{booktabs}
\usepackage{indentfirst}
\usepackage[figuresright]{rotating}
\usepackage{fancybox}
\usepackage{fontawesome}
\usepackage{threeparttable}
\usepackage{mathtools}
\usepackage{MnSymbol,wasysym}
\usepackage{rotating}
\usepackage{adjustbox}
\usepackage{graphicx}
\usepackage{xcolor}
\usepackage{tcolorbox}
\newtcolorbox{mybox}[2][]
{colback = white, colframe = black, fonttitle = \bfseries,
    colbacktitle = gray, enhanced,
    attach boxed title to top left={yshift=-3mm, xshift=3mm},
    title=#2, #1}

\usepackage{framed}
\definecolor{Gray}{gray}{0.9}
\definecolor{shadecolor}{gray}{0.95}
\usepackage{tikz}
\usetikzlibrary{trees,positioning,shapes,shadows,arrows}
\tikzset{
  basic/.style  = {draw, text width=2cm, drop shadow, font=\sffamily, rectangle},
  root/.style   = {basic, rounded corners=2pt, thin, align=center, fill=white},
  level-2/.style = {basic, rounded corners=6pt, thin,align=center, fill=white, text width=3cm},
  level-3/.style = {basic, thin, align=center, fill=white, text width=1.8cm}
}
\newcommand{\todo}[1]{}
\renewcommand{\todo}[1]{{\color{red} TODO: {#1}}}
\newcommand{\fakesection}[2][0.5em]{\vspace{#1}\noindent\textit{\textbf{#2}}}

\begin{document}
\title{Towards Automated Identification of Violation Symptoms of Architecture Erosion}

\author{Ruiyin Li}
\orcid{0000-0001-8536-4935}
\email{ryli_cs@whu.edu.cn}
\affiliation{%
  \institution{School of Computer Science, Wuhan University}
  \city{Wuhan}
  \country{China}
}

\author{Peng Liang}
\orcid{0000-0002-2056-5346}
\email{liangp@whu.edu.cn}
\affiliation{%
  \institution{School of Computer Science, Wuhan University}
  \city{Wuhan}
  \country{China}
}

\author{Paris Avgeriou}
\orcid{0000-0002-7101-0754}
\email{p.avgeriou@rug.nl}
\affiliation{%
  \institution{Department of Computing Science, University of Groningen}
  \city{Groningen}
  \country{The Netherlands}
}

\author{Yifei Wang}
\orcid{0000-0003-0100-6896}
\email{whiten@whu.edu.cn}
\affiliation{
  \institution{School of Computer Science, Wuhan University}
  \city{Wuhan}
  \country{China}
}

\thanks{We would like to thank all the practitioners who participated in our survey and interviews. This work has been partially supported by the National Natural Science Foundation of China (NSFC) with Grant No. 62402348 and 62172311.}

\acmJournal{TOSEM}
\acmVolume{0}
\acmNumber{0}
\acmArticle{0}
\acmMonth{0}

\renewcommand{\shortauthors}{Li et al.}




\begin{abstract}
Architecture erosion has a detrimental effect on maintenance and evolution, as the implementation deviates from the intended architecture. To prevent this, development teams need to understand early enough the symptoms of erosion, and particularly violations of the intended architecture. 
One feasible way is through the automated identification of architecture violations from textual artifacts, and particularly code reviews. 
\textcolor{black}{In this paper, we investigate the feasibility for automatically identifying such violation symptoms from code review comments using both traditional machine learning (ML)/deep learning (DL) techniques and state-of-the-art large language models (LLMs). We developed and evaluated 15 ML-based and 4 DL-based classifiers with three pre-trained word embeddings (word2vec, fastText, and GloVe) on code reviews from four open-source projects (OpenStack's Nova and Neutron, and Qt's Base and Creator). Additionally, we constructed state-of-the-art LLM-based classifiers using GPT-4o, Qwen-3, and DeepSeek-R1 for the same task. Our results demonstrate that for ML/DL approaches, an SVM classifier with word2vec achieved the best performance (F1-score: 0.808), while 200-dimensional word embeddings generally outperformed other dimensionalities. Ensemble classifiers using majority voting further improved the performance. Additional evaluations on a naturally imbalanced test set showed that the proposed approaches remain effective under realistic low-prevalence conditions, with SVM achieving the best Macro-F1 among ML/DL approaches. Statistical significance analysis further confirmed that the observed performance differences among classifiers are unlikely to be caused by random variation. The best LLM-based classifiers outperformed traditional ML/DL approaches, with GPT-4o achieving the highest F1-score (0.851). Validation through practitioner surveys and interviews confirmed that automatically identified violation symptoms provide practical value and early warnings of architectural issues. Furthermore, a controlled experiment quantitatively validated the approach's effectiveness: providing violation symptom hints significantly improved developers' detection rate of architecture violations from 25.9\% to 64.7\%, corroborating the positive perceptions reported in the qualitative study.  
Our work contributes an effective automated approach for identifying architecture violations, a comprehensive comparison between traditional ML/DL and LLM-based techniques, and practitioner-validated insights for improving architectural conformance and sustainability in software systems.}
\end{abstract}


\maketitle

\section{Introduction}\label{sec:Introduction}
Software architecture is critical in developing large and complex systems, as it reflects the system structure and behavior and acts as a bridge between business goals and the implemented system \cite{Bass2021sap}. However, the implemented architectures often diverge from their intended architectures over time, which is the phenomenon of \textit{architecture erosion} \cite{Li2022SMS, PerryWolf1992AEr}. This phenomenon undermines maintainability, performance, and overall sustainability \cite{Li2022SMS, Le2016rad, Venters2018ss} and often described by various terms in the literature \cite{Li2022SMS} and practice \cite{Li2021uae}, such as architectural decay, degradation, and deterioration. Various symptoms or signs can be observed when architecture erosion occurs during a software system's life cycle \cite{Li2021uae, Li2022SMS, Herold2015dvc, Fontana2016aer, Macia2012anomalies}. Our recent systematic mapping study \cite{Li2022SMS} reported four categories of such architecture erosion symptoms: \textit{structural symptoms} (e.g., cyclic dependencies), \textit{violation symptoms} (e.g., layering violation), \textit{quality symptoms} (e.g., high defect rate), and \textit{evolution symptoms} (e.g., extensive ripple effects of changes). 

In this study, we specifically focus on violation symptoms, as those are the most immediate symptoms of architecture erosion and require the most attention from practitioners \cite{Li2022sae, Li2023wvs}. This symptom type stems from the definition of architecture erosion \cite{Li2022SMS}, as a violation causes the implemented architecture to diverge from the intended one. Architecturally-relevant violations in software systems include violations of design principles, architecture patterns, decisions, and requirements \cite{Li2022SMS}. The accumulation of violation symptoms can render the architecture completely untenable \cite{DeSilva2012AErsurvey}. For brevity, we refer to \textit{violation symptoms of architecture erosion} as \textit{violation symptoms} throughout the remainder of this paper.

While a few temporary violation symptoms might be innocuous and might not result in software errors, the accumulation of architecture violations result in architecture erosion \cite{PerryWolf1992AEr, DeSilva2012AErsurvey, Mendoza2021avd}; consequently, erosion negatively affects quality attributes, including maintainability and performance~\cite{Li2021uae, Mendoza2021avd}. Thus, it is essential to identify and monitor violation symptoms to facilitate the maintenance of the system architecture and further reveal architecture inconsistencies. This is the first step towards, eventually, repairing or mitigating architecture erosion \cite{DeSilva2012AErsurvey, Li2022SMS}. 

Violation symptoms can be identified static analysis of source code, but that has certain limitations. As mentioned in previous studies \cite{Sharma2018sss, Li2022SMS, Li2023wvs}, a number of factors hinder the effective utilization of static code analysis tools in precisely identifying violation symptoms through source code. For example, such tools only detect limited violation types and narrow programming language support, and cannot capture semantics of violation symptoms by relying solely on code metrics. Alternatively, violations symptoms can be manually identified by analyzing textual artifacts that contain information related to system architecture and design, such as code comments, code reviews, and issue trackers \cite{Li2022sae}. Recent studies \cite{Li2022sae, Li2023wvs} have proposed taxonomies on violation symptoms, that can be used as a checklist to detect such symptoms during code review. However, it could be tedious, time-consuming, and potentially error-prone to manually identify potential architecture violations \cite{Li2023wvs}. Automated techniques could address this problem, but to the best of our knowledge, no prior study has investigated the automated identification of violation symptoms in textual artifacts, such as code review comments. 

To address this gap, our \textbf{goal} is to assess the feasibility of automatically identifying violation symptoms of architecture erosion from code review comments. The automated identification of violation symptoms could act in a complementary way to modern code review. Modern code review usually addresses code changes in general through knowledge sharing~\cite{badampudi2023modern}, while automated identification of violation symptoms can offer an extra layer of scrutiny. For example, such an automated approach can issue warnings for potential or ignored violation symptoms, preventing the integration of code changes with architectural violations into the code base, and ensuring a more robust and reliable development process. Moreover, automatically identifying violation symptoms from code review comments could offer valuable insights at the architecture level and facilitate the corresponding knowledge sharing between developers and reviewers. The ultimate purpose of our approach is to complement the modern code review process, through warning of and guarding against architectural violation issues, eventually maintaining architectural integrity.

In this work, we first developed 15 classifiers based on Machine Learning (ML) and 4 classifiers based on Deep Learning (DL), to identify violation symptoms from developer discussions in code reviews of four large open-source software projects from the OpenStack and Qt communities. Our results show that the SVM classifier based on \textit{word2vec} performed the best with a Precision of 0.789, a Recall of 0.828, an F1-score of 0.808, and an Accuracy of 0.803. We also found out that utilizing \textit{fastText} pre-trained word embedding can achieve relatively good results and it helps to improve the classifiers' performance. Moreover, 200-dimensional pre-trained word embedding models outperform classifiers that use 100- and 300-dimensional models. Finally, an ensemble classifier based on the majority voting strategy can further enhance the classifier and outperform the individual classifiers. Then, we conducted an online survey and semi-structured interviews to validate the effectiveness of automated identification of violation symptoms in practice. We received responses from 24 survey participants and 6 interviewees, all of whom were involved in the discussion of the collected review comments concerning architecture violations. Their feedback confirmed the practical value of automated identification of violation symptoms, as such classifiers can support practitioners in finding, prioritizing, and handling architecturally-related issues.

Furthermore, the advent of Large Language Models (LLMs) fundamentally reshaped Software Engineering (SE) paradigms~\cite{hou2024large, li2025ChatGPT}, existing studies explored the capacities of LLMs on popular SE tasks (e.g., code generation), but no studies investigate the fine-grained task, identifying violation symptoms in code review comments. Therefore, we further developed LLM-based classifiers using three state-of-the-art LLMs (i.e., \textit{GPT-4o}, \textit{Qwen-3}, and \textit{DeepSeek-R1}) and compared their performance with the traditional ML/DL techniques on the same classification task. The results show that the best LLM-based classifiers typically outperform the traditional ML/DL-based classifiers, and GPT-4o-based classifier with 2-shot setting comparatively achieves better performance (F1-score 0.851) than Qwen-3 and DeepSeek-R1 on identifying violation symptoms. Notably, ensembling LLM-based classifiers does not consistently improve the performance.

Unlike previous studies that relied on source code analysis to identify architectural issues (see Section~\ref{sec:B1}), our work takes a complementary approach by identifying violation symptoms from textual artifacts, which better aligns with practitioners' perspectives. We also explore ensemble learning techniques to detect architecture erosion symptoms, showing advantages over single ML/DL algorithms. 
The \textbf{main contributions} of this work are the following:
\begin{itemize}
    \item We conducted exploratory experiments to demonstrate the feasibility of ML/DL, LLM-based classifiers for automatically identifying violation symptoms in code review comments, and we compared their performance under different settings, respectively.
    \item We investigated the impact of pre-trained word embeddings and shot settings on classification performance. For ML/DL-based classifiers, the SVM-based classifier with a 200-dimensional pre-trained \textit{word2vec} performs best. Besides, the best LLM-based classifiers outperform ML/DL-based classifiers, and the GPT-4o-based classifier with a 2-shot setting achieves the best performance. 
    \item We gained insights from the survey and interview participants regarding the usefulness of trained classifiers in identifying architecture violations during code review. We provide the implications and suggestions based on our empirical study.
\end{itemize}

The remainder of the paper is organized as follows. Section~\ref{sec:Background} introduces background information. Section~\ref{sec:Study Design} describes the research questions, study design, and experimental setup. The answers to the research questions are presented in Section~\ref{sec:Results}. We discuss the implications of the study results in Section~\ref{sec:Discussion} and the threats to validity in Section~\ref{sec:Threats to Validity}. Section~\ref{sec:Related Work} reviews related work, and Section~\ref{sec:Conclusion} concludes this work with future directions.

\section{Background}\label{sec:Background}

In this section, we introduce a) the background of architecture erosion and its symptoms; b) architecture conformance checking,  the most commonly used approach to detect architecture violations; and c) the modern code review process in Gerrit\footnote{https://www.gerritcodereview.com/}. 

\subsection{Architecture Erosion and Related Symptoms}\label{sec:B1}
In the past decades, a wide variety of studies are concerned with the architecture erosion phenomenon, which has been extensively discussed and described with various terms \cite{Li2021uae, Li2022SMS}, such as architecture decay \cite{Hassaine2012ADvISE, Le2018emad}, degradation \cite{Herold2020asd}, degeneration \cite{Hochstein2005cad}.

Architecture erosion manifests in various symptoms during development and maintenance. A symptom is a partial sign or indicator of the emergence of architecture erosion \cite{Le2016rad, Li2022SMS, Li2022sae}. According to a recent mapping study \cite{Li2022SMS}, erosion symptoms can be classified into four categories (i.e., structural symptoms, violation symptoms, quality symptoms, and evolution symptoms). 

Previous studies investigated different symptoms of architecture erosion. Mair \textit{et al}. \cite{Mair2013tesa} proposed a formalization method regarding the process of repairing eroded architecture by finding violation symptoms and recommending optimal repair sequences. Le \textit{et al}. \cite{Le2016rad, Le2018emad} regarded architectural smells as structural symptoms and they provided metrics to detect instances of architecture erosion by analyzing the detected smells. Martin \cite{Martin2000dpdp} deemed that the evolution symptoms of architecture erosion include rigidity (a tendency to resist changes), fragility (a tendency to break frequently when modifications occur), immobility (inability to be reused), and viscosity (reduced effectiveness and efficiency due to design or environment issues).

\begin{table}[htbp]
\centering
\small
\setlength\tabcolsep{3.4pt}
\renewcommand{\arraystretch}{1.0}
\caption{Examples of violation symptoms collected from Nova, Neutron, Qt Base, and Qt Creator}\label{T:examples}
    \begin{tabularx}{\textwidth}{>{\centering\arraybackslash}m{0.12\textwidth}X}\toprule
    \textbf{Project} & \textbf{Example of violation symptoms of architecture erosion} \\\hline
    \multirow{4}{*}{\textbf{Nova}} & ``\textit{But here don't we have to make a upcall from compute to api db, which will violate api/cell isolation rules. Is there any workaround in this case?}''\\\cline{2-2}
    ~ & ``\textit{We could get some race conditions when starting the scheduler where it would not know the allocation ratios and would have to call the computes, which is a layer isolation violation to me}.''\\\hline
    \multirow{4}{*}{\textbf{Neutron}} & ``\textit{Having said all of that: I get that I'm violating an abstraction layer in LinkLocalAddressPair and that this is surprising (and therefore bad).}''\\\cline{2-2}
    ~ & ``\textit{As you pointed out, since this patch is about the router availability zone it seems like a layer violation} doing it in the mech driver.''\\\hline
    \multirow{3}{*}{\textbf{Qt Base}} & ``\textit{I think it's a layer isolation violation to just make the call here. It should be fixed at the Compute API level rather IMHO.}''\\\cline{2-2}
    ~ & ``\textit{Feels like this DB work violates the level of abstraction we are expecting here.}''\\\hline
    \multirow{3}{*}{\textbf{Qt Creator}} & ``\textit{The unit tests are designed to not depend on any qt creator library to make the dependency breaking easier. Sometimes it works but it is not designed that way.}''\\\cline{2-2}
    ~ & ``\textit{This looks like we break the design. Why do we can not hold a pointer to the plugin?}''\\\bottomrule
    \end{tabularx}
\end{table}

As mentioned in the Introduction section, we focus on the most direct erosion symptoms of architecture erosion (i.e., violation symptoms). Table~\ref{T:examples} presents several examples of violation symptoms of architecture erosion from the used dataset (see Section \ref{sec:Study Design}). To facilitate the examination of additional examples, we have made our dataset publicly available online \cite{onlinepackage}; a detailed comparison of our work and related work is given in Section \ref{sec:Related Work}.

\subsection{Architecture Conformance Checking}\label{sec:B2}
Architecture Conformance Checking (ACC) techniques are the most commonly-used approaches to detect architecture violations \cite{Li2022SMS}. They either perform static or dynamic analysis, by comparing the structure of the intended architecture (provided by the architects) with the extracted architecture information from source code. For example, Miranda \textit{et al}. \cite{Miranda2016acc} presented an ACC approach based on static code analysis techniques with a supporting tool ArchRuby. Their tool provides means to control the erosion process by reporting and visualizing architectural violations in two high-level architectural models, namely reflexion models and Dependency Structure Matrices (DSMs).

In addition, rule-based conformance checking approaches are also employed to identify architecture violations. For example, previous studies identified architecture violations by checking explicitly defined architectural rules \cite{Mendoza2021avd}. Moreover, it is viable to check architecture conformance and identify architecture violations by defining and describing the systems through Architecture Description Languages (ADLs), or Domain-Specific Languages (DSLs). 

Despite their potential, the previously mentioned approaches for identifying architecture violations have clear limitations. For instance, much effort is required to address the challenges of understanding the architecture design (e.g., concepts and relations), defining architectural rules (or description languages) in advance, and establishing a mapping between architectural elements and source code \cite{Li2022SMS, Li2023wvs}. Additionally, other limitations, such as a lack of generalizability and architectural views visualization, insufficient tooling support, have impeded the wide adoption of these approaches in practice \cite{Li2022SMS}. 

During software development, textual artifacts (e.g., code review comments) contain rich and valuable information regarding architectural design and changes apart from source code. Mining architectural violations from textual artifacts can significantly facilitate code comprehension for software engineers \cite{Li2023wvs}. In contrast to the previously mentioned approaches, our study focuses specifically on mining violation symptoms from textual artifacts.

\subsection{Code Review in Gerrit}\label{sec:B3}
Code review is a crucial software development activity that involves the systematic examination of assigned code to identify defects and improve software quality \cite{Bacchelli2013eoc}. A methodical code review process not only enhances the quality of software systems, but also facilitates the sharing of development knowledge and prevents the release of unstable and defective products. Over time, code review practices have become increasingly important, have been widely adopted in modern software development, and have received extensive tool support. In fact, tool-based code review has become the norm in both industry and open-source communities, with a variety of available code review tools, such as Gerrit, GitHub's Pull Requests, Me's Phabricator\footnote{https://www.phacility.com/}, and VMware's Review-Board\footnote{https://www.reviewboard.org/}.

\begin{figure}[t]
	\centering
	\includegraphics[width=0.48\linewidth]{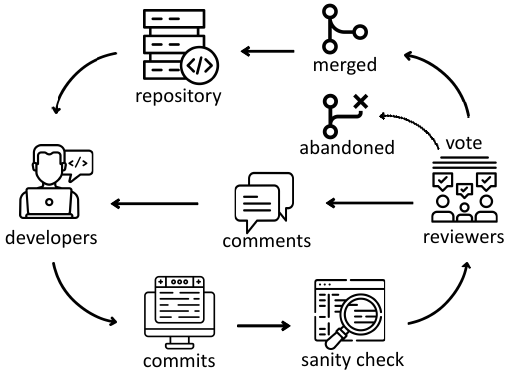}
	\caption{Code review process in Gerrit}\label{F:Code review}
\end{figure}

The two communities (i.e., OpenStack and Qt), from which the four open-source software (OSS) projects were selected in this work, use Gerrit as the code review tool. Figure~\ref{F:Code review} illustrates the modern code review process in Gerrit, which is a well-known code review tool and widely used in modern code review \cite{Davila2021SLRmcr}. Gerrit allows for efficient collaboration on code changes before they are merged into the main code base. Once a developer commits new code changes (e.g., software patches) and their descriptions to Gerrit, the system creates a page to record all changes and relevant descriptions (e.g., commit messages). Then, the system conducts a sanity check to verify whether the patch is compliant and to ensure that the code does not contain obvious compilation errors. After the submitted patch passes the sanity check, code reviewers manually examine the patch and provide comments to correct any potential errors, followed by giving a voting score. The comments are subsequently used by the developers to improve the patch and submit it for another review. Finally, the submitted patch is merged into the code repository only after passing integration tests (potentially over multiple iterations), which confirms that there are no known issues or conflicts. Alternatively, the patch can be abandoned if the voting is negative.

The above process is used, among others, in revisions that enable the analysis of potentially significant architectural changes. Code review comments provide developers a \textit{finer granularity} for investigating architectural changes and violations in their daily development routines \cite{Li2021mpl}. Meanwhile, developers provide reasoning and rationale for the changes they make in code review comments, which enables code review data to be a valuable source of knowledge for explaining changes \cite{Paixao2021crac}, offering insights into developers' concerns, suggestions, and potential violations.

\section{Study Design}\label{sec:Study Design}
This section describes the research questions that motivated this research, followed by a detailed explanation of the data collection and the ML/DL models to automatically identify architecture violations from textual artifacts. Then, we provide details of the validation survey and interview conducted to assess the effectiveness of automated identification of violation symptoms in practice. Furthermore, we introduce and compare three state-of-the-art LLMs on the same task with traditional ML/DL approaches.

\begin{figure*}[t]
	\centering
	\includegraphics[width=\linewidth]{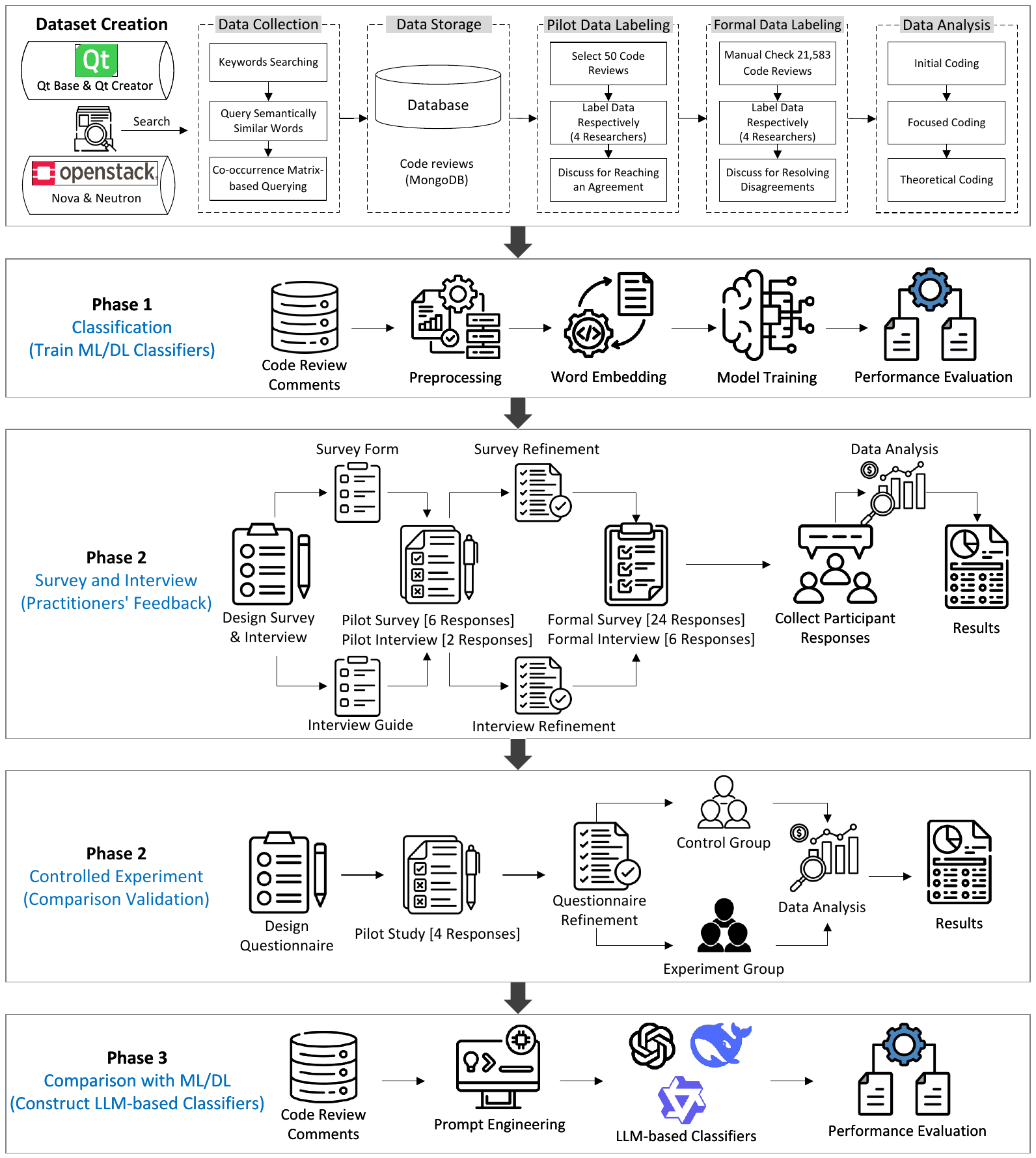}
	\caption{\textcolor{black}{An overview of the research process}}\label{F:Overview}
\end{figure*}

Figure~\ref{F:Overview} provides an overview of the study design. As illustrated in this figure (see the top box, \textbf{Dataset Creation}), we collected data and established a dataset on architecture violations in our previous work \cite{Li2023wvs}, which contains code review comments related to architecture violations During \textbf{Phase 1}, we explored the possibility of using this dataset to train ML/DL-based classifiers to automatically identify violation symptoms in code review comments. During \textbf{Phase 2}, we sent the survey questionnaire to 200 involved participants and received 24 responses that were used to validate the usefulness of automated identification of violation symptoms in practice. \textcolor{black}{Additionally, we conducted a controlled experiment to further quantitatively validate the usefulness of our approaches during code review.} During \textbf{Phase 3}, we constructed LLM-based classifiers through prompts using state-of-the-art LLMs and compared their performance against the ML/DL methods from Phase 1.

\subsection{Research Questions}\label{sec:RQ}
In this work, we aim to explore the feasibility of automatically identifying violation symptoms of architecture erosion from code review comments. Specifically, we formulated the goal of this study by following the Goal-Question-Metric method \cite{Basili1994gmq}: \textbf{analyze} \textit{code review comments} \textbf{for the purpose of} \textit{exploring the feasibility of }automatically identifying violation symptoms of architecture erosion \textbf{with respect to} \textit{the performance on architecture erosion identification} \textbf{from the point of view of} \textit{developers} \textbf{in the context of} \textit{open source software development}. To achieve our goal, we formulate four Research Questions (RQs):

\begin{tcolorbox}[colback = white!25!white, colframe = black!75!black]
\textbf{RQ1: How accurately can classifiers identify violation symptoms from code review comments?}
    \begin{itemize}
        \item \textbf{RQ1.1: Which classifier performs best in identifying the violation symptoms?}
        \item \textbf{RQ1.2: Which word embedding models can help to improve the performance of classifiers?}
        \item \textbf{RQ1.3: To what extent, does the dimensionality of word embedding models impact the performance of classifiers?}
    \end{itemize}
\end{tcolorbox}

\textbf{Rationale}: This RQ focuses on training feasible classifiers to automatically identify violation symptoms from textual artifacts. The percentage of violation symptoms (belonging to architecture-level issues) is relatively lower than code-level issues. We thus expect that the description of violation symptoms is scattered in textual artifacts during development. RQ1.1 focuses on finding out the classifier with the best performance to distinguish violation symptoms from natural language by comparing several popular ML/DL algorithms. RQ1.2 aims at employing several well-known word embedding models to generate word vectors as features of labeled data, in order to choose the word embedding models that can help to improve the accuracy of classifiers. The word embedding models are a common ground for both ML/DL-based classifiers in our study. Finally, with RQ1.3, we seek to compare the differences in dimension size among different word embedding models, as it is one of the most important hyperparameters affecting the model performance\cite{hutter2019aml}. Specifically, RQ1.3 endeavors to shed light on the extent to which the selection of dimensionality impacts the performance of the classifiers in automatically identifying violation symptoms. 


\begin{tcolorbox}[colback = white!25!white, colframe = black!75!black]
\textbf{RQ2: To what extent, a voting strategy can help to improve the performance of identifying violation symptoms from code review comments?}
\end{tcolorbox}

\textbf{Rationale}: After training individual classifiers based on ML/DL algorithms, we want to know whether, and to what extent, an ensemble classifier composed of trained classifiers can help to improve the performance of identifying violation symptoms from code review comments. To answer this RQ, we compared the performance of individual classifiers with that of a combined classifier. More specifically, we employed a voting strategy to integrate the prediction values of each classifier and generated results after voting among individual classifiers. Answering this RQ can help to explore the possibility of further improving the performance of automated identification of violation symptoms.

\begin{tcolorbox}[colback = white!25!white, colframe = black!75!black]
{\textbf{RQ3: Do practitioners find automated identification of violation symptoms useful in practice?}}
\end{tcolorbox}

\textbf{Rationale}: If the results generated by the classifiers based on ML/DL algorithms are acceptable regarding the effectiveness of identified violation symptoms, automatically identifying violation symptoms from textual artifacts would be helpful in practice. Such automated techniques could facilitate tasks such as locating architecture violations and inspiring further investigation and remediation of similar architectural issues. To answer this RQ, we conducted a survey and semi-structured interviews to investigate how practitioners (both developers and reviewers) assess the effectiveness and applicability of the trained classifiers. \textcolor{black}{Additionally, we performed a controlled experiment to quantitatively evaluate whether providing automatically identified violation symptoms as hints can improve developers' actual performance in detecting and fixing architecture violations during code review. Together, the qualitative perceptions and quantitative outcomes provide complementary evidence for assessing the practical usefulness of our approach.}

\begin{tcolorbox}[colback = white!25!white, colframe = black!75!black]
{\textbf{RQ4: Do state‑of‑the‑art LLM-based classifiers truly outperform traditional ML/DL classifiers on the same classification tasks, and if so, to what extent do they offer performance gains?}}
\end{tcolorbox}

\textbf{Rationale}: Considering the remarkable performance of LLMs on various SE tasks \cite{hou2024large, li2025ChatGPT}, we aim to further validate our conjecture whether LLM-based classifiers truly outperform traditional ML/DL-based classifiers regarding the automated identification of violation symptoms in code review comments. To answer RQ4, we developed several LLM-based classifiers using different prompt strategies to perform the same classification tasks in RQ1 and RQ2, enabling a fair comparison of performance gains.


\subsection{Data Collection}\label{sec:Data Collection}


We relied on the dataset that we mined and manually identified in our previous study \cite{Li2023wvs}. As shown in Figure \ref{F:Overview} (see \textbf{Dataset Creation}), we followed five steps to establish a specific dataset on architecture violations mined from code review comments. More specifically, we mined code review comments of four OSS projects in the OpenStack (i.e., Nova and Neutron) and Qt (i.e., Qt Base and Qt Creator) communities between 2014 and 2020. We mined the data through the REST API\footnote{\url{https://gerrit-review.googlesource.com/Documentation/rest-api.html}} supported by Gerrit. We first conducted experiment-based keyword refinement to improve our keyword search, as our previous experience indicated that a keyword-based approach could only capture very small proportions of code review comments containing violation symptoms~\cite{Li2023wvs}. The dataset collection process comprised three main steps: keywords searching, semantically-similar words querying (identifying potentially semantic-related data), and co-occurrence matrix-based querying (capturing potentially co-occurred and associated data). The whole manual labeling and analysis process took the researchers around one and a half months. Subsequently, we proceeded with manual identification, and pilot and formal data labeling of review comments related to architecture violations. In total, we manually identified and labeled 606 code review comments pertaining to architecture violations. Code review comments were created for patches, and comments were made by code reviewers and developers. In terms of the data labeling process, four researchers first conducted a pilot data labeling by randomly selecting 50 review comments, to reach a consensus (the inter-rater agreement between the researchers measured in the Cohen's Kappa coefficient value \cite{Cohen1960cans} was 0.857) and ensure that we have the same understanding of violation symptoms. Any disagreements were discussed between the four researchers to reach a consensus. After that, the four researchers started the formal data labeling by dividing the retrieved 21,583 code review comments into four parts (each researcher manually labeled around 5,400 comments). After the formal data labeling, the first author checked the data labeling results from the other three researchers. Meanwhile, to mitigate potential bias, the four researchers discussed all the conflicts in the labeling results until we reached an agreement. To summarize, all data labeling results were checked by at least two researchers and conflicts were discussed among four researchers. Subsequently, all the collected code review comments regarding architecture violations (i.e., 606 code review comments) from the four projects were combined into an integrated dataset used in this work. The dataset is available in our replication package~\cite{onlinepackage}.

\subsection{Phase 1 - Automated Classification of Violation Symptoms based on ML/DL Techniques}\label{sec:Automatic Classification}
\begin{figure*}[t]
	\centering
	\includegraphics[width=0.95\linewidth]{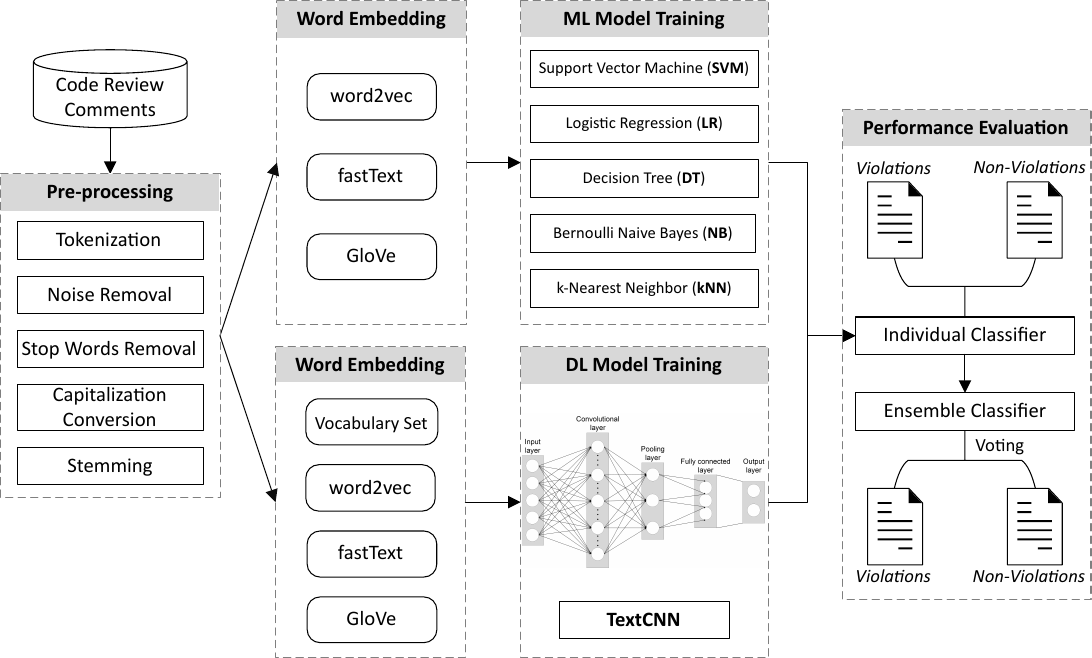}
	\caption{The framework of the experimental setup for classifying violation symptoms of architecture erosion from code review comments}\label{F:Framework}
\end{figure*}

In this section, we describe the experimental setup that we followed to evaluate the performance of the trained classifiers (see Figure~\ref{F:Framework}). The steps in this setup correspond to the steps of Phase 1 in Figure~\ref{F:Overview}. First, we pre-processed the collected code review comments to generate structured word sets, and then utilized word embedding techniques to encode the words and generate vectors for presenting the words in review comments. Subsequently, the generated vectors acted as the input to train the classifiers based on ML/DL models, in order to learn to classify review comments as violations and non-violations. The experimental environment is a computer equipped with Intel(R) Core(TM) i7-10510U CPU and 16GB RAM, running Windows 11 (64-bit). Strictly speaking, DL is a subset of ML. However, to clearly differentiate between traditional ML and DL approaches, we provide a detailed explanation of these steps for ML/DL models separately in the following subsections. More details of the experiments (such as hyperparameters and experimental environment) can be found in the replication package \cite{onlinepackage}.

\subsubsection{Machine Learning Models}\label{sec:MLM}
Before training ML models, we need to pre-process our collected data, which involves transforming raw data into a structural format for model training. The \textbf{pre-processing} includes five steps listed as follows:

\textit{Step 1. Tokenization}: This process breaks a stream of text into words, punctuation, and other meaningful elements called \textit{tokens}. In this work, we tokenized the review comments by splitting the text into its constituent set of words.

\textit{Step 2. Noise Removal}: Developers usually use plain text when they discuss violation symptoms during code review, which contain various noise data such as punctuation, numbers, and special characters (e.g., ``$\backslash$'', ``*''). Noise data usually does not contain valuable semantic information, and we therefore removed it.

\textit{Step 3. Stop words Removal}: Stop words are frequently observed in text, for the purpose of distinguishing between texts, such as ``the'', ``are'', and ``is'', but they are typically devoid of valuable semantic content. Thus, we removed stop words in our dataset by using the Natural Language Toolkit (NLTK) \cite{Bird2010nlp}.

\textit{Step 4. Capitalization Conversion}: In order to ensure the consistency of word form and to avoid redundant counts of words, all textual data was converted to lowercase.

\textit{Step 5. Stemming}: Stemming can be used to convert the words to their base forms. For instance, ``architecture'' and ``architectural'' have the same stem (i.e., token) ``architectu''. We conducted a stemming process (using Snowball Stemmer from the NLTK toolkit \cite{Bird2010nlp}) to obtain the stem of each token, since it is a more comprehensive and aggressive version of the Porter Stemmer \cite{SnowballStemmer}.

\textbf{Word Embedding}. Feature extraction techniques are widely utilized to transform unstructured text sequences into structured feature spaces (e.g., matrices, vectors, or encodings). In this work, we employ word embedding techniques to extract textual features for generating interpretable input for machine learning algorithms. Word embedding is a prominent method employed in natural language processing to represent words through dense vectors. In this work, we utilized the three prevalent and popular techniques (i.e., word2vec, fastText, and GloVe) to generate word embeddings. These techniques have gained widespread use in machine learning and deep learning-based tasks, owing to their efficacy in capturing the semantic and syntactic relationships among words.

\begin{itemize}
\item \textit{word2vec}: word2vec was developed in 2013 \cite{Mikolov2013word2vec} and it takes words as input and provides a word embedding matrix that contains high-dimensional unique vectors to represent each distinct word in given training corpora. word2vec can generate word embeddings through skip-gram and Continuous Bag-of-Words (CBOW) models. In this study, we adopted the pre-trained word2vec model proposed by Efstathiou \textit{et al}. \cite{Efstathiou2018weso}. This model was trained based on 
the corpus comprising over 15 GB of textual data collected from Stack Overflow posts, which contains a plethora of textual information in the SE domain.

\item \textit{fastText}: fastText is an open-source and lightweight library proposed by Meta AI Research lab in 2017 \cite{Bojanowski2017fastText} and it is capable of learning text representations and training text classifiers. Based on the fastText model, Meta published pre-trained word vectors with 300 dimensions for 157 languages constructed on Common Crawl and Wikipedia\footnote{\url{https://fasttext.cc/docs/en/crawl-vectors.html}}. In this study, we employed the pre-trained fastText model, which can be transformed into 100 and 200-dimensional models, to control variables and avoid the impact of dimensionality on classifiers' performance.

\item \textit{GloVe}: Global Vectors for Word Representation (GloVe) is another common word embedding technique \cite{Pennington2014GloVe}. A GloVe model is inspired by aggregated word co-occurrence statistics that may encode global information for words. Compared with word2vec, GloVe avoids the weakness regarding using local word co-occurrence information. In this study, we adopted the GloVe pre-trained word vectors on Twitter data (two billion tweets) with 200 dimensions.
\end{itemize}

\textbf{Model Training}. Given the binary classification task in this study, we selected five common ML algorithms that are widely used in classification tasks~\cite{Han2022dm}, including Support Vector Machine (SVM), Logistic Regression (LR), Decision Tree (DT), Bernoulli Naive Bayes (NB), and k-Nearest Neighbor (kNN). These algorithms are also commonly applied in areas of SE studies \cite{Yang2022pmse}.


For the ML classifiers, we performed a \textit{grid search} to tune the hyperparameters of each model and picked the best-performing ML models with specific hyperparameters on the validation set. Grid search, an exhaustive search technique, explores every possible combination within a predefined search space to identify the optimal configuration. Specifically, it is widely employed in the realms of machine learning and optimization and it involves a systematic evaluation that examines all of the combinations of a set of candidate settings to find the best hyperparameter combination \cite{hutter2019aml}. 

\subsubsection{Deep Learning Models}\label{sec:DLM}

Similarly to the ML models, for the DL models we pre-processed the collected data before training DL classifiers, using the same \textbf{pre-processing} steps described in Section~\ref{sec:MLM}. For the pre-processing of DL classifiers, we needed to ensure the equal size of each paragraph length; therefore, we utilized the zero-padding strategy before dimension transformation to equalize all input text (i.e., the maximum length is set to 2000 words).

\textbf{Model Training}. In recent years, DL-based models are increasingly used in various text classification tasks and Convolutional Neural Network (CNN) has become one of the most popular model architectures for text classification \cite{Minaee2021dltc}. We selected the TextCNN model to train the DL-based classifiers, since TextCNN, as an adapted CNN-based model, is reported to improve the state of the art on text classification \cite{Minaee2021dltc}. TextCNN is a CNN-based architecture designed for processing textual data, and was introduced by Yoon Kim in 2014 \cite{Kim2014TextCNN}. It can be used for a variety of Natural Language Processing (NLP) tasks, such as text classification and sentiment analysis. Compared with traditional ML models regarding text processing, the key advantage of TextCNN is its ability to automatically learn features from raw textual data, without relying on manual feature engineering.

We built the TextCNN for identifying violations and non-violations in code review comments through several processing layers, including an input layer, a convolutional layer, a pooling layer, a fully connected layer, and an output layer. Specifically, the TextCNN model applies one layer of convolutions over n-dimensional word embeddings to capture n-gram features at different scales. The convolutional layer learns to detect local patterns in the input text. The resulting feature maps are then fed through one max-pooling layer, which extracts the most salient features from each map and concatenates them to produce a fixed-length vector representation of the input text. This vector is then passed through one or more fully connected layers to produce the final output, i.e., the predicted labels for the input review comments.

TextCNN is a convolutional neural network model that is commonly used for text classification tasks. This model takes a sentence as input and applies multiple convolutional filters to extract important features from the text. A TextCNN classifier based on the vocabulary set of a dataset uses a one-hot encoding scheme to represent words and generate the word embedding, while a TextCNN classifier can also utilize pre-trained word embedding models to capture the semantic information of words. Additionally, the TextCNN model requires a proper setting to construct neural network models, such as hyperparameters and word embedding parameters. In this work, we set the embedding dimension to 200, the learning rate to 0.001, and the dropout to 0.25 to handle the overfitting, while the filter region size is (3, 4, 5). Each DL model's batch size is set to 16, and the network is trained for 100 epochs with the early stopping strategy with the purpose of avoiding overfitting. We configured the patience parameter to 8, indicating that the training process would terminate after 8 consecutive epochs with no observed improvement in model performance. These approaches and settings are common practices in work related to ML/DL and text classification, aimed at preventing overfitting and optimizing the model's convergence \cite{Watson2022SLRDL}.

\subsubsection{Performance Evaluation}\label{sec:Evaluation}
For the performance evaluation of ML/DL classifiers, we adopted four commonly used metrics, namely, Precision, Recall, F1-score (i.e., the harmonic mean of Precision and Recall), and Accuracy.

\textbf{Precision} presents the proportion of correctly classified violation symptom comments to all comments classified as violation symptoms. \textbf{Recall} represents the proportion of all violation symptom comments that are correctly identified. \textbf{F1-score} is the harmonic mean between precision and recall, assessing if an increase in one compensates for a decrease in the other. \textbf{Accuracy} is the percentage of correct classifications by a trained learning model. The above metrics are defined and calculated by the following equations.

\begin{equation}\label{equ:Precision}
\rm Precision = \frac{TP}{TP+FP}
\end{equation}

\begin{equation}\label{equ:Recall}
\rm Recall = \frac{TP}{TP+FN}
\end{equation}

\begin{equation}\label{equ:F1}
\rm F_1-score = 2\times \frac{Precision\times Recall}{Presion+Recall}
\end{equation}

\begin{equation}\label{equ:Accuracy}
\rm Accuracy = \frac{TP+TN}{TP+TN+FP+FN}
\end{equation}

Here, we need to use four statistics to calculate the metrics, namely, False Positives (FP) represent the number of non-violation comments that are classified as violation comments; False Negatives (FN) represent the number of violation comments that are classified as non-violation comments; True Positives (TP) represent the number of violation comment that are classified as violation comments; True Negatives (TN) represent the number of non-violation comments that are classified as non-violation comments.

To avoid the class imbalance problem \cite{he2009imbalance, Buda2018SMSCNN}, a data distribution problem that significantly impacts classifiers' performance, we constructed a balanced dataset containing equal numbers of review comments labeled as ``\textit{violation}'' and ``\textit{non-violation}'', respectively. Specifically, from our corpus of code review comments, we randomly selected 606 ``\textit{non-violation}'' comments to balance the dataset. \textcolor{black}{This approach was deliberately chosen because: (1) it prevents classifiers from being biased toward the majority class; (2) it ensures equal representation of both classes throughout all experimental phases; and (3) it aligns with widely adopted strategies for handling imbalanced classification problems in software engineering contexts \cite{he2009imbalance, Buda2018SMSCNN}.} After creating this balanced dataset, we evaluated the classifiers on a held-out test set. For ML models, we additionally used 10-fold stratified cross-validation on the balanced data for fold-level validation and statistical comparison. For DL models, we used a train/validation split on the balanced data and a separate held-out test set for final reporting.

\subsubsection{Ensemble Classifier}\label{sec:Ensemble Classifier}
To identify architecture violations from textual artifacts (i.e., code review comments), we trained several ML/DL-based classifiers to predict the labels of new review comments. However, the trained classifiers have varying performances in predicting the labels of target artifacts. In RQ2, we asked whether composing the individual classifiers into an ensemble classifier can help improve the overall performance. This composition can be done by utilizing the ensemble learning technique, i,e., majority voting strategy \cite{Dietterich2000emml}. Specifically, a combined classifier of several individual classifiers is generated with the purpose of building an improved classifier, which predicts an output (class) based on the highest probability of chosen class as the output \cite{Dietterich2000emml}. This aggregating criterion is like an election process and can produce combined voting results from each classifier's output.

The voting criteria include \textit{Hard Voting} (voting is calculated on the predicted output class) and \textit{Soft Voting} (voting is calculated on the predicted probability of the output class). Considering the computational and memory requirements, we employed the \textit{Hard Voting} strategy (a cost-efficient majority voting strategy) in this work. In other words, the final predicted results depend on the majority output (i.e., labels) of the individual classifiers.

\subsection{Phase 2 - Validation: Survey and Interview}\label{sec:Survey}

To answer RQ3, we designed and executed a survey and semi-structured interviews by following the guidelines proposed by Kitchenham and Pfleeger's personal opinion surveys \cite{gaese2008CH3}; in terms of the survey, we used an anonymous, web-based questionnaire to increase the response rate.

\textbf{(1) Survey}. Surveys can be conducted in different ways, such as online questionnaires, phone surveys \cite{Lethbridge2005sse}. Our survey is an online questionnaire of the \textit{cross-sectional} type, which allows us to collect data and compare differences between participants at a single point in time. We adopted self-administered questionnaires to collect responses, since (a) respondents can answer the survey at their convenience (reducing their time investment and increasing the response rate); (b) questionnaires can reduce research bias since the researcher does not have any contact with the respondents when they fill out the survey; and (c) respondents enjoy better privacy than interviews.

We invited all the developers involved in our collected dataset through customized emails (see the replication package \cite{onlinepackage}) attaching the links to their code review comments that contain violation symptoms and are identified by our approach (i.e., the best-performing classifier), as well as the survey form (also see the replication package \cite{onlinepackage}). Our survey was constructed based on four parts: \textit{Problem Statement}, \textit{Our approach}, \textit{Survey Goal}, and \textit{Survey Questions}. \textit{Problem Statement} introduces the research background regarding what is architecture erosion and its violation symptoms, as well as why it is important to investigate violation symptoms. \textit{Our approach} describes the approach that we developed to automatically identify violation symptoms with high performance from a large number of code reviews as well as two examples of violation symptoms of architecture erosion detected by our approach from code review comments. In \textit{Survey Goal}, we explain the purpose of this survey. In \textit{Survey Questions}, we ask five general questions and present six statements concerning the usefulness of our approach. Note that, before the formal survey, we did a pilot survey (receiving 6 responses) with the purpose of revising and improving our survey questions and examples. 
    
The five general survey questions include demographic information (e.g., their countries and work experience). The six statements are assessed using a five-point Likert scale (Strong Agree, Agree, Neutral, Disagree, and Strongly Disagree) with one option (``\textit{I Don't Know}''). This scale enables participants to rate the extent that they agree or disagree with the statements. At the end of the survey, we ask an open-ended question to let the respondents freely provide comments and suggestions about our survey. To formulate the statements, we took inspiration from the survey questions of two survey studies \cite{Nasab2021ais, Khalajzadeh2022sda} that evaluate the usefulness of automatic classifiers based on ML/DL techniques in the SE community. We chose to use declarative statements in our validation survey since declarative statements are deemed suitable for exploratory studies to convey simple and clear judgments; in contrast, descriptive statements are more suitable in explanatory studies to provide detailed judgments in various aspects. The statements are as follows:

\begin{itemize}
    \item Statement 1. ``\textit{Violation symptoms in code review comments identified by the approach, represent potential architectural violations}''.
    \item Statement 2. ``\textit{Violation symptoms in code review comments identified by the approach can be used to improve the code quality}''.
    \item Statement 3. ``\textit{I (as a practitioner) can find potentially constructive and useful architectural information from violation symptoms in code review comments identified by the approach}''.
    \item Statement 4. ``\textit{Violation symptoms in code review comments identified by the approach can help us identify violation-related issues faster than if we do this manually}''.
    \item Statement 5. ``\textit{Violation symptoms in code review comments identified by the approach might help us locate, identify, and prioritize potential architectural violations in our systems}''.
    \item Statement 6. ``\textit{Violation symptoms in code review comments identified by the approach can provide us with input to find other violations of similar nature or otherwise}''.
\end{itemize}

\textbf{(2) Interview}.
\noindent As a supplement to the survey, we also invited the same practitioners to participate in semi-structured interviews for more in-depth discussions. Specifically, the interview focused on the developers' opinions on classifier effectiveness, as well as the role of such automated techniques in the code review process during software development. The semi-structured interviews were conducted by the following three open-ended Interview Questions (IQs):

\begin{itemize}
    \item \textit{\textbf{IQ1}: How do you think the integration of an automated system for identifying violation symptoms of architecture erosion from code review comments would fit within the code review workflow? This system would analyze reviewer comments in real-time to flag potential or overlooked violation symptoms. By doing so, it would prevent the incorporation of code changes that violate system architecture, thereby contributing to a more robust and reliable development process. What are your thoughts on the potential benefits and challenges of such integration?} This question aims to assess the practical feasibility and acceptance of integrating automated violation symptoms identification into existing code review workflows.
    \item \textit{\textbf{IQ2}: Based on your experience, what strategies would you recommend to effectively use automated identification of violation symptoms in the code review process? How can this technology be utilized to maximize its benefits and efficiency in identifying and addressing architectural issues during code review?} This question seeks to gather expert insights on optimal deployment strategies and best practices for automated violation symptoms identification.
    \item \textit{\textbf{IQ3}: In addition to code review scenarios, can you envision other contexts or scenarios where our approach for automated identification of violation symptoms could be beneficial in software development practice? What potential applications or integrations do you see as valuable in enhancing code quality and development workflows?} This question explores the broader applicability and extensibility of automated violation symptoms identification beyond code review contexts.
\end{itemize}

\textbf{Participants}. 
Our collected data from Gerrit contains the developers' information (i.e., names and email addresses); this information is not included in the replication package for privacy reasons. For the \textit{survey}, we invited the potential participants to fill out our survey by sending emails to all developers whose review comments are included in our collected dataset. During the pilot survey, we got responses from 6 developers. The pilot survey helped us to evaluate and refine our survey design. For example, we added more precise examples of violation symptoms from code review comments in our survey in order to enhance practitioners' understanding, and we refined the content of the customized emails based on the received pilot feedback. Then, we invited the remaining developers involved in the discussions on violation symptoms to fill in the survey. Finally, for the semi-structured \textit{interviews}, we invited the above-mentioned participants to give feedback on our trained classifiers and the role of such automated techniques during code review. We first conducted two pilot interviews and found that, the interview questions are suitable. Subsequently, we interviewed the remaining participants who accepted our invitation, resulting in a total of six interviews.

\textbf{Sample and Population}. 
The target population is limited to developers whose code review comments are identified by our approach. We used a Non-Probabilistic Sampling method in this survey, namely, Convenience Sampling. As a type of non-probability sampling method, Convenience Sampling can help conveniently reach a suitable number of developers who are willing to participate in our survey. The reason we chose this sampling method is that we could not intentionally choose a sample due to the limitations of geographical constraints and scheduling conflicts.

\textbf{Data Analysis}. 
For the survey questions, we applied descriptive statistics to analyze the responses to the demographic and Likert scale questions. For the open-ended interview questions, interviews were recorded and transcribed; subsequently, we used the Open Coding and Constant Comparison method \cite{Adolph2011CC} (a fundamental qualitative analysis technique widely used in empirical software engineering research) to analyze and categorize the qualitative responses.


{\color{black}
\subsection{Phase 2 - Validation: Controlled Experiments}\label{sec:CtrExp}

To further evaluate the effectiveness of our approach in realistic code review scenarios, we designed a controlled experiment to assess whether providing developers with classifier-identified violation symptom hints improves their ability to detect and address architecture violations during code review.

We plan to recruit 30 developers with diverse industrial and academic backgrounds from multiple countries. To ensure an adequate level of expertise, participants must satisfy the following eligibility criteria: (1) at least one year of software development experience, (2) familiarity with modern code review practices (e.g., Gerrit, GitHub Pull Requests), and (3) a basic knowledge of software architecture concepts. To ensure comparable baselines between experimental conditions, participants will be grouped by years of experience and randomly assigned to two groups: a \textit{Control Group} and an \textit{Experimental Group}, with at least 15 participants per group. Demographic information and professional background data will be collected via a pre-experiment questionnaire to verify group comparability.

\begin{itemize}
    \item \textit{Control Group}: Participants perform manual code review based solely on the original code snippets and their associated review comments, without any additional support.
    \item \textit{Experimental Group}: Participants perform the same tasks, but are additionally provided with hints generated by classifiers, highlighting review comments that potentially indicate architecture violation symptoms.
\end{itemize}

To protect confidentiality, all code snippets are anonymized and hosted on GitHub Gist. Participants access the tasks through a web-based survey (Google Forms), with strict protocols ensuring that no personally identifiable information or original author details were disclosed. Before the formal study, we conduct a pilot survey to refine the questionnaire and validate question clarity. Before completing the tasks, all participants reviewed a tutorial detailing the definitions of architecture violation symptoms, the study goal, and experimental guidelines (e.g., independent completion and a strict prohibition of AI assistance).

\begin{table}[h]
    \centering
    \footnotesize
    \color{black}  
    \renewcommand{\arraystretch}{1.0} 
    \caption{\textcolor{black}{Variables involved in the controlled experiment}}
    \label{tab:variables}
    \begin{tabular}{lp{10cm}}
    \toprule
    \textbf{Type} & \textbf{Variable} \\
    \midrule
    \textbf{Dependent variable} & Proportion of target violations correctly identified. \\
    \textbf{Independent variable} & Presence of classifier-identified violation symptom hints in code snippets (with hints vs. without hints) \\
    \textbf{Control variables} & Developers, tasks, code snippets, experience distribution \\
    \bottomrule
    \end{tabular}
\end{table}

Table~\ref{tab:variables} summarizes the variables involved in the controlled experiment. The independent variable is the presence of classifier-identified violation symptom hints, and the dependent variable captures participants' effectiveness in identifying and addressing violations. Control variables include the task set, code snippets, and participant background distribution. The dependent variable (i.e., proportion of target violations correctly identified) is evaluated by comparing their responses against ground-truth refactoring solutions derived from expert-reviewed code review histories. These reference solutions correspond to accepted refactorings that were previously reviewed and integrated, thereby providing an objective basis for evaluation. Specifically, participant performance is assessed along two criteria: (1) \textit{Violation Identification}: whether the participant correctly identifies the presence of architecture violations. (2) \textit{Fix Quality}: whether the proposed solution correctly resolves the violation and covers all relevant code locations. Based on the criteria, each response is categorized into one of the following three outcomes:
\begin{itemize}
    \item \textbf{Category 1 (Identified \& Fixed):} The participant correctly identifies the violation symptom and proposes a semantically and structurally valid refactoring that aligns with the ground-truth solution.
    \item \textbf{Category 2 (Identified but Incorrect Fix):} The participant correctly identifies the violation symptom, but the proposed fix is incomplete, incorrect, or deviates significantly from the accepted refactoring pattern.
    \item \textbf{Category 3 (Not Identified):} The participant fails to recognize the violation symptom or attributes it to non-architectural concerns (e.g., coding style, performance concerns).
\end{itemize}

}

\subsection{Phase 3 - Comparison: LLMs vs. ML/DL Techniques}\label{sec:Comparison}

Unlike traditional ML/DL approaches, LLMs eliminate the need for task-specific training from scratch. Therefore, we leveraged pre-trained and open-source LLMs using techniques like prompt engineering to perform the same classification tasks in Phases 1 and 2. 
\textcolor{black}{Our evaluation strategy included: (1) Conducting preliminary experiments to identify high-performing prompts; (2) Adopting the best-performing prompt for formal evaluations; (3) Employing both zero-shot and few-shot learning approaches (with 0, 2, 4, 6, and 10 examples); (4) Assessing performance variations across three representative LLMs: \textit{GPT-4o}, \textit{Qwen-3}, and \textit{DeepSeek-R1}.}

\textbf{LLM Selection}. We selected three representative and high-performance LLMs based on their technical diversity, capabilities, and availability.

\begin{itemize}
\item \textit{GPT-4o}: A proprietary multimodal LLM developed by OpenAI \cite{GPT-4o}. Although its architecture is undisclosed, industry speculation suggests that it involves approximately 200 billion parameters. GPT-4o demonstrates state-of-the-art performance in code generation (HumanEval) and mathematical reasoning (MATH), and supports real-time multimodal interactions with an average response latency of 320 ms. Primary applications include enterprise AI, educational systems, and multimedia content generation.

\item \textit{Qwen-3}: An open-weight LLM series developed by Alibaba's Qwen team. In this study, we used Qwen3-235B-A22B-Instruct, a Mixture-of-Experts model from the Qwen3 family with 235 billion total parameters and 22 billion activated parameters \cite{qwen3techreport}. Its multilingual and code-oriented capabilities make it suitable for evaluating LLM-based classifiers on SE tasks.

\item \textit{DeepSeek-R1}: An open-source 671-billion-parameter model \cite{deepseekai2025} using a Mixture of Experts (MoE) design with only 8 billion active parameters per inference. This architecture offers a 60\% reduction in training cost and over 2.3× inference throughput compared to dense models like LLaMA 3-70B. Its efficient MoE architecture and strong reasoning capabilities make it particularly suitable for evaluating performance across diverse SE tasks while maintaining computational efficiency.
\end{itemize}

\textbf{Zero-Shot and Few-Shot Learning.} LLMs inherently support zero-shot and few-shot learning due to their robust generalization and natural language understanding capabilities.

\begin{itemize}
    \item \textit{Zero-shot learning} involves presenting no labeled examples. Instead, the model receives only a conceptual description (e.g., ``violation symptoms'') and must perform classification based on this abstract guidance alone.
    \item \textit{Few-shot learning} supplements the conceptual definition with a small set of labeled examples. The model uses these demonstrations to infer classification strategies, blending prior knowledge with task-specific cues.
\end{itemize}

To investigate performance variability, we tested four different few-shot settings (2, 4, 6, and 10 examples), ensuring class balance by selecting even numbers of samples.

\textbf{Prompt Engineering.} Prompt engineering is critical in guiding LLMs toward accurate and reliable outputs. \textcolor{black}{Since optimal prompt design can be challenging due to the large design space, we adopted an empirical evaluation approach: (1) Generated multiple prompt candidates for both zero-shot and few-shot learning; (2) Evaluated their performance in preliminary trials; (3) Selected the best-performing prompt configuration. Our final prompt structure defined the LLM's assumed role, introduced the concept of violation symptoms, provided demonstrations (in few-shot settings), and specified the required output format. For reproducibility, all prompt templates used in our experiments are provided in our replication package under \texttt{scripts/classifiers/prompt\_templates/}.}

\section{Results}\label{sec:Results}
\subsection{RQ1: Identifying Violation Symptoms}\label{sec:RQ1_results}

\subsubsection{\textcolor{black}{Performance of Different Classifiers}}\label{sec:Model_performance}
To answer RQ1, we trained five ML classifiers and one DL classifier based on three word embedding techniques (i.e., word2vec, fastText, and GloVe). Moreover, we evaluated the performance of these classifiers and compared the differences between their performance. In general, we use the F1-score as the overall evaluation of classifiers' performance.

\begin{table}[htbp]
\centering
\setlength\tabcolsep{1.5pt}
\renewcommand{\arraystretch}{1.4}
\caption{\textcolor{black}{Performance comparison among ML-based classifiers with different word embeddings}}\label{T: Performance of ML classifiers}
\scalebox{1.03}{\scriptsize 
\begin{threeparttable}
\begin{tabular}{cccccccccccccccccccc}\toprule
\multirow{2}{*}{\textbf{Classifier}} & \multicolumn{4}{c}{\textbf{Precision}}&  & \multicolumn{4}{c}{\textbf{Recall}}&  & \multicolumn{4}{c}{\textbf{F1-score}}&  &\multicolumn{4}{c}{\textbf{Accuracy}}\\ \cline{2-5} \cline{7-10} \cline{12-15} \cline{17-20}
 & w2v & FT & GloVe & Average & & w2v & FT & GloVe & Average & & w2v & FT & GloVe & Average & & w2v & FT & GloVe & Average\\\hline
    
\textbf{SVM} & \underline{0.789} & 0.728 & 0.759 & 0.759 && \underline{\textbf{0.828}} & \textbf{0.746} & \underline{\textbf{0.828}} & \textbf{0.801} && \underline{\textbf{0.808}} & 0.737 & \textbf{0.792} & \textbf{0.779} && \underline{\textbf{0.803}} & 0.734 & \textbf{0.783} & \textbf{0.773}\\
\textbf{LR}  & \underline{0.832} & 0.777 & \textbf{0.798} & 0.802 && \underline{0.730} & 0.713 & 0.713 & 0.719 && \underline{0.777} & \textbf{0.744} & 0.753 & 0.758 && \underline{0.791} & \textbf{0.754} & 0.766 & 0.770\\
\textbf{NB}  & 0.728 & \underline{0.743} & 0.717 & 0.729 && 0.615 & 0.615 & \underline{0.623} & 0.618 && 0.667 & \underline{0.673} & 0.667 & 0.669 && 0.693 & \underline{0.701} & 0.689 & 0.694 \\
\textbf{DT}  & \underline{0.667} & 0.657 & 0.643 & 0.656 && 0.689 & \underline{0.721} & 0.664 & 0.691 && 0.677 & \underline{0.688} & 0.653 & 0.673 && \underline{0.672} & \underline{0.672} & 0.648 & 0.664 \\
\textbf{kNN} & \underline{\textbf{0.863}} & \textbf{0.788} & 0.797 & \textbf{0.816} && 0.516 & \underline{0.549} & 0.451 & 0.505 && 0.646 & \underline{0.647} & 0.576 & 0.623 && \underline{0.717} & 0.701 & 0.668 & 0.695 \\\hline
\textbf{Average} & 0.776 & 0.739 & 0.743 & 0.753 && 0.676 & 0.669 & 0.656 & 0.667 && 0.715 & 0.698 & 0.688 & 0.700 && 0.735 & 0.712 & 0.711 & 0.719 \\\bottomrule 
\end{tabular}
\end{threeparttable}}
\end{table}

\begin{table}[htbp]
\centering
\footnotesize
\setlength\tabcolsep{1.5pt}
\renewcommand{\arraystretch}{1.2}
\caption{Performance comparison among DL-based classifiers}\label{T: Performance of DL classifiers}
\begin{tabular}[width=\textwidth]{ccccc}\toprule
\textbf{Classifier} & \textbf{Precision} & \textbf{Recall} & \textbf{F1-score}& \textbf{Accuracy}\\\hline
\textbf{TextCNN\_voc}     & 0.475 & 0.541 & 0.469 & 0.471\\
\textbf{TextCNN\_w2v}      & 0.584 & 0.680 & 0.596 & 0.598\\
\textbf{TextCNN\_FT} & \textbf{0.728} & 0.549 & \textbf{0.667} & \textbf{0.672} \\
\textbf{TextCNN\_GloVe}    & 0.507 & \textbf{0.885} & 0.434 & 0.512\\\hline
\textbf{Average} & 0.574 & 0.664 & 0.542 & 0.563\\\bottomrule
\end{tabular}
\end{table}

\begin{table*}[htbp]
\centering
\setlength\tabcolsep{1.8pt}
\renewcommand{\arraystretch}{1.4}
\caption{Performance comparison of the classifiers based on the fastText model with different dimension size}\label{T: Dimensionality}
\scalebox{1.08}{\scriptsize 
\begin{tabular}{cccccccccccccccc}
\toprule
\multirow{2}{*}{\textbf{Classifier}} & \multicolumn{3}{c}{\textbf{Precision}}&  & \multicolumn{3}{c}{\textbf{Recall}}&  & \multicolumn{3}{c}{\textbf{F1-score}}&  &\multicolumn{3}{c}{\textbf{Accuracy}}\\ \cline{2-4} \cline{6-8} \cline{10-12} \cline{14-16} & 100-dim & 200-dim & 300-dim & & 100-dim & 200-dim & 300-dim & & 100-dim & 200-dim & 300-dim & & 100-dim & 200-dim & 300-dim\\\hline
\textbf{SVM} & \underline{0.745} & 0.728 & 0.657 &  & 0.672 & 0.746 & \underline{0.754} & & 0.707 & \underline{0.737} & 0.702 &  & 0.721 & \underline{0.734} & 0.680\\
\textbf{LR} & 0.704 & \underline{0.777} & 0.745 &  & 0.664 & \underline{0.713} & 0.648 &  & 0.684 & \underline{0.744} & 0.693 &  & 0.693 & \underline{0.754} & 0.713\\
\textbf{NB} & \underline{0.745} & 0.743 & 0.712 &  & \underline{0.623} & 0.615 & 0.607 &  & \underline{0.679} & 0.673 & 0.655 &  & \underline{0.705} & 0.701 & 0.680\\
\textbf{DT} & 0.641 & \underline{0.657} & 0.612 &  & 0.615 & \underline{0.721} & 0.648 &  & 0.628 & \underline{0.688} & 0.629 &  & 0.635 & \underline{0.672} & 0.619\\
\textbf{kNN} & \underline{0.831} & 0.788 & \underline{0.831} &  & 0.525 & \underline{0.549} & 0.443 &  & 0.643 & \underline{0.647} & 0.578 &  & \underline{0.709} & 0.701 & 0.676\\\hline
\textbf{TextCNN} & 0.577 & \underline{0.728} & 0.494 & &  \underline{0.648} & 0.549 & 0.730 &  & 0.585 & \underline{0.667} & 0.461 &  & 0.586 & \underline{0.672} & 0.492\\  \bottomrule
\end{tabular}
}
\end{table*}

\begin{sidewaystable}[hp]
\vspace*{0.66\textheight} 
\centering
\small
\setlength\tabcolsep{1.5pt}
\renewcommand{\arraystretch}{1.6}
\caption{Performance comparison of ensemble ML/DL-based classifiers}\label{T:ensemble classifiers}
\begin{minipage}[t]{\textwidth} 
\centering
\resizebox{\textwidth}{!}{
\begin{tabular}{cccccccccccccccccccccccc}
\toprule
\multirow{2}{*}{\textbf{Classifier}} & \multicolumn{5}{c}{\textbf{Precision}}&  & \multicolumn{5}{c}{\textbf{Recall}}&  & \multicolumn{5}{c}{\textbf{F1-score}}&  &\multicolumn{5}{c}{\textbf{Accuracy}}\\ \cline{2-6} \cline{8-12} \cline{14-18} \cline{20-24}
 & Mean & Best & Voting & Imp\_M & Imp\_B & & Mean & Best & Voting & Imp\_M & Imp\_B & & Mean & Best & Voting & Imp\_M & Imp\_B & & Mean & Best & Voting & Imp\_M & Imp\_B\\\hline
\textbf{SVM}     & 0.759 & 0.789 & 0.795 & 4.74\%  & 0.76\%   &  & 0.801 & 0.828 & 0.828 & 3.37\%  & 0.00\%  &  & 0.779 & 0.808 & 0.811 & 4.11\%  & 0.37\%  &  & 0.773 & 0.803 & 0.807 & 4.40\%  & 0.50\%   \\
\textbf{LR}      & 0.802 & 0.832 & 0.833 & 3.87\%  & 0.12\%   &  & 0.719 & 0.730 & 0.738 & 2.64\%  & 1.10\%  &  & 0.758 & 0.777 & 0.783 & 3.30\%  & 0.77\%  &  & 0.770 & 0.791 & 0.795 & 3.25\%  & 0.51\%   \\
\textbf{NB}      & 0.729 & 0.743 & 0.738 & 1.23\%  & -0.67\%  &  & 0.618 & 0.615 & 0.623 & 0.81\%  & 1.30\%  &  & 0.669 & 0.673 & 0.676 & 1.05\%  & 0.45\%  &  & 0.694 & 0.701 & 0.701 & 1.01\%  & 0.00\%   \\
\textbf{DT}      & 0.656 & 0.657 & 0.689 & 5.03\%  & 4.87\%   &  & 0.691 & 0.721 & 0.746 & 7.96\%  & 3.47\%  &  & 0.673 & 0.688 & 0.717 & 6.54\%  & 4.22\%  &  & 0.664 & 0.672 & 0.705 & 6.17\%  & 4.91\%   \\
\textbf{kNN}     & 0.816 & 0.863 & 0.836 & 2.45\%  & -3.13\%  &  & 0.505 & 0.516 & 0.500 & -0.99\% & -3.10\% &  & 0.623 & 0.646 & 0.626 & 0.48\%  & -3.10\% &  & 0.695 & 0.717 & 0.701 & 0.86\%  & -2.23\%  \\\hline
\textbf{TextCNN} & 0.574 & 0.728 & 0.537 & -6.45\% & -26.24\% &  & 0.664 & 0.549 & 0.713 & 7.38\%  & 29.87\% &  & 0.542 & 0.667 & 0.613 & 13.10\% & -8.10\% &  & 0.563 & 0.672 & 0.549 & -2.49\% & -18.30\%  \\
\bottomrule
\end{tabular}
}
\end{minipage}

\vspace{2.3cm} 

\begin{minipage}[t]{\textwidth} 
\centering
\small
\setlength\tabcolsep{1.5pt}
\renewcommand{\arraystretch}{1.5}
\caption{Performance comparison among ensemble ML-based classifiers with three word embeddings}\label{T:voting_embeddings}
\resizebox{\textwidth}{!}{
\begin{tabular}{cccccccccccccccccccccccc}
\toprule
\multirow{2}{*}{\textbf{Classifier}} & \multicolumn{5}{c}{\textbf{Precision}}&  & \multicolumn{5}{c}{\textbf{Recall}}&  & \multicolumn{5}{c}{\textbf{F1-score}}&  &\multicolumn{5}{c}{\textbf{Accuracy}}\\ \cline{2-6} \cline{8-12} \cline{14-18} \cline{20-24}
 & Mean & Best & Voting & Imp\_M & Imp\_B & & Mean & Best & Voting & Imp\_M & Imp\_B & & Mean & Best & Voting & Imp\_M & Imp\_B & & Mean & Best & Voting & Imp\_M & Imp\_B\\\hline
\textbf{ML\_word2vec} & 0.776 & 0.789 & 0.810 & 4.38\% & 2.66\% &  & 0.676 & 0.828 & 0.697 & 3.11\% & -15.82\% &  & 0.715 & 0.808 & 0.749 & 4.76\% & -7.30\% &  & 0.735 & 0.803 & 0.766 & 4.22\% & -4.61\% \\
\textbf{ML\_fastText} & 0.739 & 0.777 & 0.798 & 7.98\% & 2.70\% &  & 0.669 & 0.713 & 0.713 & 6.58\% & 0.00\%   &  & 0.698 & 0.744 & 0.753 & 7.88\% & 1.21\%  &  & 0.712 & 0.754 & 0.766 & 7.58\% & 1.59\%  \\
\textbf{ML\_GloVe}    & 0.743 & 0.759 & 0.768 & 3.36\% & 1.19\% &  & 0.656 & 0.828 & 0.705 & 7.47\% & -14.86\% &  & 0.688 & 0.792 & 0.735 & 6.83\% & -7.20\% &  & 0.711 & 0.783 & 0.746 & 4.92\% & -4.73\%\\\bottomrule
\end{tabular}
}
\end{minipage}
\end{sidewaystable}

\fakesection{RQ1.1: Performance of Classifiers}

For ML-based classifiers, as shown in Table~\ref{T: Performance of ML classifiers}, we trained the five types of ML classifiers with three word embedding techniques (i.e., word2vec (w2v), fastText (FT), and GloVe), leading to 15 classifiers (i.e., the five ML models times three word embeddings). We calculated the average values of the classifiers over the three pre-trained word embeddings. For each classifier based on three word embeddings, we underlined the best result of each metric (horizontal comparison); for each metric of the classifiers (precision, recall, F1-score, and accuracy), we marked the best metric result of each classifier on three word embeddings in bold (vertical comparison). From the results, we can see that the kNN classifiers have better scores of Precision than other classifiers with an average Precision of 0.816. As for the remaining metrics, the SVM classifiers achieve relatively better performance on nearly all metrics than other classifiers, with an average Recall of 0.801, F1-score of 0.779, and Accuracy of 0.773. Specifically, the SVM classifier based on \textit{word2vec} performed the best with a Precision of 0.789, a Recall of 0.828, an F1-score of 0.808, and an Accuracy of 0.803.

For DL-based classifiers, as shown in Table~\ref{T: Performance of DL classifiers}, we trained one type of classifier, i.e., TextCNN, based on the vocabulary set from our dataset and three word embedding techniques (i.e. in total four classifiers) as mentioned in Section~\ref{sec:MLM}. We can see that TextCNN\_FT performs the best compared with the other three DL models.


The ML-based classifiers demonstrated superior performance, achieving, on average, a Precision of 0.753, a Recall of 0.667, an F1-score of 0.700, and an Accuracy of 0.719. In contrast, the DL-based classifiers exhibited lower performance, with an average Precision of 0.574, Recall of 0.664, F1-score of 0.542, and Accuracy of 0.563. These findings suggest that the ML-based classifiers are more effective in accurately classifying the target variable than the DL-based classifiers in our case. 
Overall, among these classifiers, the SVM classifiers outperform other classifiers, exhibiting the best average results on all metrics, except for the Precision metric.

\fakesection{RQ1.2: Comparison of Embedding Models}

From the results in Tables \ref{T: Performance of ML classifiers} and \ref{T: Performance of DL classifiers}, we can also observe the differences between the classifiers based on three pre-trained word embedding models. As mentioned before, for each classifier based on three word embeddings, we underlined the best result of each classifier on each metric (horizontal comparison). When using the \textit{word2vec} pre-trained word embedding, most of the ML classifiers have relatively better performance on Precision and Accuracy. Most of the ML classifiers with the \textit{fastText} model have fairly better F1-scores. Besides, we found that the TextCNN classifier based on the \textit{fastText} pre-trained word embedding achieves the best performance on all metrics except for Recall, with a Precision of 0.728, a Recall of 0.549, an F1-score of 0.667, and an Accuracy of 0.672.

In addition, we can see in Table \ref{T: Performance of DL classifiers} that the overall performance of the three TextCNN classifiers with pre-trained word embeddings outperforms the TextCNN classifier based on the vocabulary set from our dataset. This is because pre-trained word embedding models contain more features due to the large textual data used for generating word vectors, and they have been proven to be invaluable for improving the performance in NLP tasks \cite{Qi2018pwe, Radford2018ilu}.

In general, the three pre-trained word embedding models have an impact on the performance of the DL-based classifiers. As shown in Table~\ref{T: Performance of DL classifiers}, when we compared their classification performance, in most cases, the classifiers with the \textit{fastText} model can achieve relatively good performance.

\fakesection{RQ1.3: Comparison of Word Embedding Model Dimensionality}

According to the results of RQ1.2, the \textit{fastText} model can help to generate better results than the other two pre-trained word embedding models. Therefore, to further explore the impact of dimension sizes of the pre-trained word embedding models, we trained the classifiers based on the \textit{fastText} model and chose three common dimension sizes: 100, 200, and 300. In Table~\ref{T: Dimensionality}, similarly as before, we underlined the best result of each classifier on each metric (horizontal comparison). We can see that five of the six ML/DL classifiers can generate better F1-scores based on the \textit{fastText} model with 200-dimensional word embedding, compared with 100 and 300-dimensional models.

{\color{black}
\subsubsection{Performance on an Imbalanced Test Set}
\label{sec:rq1_imbalanced}

The original evaluation results of the classifiers' performance in Section~\ref{sec:Model_performance} are conducted on a class-balanced test setting for controlled comparison among classifiers. However, such a setting does not reflect the naturally skewed distribution encountered in practice. In our collected corpus, only 606 comments are labeled as architecture violations among 21,583 comments, i.e., approximately 2.8\% positives (violation symptoms are treated as positive classes). As a supplementary low-prevalence robustness check under a controlled evaluation budget, we construct a fixed held-out imbalanced test set containing 150 comments, including 7 violation comments and 143 non-violation comments (4.7\% positives). This ratio is still highly imbalanced, although it is slightly higher than the corpus-level prevalence, ensuring that the fixed test set contains enough positive cases for metric calculation. All models are evaluated on this same test set, and the reported metrics are computed from the corresponding confusion matrices. 

Importantly, no resampling, re-weighting, or class balancing is applied to the held-out test set. Any over-sampling strategy is applied exclusively to the training data. Accordingly, we do not interpret class imbalance itself as problematic; rather, the concern is that standard metrics such as Accuracy may become misleading under strong class skew \cite{he2009imbalance}. Under highly imbalanced distributions, Accuracy can remain high even for trivial majority-only predictions while failing to identify rare violations effectively. Therefore, besides \textit{Accuracy}, we emphasize Precision, Recall, F1-score, and Macro-F1 for a more informative assessment.

In particular, we additionally report \textit{Macro-F1}, i.e., the unweighted mean of per-class F1-scores. Macro-F1 is widely recommended for imbalanced classification because it penalizes degenerate ``all-majority'' prediction behaviors that can still achieve high Accuracy \cite{he2009imbalance, sokolova2009systematic, powers2011evaluation}. Unlike positive-class F1-score alone, Macro-F1 additionally reflects performance on the majority non-violation class, thereby reducing the risk of overly optimistic conclusions driven solely by minority-focused optimization. In our context, Macro-F1 is especially informative because a practical reviewer should not only retrieve rare violations but also maintain a low false alarm rate on the dominant non-violation class to avoid excessive false alarms.

\textbf{Cost-sensitive learning for ML classifiers.} 
To mitigate majority-class bias, we incorporate cost-sensitive learning by assigning higher cost to false negatives than false positives \cite{elkan2001foundations}. Across ML-based classifiers, we implement cost sensitivity through (i) training-time reweighting when supported, (ii) optional resampling applied only to the training partition, and (iii) validation-time decision tuning (threshold moving or minimum expected-cost decision) \cite{zadrozny2001costs}. All cost-related choices are made using only the training/validation data and are then applied unchanged to the fixed held-out imbalanced test set, thereby avoiding information leakage. For SVM/LR/DT, class weights increase the penalty of misclassifying violation samples during training, which tends to shift the decision boundary toward higher minority-class recall. For NB, we use a minimum expected-cost rule at prediction time instead of a symmetric \texttt{argmax} over posteriors. For kNN, we apply cost-aware weighted voting followed by a validation-selected threshold; this follows established distance-weighting ideas for kNN decision rules \cite{dudani1976distance}. When enabled, \texttt{oversample\_method} (e.g., SMOTE) is applied only to the training partition~\cite{chawla2002smote}.

\providecommand{\ttsplit}[2]{\texttt{#1}\allowbreak\texttt{#2}}

\begin{table}[htbp]
\centering
\color{black}  
\caption{\textcolor{black}{Cost-sensitive hyperparameters for ML classifiers on the imbalanced setting.}}
\label{tab:cs_ml_hparams}
\scriptsize
\setlength{\tabcolsep}{3.0pt}
\renewcommand{\arraystretch}{1.10}
\begin{threeparttable}
\begin{tabular}{
  >{\raggedright\arraybackslash}p{1.15cm}
  >{\raggedright\arraybackslash}p{3.5cm}
  >{\raggedright\arraybackslash}p{5.3cm}
  >{\raggedright\arraybackslash}p{2.80cm}
}
\toprule
\textbf{Classifier} &
\textbf{Training-time cost sensitivity} &
\textbf{Test-time decision policy} &
\textbf{Resampling} \\
\midrule
\textbf{SVM} &
\ttsplit{class\_}{weight}$=\{0{:}1.0,\,1{:}10.0\}$ &
\ttsplit{decision\_}{threshold} $\tau^\star=0.70$ (validation-selected) &
\ttsplit{oversample\_}{method}$=$\texttt{smote} \\

\textbf{LR} &
\ttsplit{class\_}{weight}$=\{0{:}1.0,\,1{:}8.0\}$ &
\ttsplit{decision\_}{threshold} $\tau^\star=0.65$ (validation-selected) &
\ttsplit{oversample\_}{method}$=$\texttt{smote} \\

\textbf{NB} &
(\emph{not supported}) &
\ttsplit{cost\_}{matrix}$=[[0,1],[10,0]]$
  (rows=true class, columns=predicted class; FN cost $>$ FP cost) &
  \ttsplit{oversample\_}{method}$=$\texttt{none} \\

\textbf{DT} &
\ttsplit{class\_}{weight}$=\{0{:}1.0,\,1{:}6.0\}$ &
\ttsplit{dt\_}{threshold} $\tau^\star=0.60$ (validation-selected cutoff) &
\ttsplit{oversample\_}{method}$=$\texttt{none} \\

\textbf{kNN} &
(\emph{not supported}) &
\ttsplit{knn\_class\_}{weights} $\{0{:}1.0,\,1{:}5.0\}$ with
\ttsplit{knn\_}{threshold} $\tau^\star=0.55$ &
\ttsplit{oversample\_}{method}$=$\texttt{none} \\
\bottomrule
\end{tabular}
\end{threeparttable}
\end{table}

\begin{table}[htbp]
\centering
\color{black}  
\setlength\tabcolsep{1.2pt}
\renewcommand{\arraystretch}{1.35}
\caption{\textcolor{black}{Performance comparison among cost-sensitive ML classifiers on the imbalanced test set.}}
\label{T: Performance of ML classifiers imbalanced}
\scalebox{1.03}{\scriptsize
\begin{threeparttable}
\begin{tabular}{cccccccccccccccccccc}
\toprule
\multirow{2}{*}{\textbf{Classifier}} & \multicolumn{4}{c}{\textbf{Precision}} & & \multicolumn{4}{c}{\textbf{Recall}} & & \multicolumn{4}{c}{\textbf{F1-score}} & & \multicolumn{4}{c}{\textbf{Accuracy}} \\
\cline{2-5}\cline{7-10}\cline{12-15}\cline{17-20}
 & w2v & FT & GloVe & Average & & w2v & FT & GloVe & Average & & w2v & FT & GloVe & Average & & w2v & FT & GloVe & Average \\
\hline
\textbf{SVM} & \underline{0.545} & 0.467 & 0.500 & 0.504 && 0.857 & \underline{1.000} & \underline{1.000} & 0.952 && \underline{0.667} & 0.636 & \underline{0.667} & 0.657 && \underline{0.960} & 0.947 & 0.953 & 0.953 \\
\textbf{LR} & 0.467 & \underline{1.000} & 0.500 & 0.656 && \underline{1.000} & 0.429 & 0.714 & 0.714 && \underline{0.636} & 0.600 & 0.588 & 0.608 && 0.947 & \underline{0.973} & 0.953 & 0.958 \\
\textbf{NB} & 0.455 & \underline{0.500} & 0.438 & 0.464 && 0.714 & 0.857 & \underline{1.000} & 0.857 && 0.556 & \underline{0.632} & 0.609 & 0.599 && 0.947 & \underline{0.953} & 0.940 & 0.947 \\
\textbf{DT} & 0.556 & 0.556 & \underline{0.750} & 0.621 && \underline{0.714} & \underline{0.714} & 0.429 & 0.619 && \underline{0.625} & \underline{0.625} & 0.545 & 0.598 && 0.960 & 0.960 & \underline{0.967} & 0.962 \\
\textbf{kNN} & \underline{0.455} & \underline{0.455} & 0.417 & 0.442 && \underline{0.714} & \underline{0.714} & \underline{0.714} & 0.714 && \underline{0.556} & \underline{0.556} & 0.526 & 0.546 && \underline{0.947} & \underline{0.947} & 0.940 & 0.945 \\
\hline
\textbf{Average} & 0.496 & 0.596 & 0.521 & 0.537 && 0.800 & 0.743 & 0.771 & 0.771 && 0.608 & 0.610 & 0.587 & 0.602 && 0.952 & 0.956 & 0.951 & 0.953 \\
\bottomrule
\end{tabular}
\end{threeparttable}
}
\end{table}

Table~\ref{T: Performance of ML classifiers imbalanced} reports performance on the fixed imbalanced test set. Overall, cost-sensitive ML models show meaningful performance under this low-prevalence setting. SVM yields the best average F1-score (0.657) while keeping Accuracy above 0.95. SVM also attains perfect Recall under FT and GloVe settings, suggesting that class weighting together with validation-time threshold tuning can be a useful operating choice when missing a violation is considered costly \cite{elkan2001foundations}. Macro-F1 trends are consistent with this observation, indicating that higher violation detection performance in these configurations is not accompanied by a severe degradation on the majority class (see Table~\ref{tab:macroF1_all_imbalanced}). 

\textbf{Cost-sensitive learning for DL classifiers.}
For TextCNN, we adopt a unified cost-sensitive strategy for imbalanced settings: cost-weighted class-balanced focal loss \cite{lin2017focal,cui2019classbalanced} combined with SMOTE-based oversampling applied only to the training partition \cite{chawla2002smote}, followed by probability calibration \cite{guo2017calibration,platt1999probabilistic} and validation-time threshold selection. This combination is intended to emphasize informative minority-class patterns while reducing the risk of an overly aggressive ``always-positive'' operating regime.

\begin{table}[htbp]
\centering
\color{black}  
\setlength\tabcolsep{4.8pt}
\renewcommand{\arraystretch}{1.15}
\caption{\textcolor{black}{Cost-sensitive hyperparameters for TextCNN on the imbalanced setting.}}
\label{tab:cs_dl_hparams}
\scalebox{0.98}{\scriptsize
\begin{threeparttable}
\begin{tabular}{p{1.6cm} p{11.8cm}}
\toprule
\textbf{Component} & \textbf{Hyperparameters / choices} \\
\midrule
Loss &
Class-balanced focal loss with explicit cost weight:
$\beta=0.9999$, $\gamma=2.0$, $\omega_{+}=6.0$, $\alpha_1=0.75$ \\
Resampling &
\texttt{resample\_strategy} $=$ \texttt{SMOTE} applied only to the training partition,
target positive fraction $r_{+}=0.20$ \\
Calibration &
temperature scaling, $T$ is optimised on the validation set by minimising NLL via L‑BFGS (initial $T=1.5$, lr=0.01, max\_iter=100) \\
Decision policy &
Validation-selected threshold $\tau^\star=0.65$ to minimize expected cost;
(optional) reject interval $[\tau_{\mathrm{low}},\tau_{\mathrm{high}}]=[0.45,0.55]$ \\
\bottomrule
\end{tabular}
\end{threeparttable}
}
\end{table}

\begin{table}[htbp]
\centering
\footnotesize
\color{black}  
\setlength\tabcolsep{1.5pt}
\renewcommand{\arraystretch}{1.2}
\caption{\textcolor{black}{Performance comparison among cost-sensitive DL-based classifiers on the imbalanced test set.}}
\label{T: Performance of DL classifiers imbalanced}
\begin{tabular}{ccccc}\toprule
\textbf{Classifier} & \textbf{Precision} & \textbf{Recall} & \textbf{F1-score}& \textbf{Accuracy}\\\hline
\textbf{TextCNN\_voc}     & \underline{0.500} & 0.714 & \underline{0.588} & \underline{0.953}\\
\textbf{TextCNN\_w2v}      & 0.318 & \underline{1.000} & 0.483 & 0.900\\
\textbf{TextCNN\_FT} & 0.400 & 0.571 & 0.471 & 0.940\\
\textbf{TextCNN\_GloVe}    & \underline{0.500} & 0.571 & 0.533 & \underline{0.953}\\\hline
\textbf{Average} & 0.430 & 0.714 & 0.519 & 0.937\\\bottomrule
\end{tabular}
\end{table}

Table~\ref{T: Performance of DL classifiers imbalanced} shows that, with cost-sensitive training and validation-time decision tuning, TextCNN obtains measurable performance on the imbalanced test set, although its performance is sensitive to the embedding configuration. TextCNN\_voc achieves the best F1-score (0.588) and the best Macro-F1 among DL variants (see Table~\ref{tab:macroF1_all_imbalanced}), indicating comparatively balanced behavior across classes. In contrast, TextCNN\_w2v attains high Recall (1.000) but low Precision (0.318), suggesting that excessive emphasis on the minority class may make the model ``over-alerting'' under strong class imbalance.

\textbf{Prompt engineering for LLM-based classifiers.} Unlike ML/DL models that can be directly re-weighted during training, LLMs are used \emph{as-is} and must be adapted only through prompts at inference time.
We therefore use cost-aware prompting to reduce the tendency of the model to favor the dominant non-violation class under naturally imbalanced settings \cite{brown2020language,liu2023pretrainpromptpredict}. Specifically, the prompt (i) states that although the in-context examples are class-balanced, real-world situations are not always balanced, without revealing the class ratio of the fixed test set, (ii) provides a decision rubric aligned with our violation taxonomy, including non-violation examples, and (iii) uses a balanced set of violation and non-violation demonstrations that the model can observe representative violations from both classes despite the rarity of violations. 

\begin{table*}[htbp]
\centering
\color{black}  
\setlength\tabcolsep{3.6pt}
\renewcommand{\arraystretch}{1.35}
\caption{\textcolor{black}{Performance comparison among LLM-based classifiers on the imbalanced test set using cost-aware prompting (\textit{k}-shot in-context examples).}}
\label{tab:LLM_comparison_imbalanced}
\scalebox{0.98}{\scriptsize
\begin{threeparttable}
\begin{tabular}{cccccccccccccccc}
\toprule
\multirow{2}{*}{\textbf{Model}} & \multicolumn{6}{c}{\textbf{Precision}} & \multicolumn{3}{c}{} & \multicolumn{6}{c}{\textbf{Recall}} \\
\cline{2-7} \cline{11-16}
 & 0-shot & 2-shot & 4-shot & 6-shot & 10-shot & Average &  &  &  & 0-shot & 2-shot & 4-shot  & 6-shot & 10-shot & Average\\
\hline
\textbf{Qwen-3} & \underline{1.000} & 0.292 & 0.600 & 0.368 & 0.400 & 0.532 &  &  &  & 0.286 & \underline{1.000} & 0.429 & \underline{1.000} & 0.571 & 0.657 \\
\textbf{GPT-4o} & 0.368 & 0.429 & 0.333 & \underline{0.750} & 0.353 & 0.447 &  &  &  & \underline{1.000} & 0.857 & \underline{1.000} & 0.429 & 0.857 & 0.829 \\
\textbf{DeepSeek-R1} & 0.280 & 0.400 & \underline{0.750} & 0.312 & 0.316 & 0.412 &  &  &  & \underline{1.000} & 0.571 & 0.429 & 0.714 & 0.857 & 0.714 \\
\hline
\textbf{Average} & 0.549 & 0.374 & 0.561 & 0.477 & 0.356 & 0.463 &  &  &  & 0.762 & 0.809 & 0.619 & 0.714 & 0.762 & 0.733 \\
\toprule
\multirow{2}{*}{\textbf{Model}} & \multicolumn{6}{c}{\textbf{F1-score}} & \multicolumn{3}{c}{} & \multicolumn{6}{c}{\textbf{Accuracy}} \\
\cline{2-7} \cline{11-16}
 & 0-shot & 2-shot & 4-shot & 6-shot & 10-shot & Average &  &  &  & 0-shot & 2-shot & 4-shot  & 6-shot & 10-shot & Average\\
\hline
\textbf{Qwen-3} & 0.444 & 0.452 & 0.500 & \underline{0.538} & 0.471 & 0.481 &  &  &  & \underline{0.967} & 0.887 & 0.960 & 0.920 & 0.940 & 0.935 \\
\textbf{GPT-4o} & 0.538 & \underline{0.571} & 0.500 & 0.545 & 0.500 & 0.531 &  &  &  & 0.920 & 0.940 & 0.907 & \underline{0.967} & 0.920 & 0.931 \\
\textbf{DeepSeek-R1} & 0.438 & 0.471 & \underline{0.545} & 0.435 & 0.462 & 0.470 &  &  &  & 0.880 & 0.940 & \underline{0.967} & 0.913 & 0.907 & 0.921 \\
\hline
\textbf{Average} & 0.473 & 0.498 & 0.515 & 0.506 & 0.478 & 0.494 &  &  &  & 0.922 & 0.922 & 0.945 & 0.933 & 0.922 & 0.929 \\
\bottomrule
\end{tabular}
\end{threeparttable}
}
\end{table*}

To facilitate threshold-based decision making, the model is instructed to output both a label and a confidence estimate. We evaluate three representative LLMs (GPT-4o \cite{GPT-4o}, DeepSeek-R1 \cite{deepseekai2025}, and Qwen-3 \cite{qwen3techreport}) under $k\in\{0,2,4,6,10\}$ in-context examples. As shown in Table~\ref{tab:LLM_comparison_imbalanced}, LLM performance under extreme imbalance is sensitive to prompting details. GPT-4o yields the best average F1-score (0.531), with the best setting at 2-shot (F1-score = 0.571). Qwen-3 exhibits the highest Precision in the 0-shot setting (1.000) but suffers from low Recall (0.286), indicating a conservative default policy; adding demonstrations increases Recall but can also introduce false positives. DeepSeek-R1 shows its strongest F1-score at 4-shot (0.545), suggesting that a small number of diverse demonstrations helps the model internalize the decision boundary.

\begin{table*}[htbp]
\centering
\color{black}  
\setlength\tabcolsep{4.0pt}
\renewcommand{\arraystretch}{1.25}
\caption{\textcolor{black}{Macro-F1 on the imbalanced test set across ML, DL, and LLM-based classifiers.}}
\label{tab:macroF1_all_imbalanced}

\scriptsize 

\newsavebox{\ta}
\newsavebox{\tb}

\sbox{\ta}{%
\begin{tabular}{c}
\textbf{(a) ML classifiers (word embeddings)}\\[-0.2em]
\begin{tabular}{lcccc}
\toprule
\textbf{Classifier} & \textbf{w2v} & \textbf{FT} & \textbf{GloVe} & \textbf{Average} \\
\hline
\textbf{SVM} & \underline{0.823} & 0.804 & 0.821 & 0.816 \\
\textbf{LR} & \underline{0.804} & 0.793 & 0.782 & 0.793 \\
\textbf{NB} & 0.764 & \underline{0.803} & 0.788 & 0.785 \\
\textbf{DT} & \underline{0.802} & \underline{0.802} & 0.764 & 0.789 \\
\textbf{kNN} & \underline{0.764} & \underline{0.764} & 0.747 & 0.758 \\
\hline
\textbf{Average} & 0.791 & 0.793 & 0.780 & 0.788 \\
\bottomrule
\end{tabular}
\end{tabular}
}

\sbox{\tb}{%
\begin{tabular}{c}
\textbf{(b) DL classifiers (TextCNN variants)}\\[-0.2em]
\begin{tabular}{lc}
\toprule
\textbf{Classifier} & \textbf{Macro-F1}\\
\hline
\textbf{TextCNN\_voc} & \textbf{0.782}\\
\textbf{TextCNN\_w2v} & 0.714\\
\textbf{TextCNN\_FT} & 0.719\\
\textbf{TextCNN\_GloVe} & 0.754\\
\hline
\textbf{Average} & 0.742\\
\bottomrule
\end{tabular}
\end{tabular}
}

\noindent\makebox[\textwidth][c]{%
\usebox{\ta}\hspace{1.2em}\usebox{\tb} 
}
\vspace{1.2em}

\begin{subtable}[t]{\textwidth}
\centering
\textbf{(c) LLM-based classifiers (\textit{k}-shot prompts)}\\[-0.2em]
\begin{tabular}{lcccccc}
\toprule
\textbf{Model} & \textbf{0-shot} & \textbf{2-shot} & \textbf{4-shot} & \textbf{6-shot} & \textbf{10-shot} & \textbf{Average} \\
\hline
\textbf{Qwen-3} & 0.714 & 0.694 & 0.740 & \underline{0.747} & 0.719 & 0.723 \\
\textbf{GPT-4o} & 0.747 & \underline{0.770} & 0.724 & 0.764 & 0.728 & 0.747 \\
\textbf{DeepSeek-R1} & 0.685 & 0.719 & \underline{0.764} & 0.694 & 0.705 & 0.713 \\
\hline
\textbf{Average} & 0.715 & 0.728 & 0.743 & 0.735 & 0.717 & 0.728 \\
\bottomrule
\end{tabular}
\end{subtable}

\end{table*}

Overall, the evaluation on the naturally imbalanced test set suggests that cost-aware adaptations can help maintain effective violation detection under realistic low-prevalence conditions. Among the ML and DL approaches, SVM achieves the best average Macro-F1 (0.816), and TextCNN\_voc is the most balanced DL variant (Macro-F1 = 0.782). For LLM-based classifiers, cost-aware prompting gives LLMs Macro-F1 values in a comparable range despite using prompt-only adaptation without parameter training. GPT-4o reaches the highest average Macro-F1 (0.747 across different $k$ settings), and the overall LLM average Macro-F1 of 0.728 indicates meaningful performance on both violation and non-violation classes.

\subsubsection{Statistical Significance and Model Ranking}
\label{sec:rq1_imbalanced_stats}

To determine whether the observed performance differences among ML/DL classifiers are statistically meaningful rather than artifacts of a particular data partition, we perform a statistical comparison based on the 10-fold cross-validation results. For clarity, we include nine representative configurations: the best-performing embedding variant (measured by Macro-F1) for each ML classifier and all TextCNN variants. Following the Nemenyi post-hoc procedure \cite{demsar2006statistical}, we conduct a Friedman test on fold-level rankings across models, followed by a Nemenyi post-hoc analysis. The resulting average ranks are visualized using a Critical Difference (CD) diagram (see Figure~\ref{fig:CD}).

Because all classifiers are evaluated on the same cross-validation folds and the distribution of performance scores cannot be assumed to be normal, the Friedman test is more appropriate than parametric alternatives. The results indicate statistically significant differences among the compared models. 
For the positive-class F1-score, the Friedman test yields $\chi^2=63.17$ ($p=1.1\times10^{-10}$), suggesting that the observed ranking differences are unlikely to be explained solely by random variation across folds. A similar conclusion holds for Macro-F1, where the Friedman statistic is $\chi^2=70.69$ ($p=3.6\times10^{-12}$), indicating systematic differences in the classifiers' ability to maintain balanced performance under the naturally imbalanced setting. 

The post-hoc analysis further reveals several consistent ranking patterns. Overall, SVM-w2v-based classifiers achieve the best average ranks, particularly when combined with word2vec embeddings. 
The CD diagram shows that SVM-w2v belongs to the top-performing group and significantly outperforms several weaker configurations, including TextCNN-GloVe, TextCNN-fastText, and TextCNN-word2vec ($p<0.05$). At the same time, the differences among the strongest ML-based classifiers are comparatively small, suggesting that multiple ML configurations achieve competitive performance on this task. These findings are consistent with the results reported in Tables~\ref{T: Performance of ML classifiers imbalanced} and~\ref{tab:macroF1_all_imbalanced}, where SVM variants consistently achieve the highest F1-score and Macro-F1 values. The ranking analysis also highlights the influence of text representations. Across classifiers, fastText- and word2vec-based configurations generally obtain higher average ranks than their GloVe counterparts, suggesting that embedding choice contributes substantially to classification effectiveness. However, the relative differences among embeddings are smaller than those observed between classifier families, indicating that performance is jointly influenced by both the learning algorithm and the representation.

LLM-based classifiers are not included in the CD diagram. Unlike ML/DL models, LLMs are evaluated through prompting rather than cross-validation and therefore do not produce paired fold-level observations suitable for Friedman/Nemenyi comparisons. Furthermore, few-shot performance depends on the selection of in-context demonstrations, making the resulting observations fundamentally different from those obtained under a fixed cross-validation protocol. To avoid mixing heterogeneous evaluation procedures and violating the assumptions of paired statistical tests, we report LLM results separately as descriptive baselines (see Tables~\ref{tab:LLM_comparison_imbalanced} and~\ref{tab:macroF1_all_imbalanced}).

\begin{figure*}[htbp]
\centering
\includegraphics[width=0.92\textwidth]{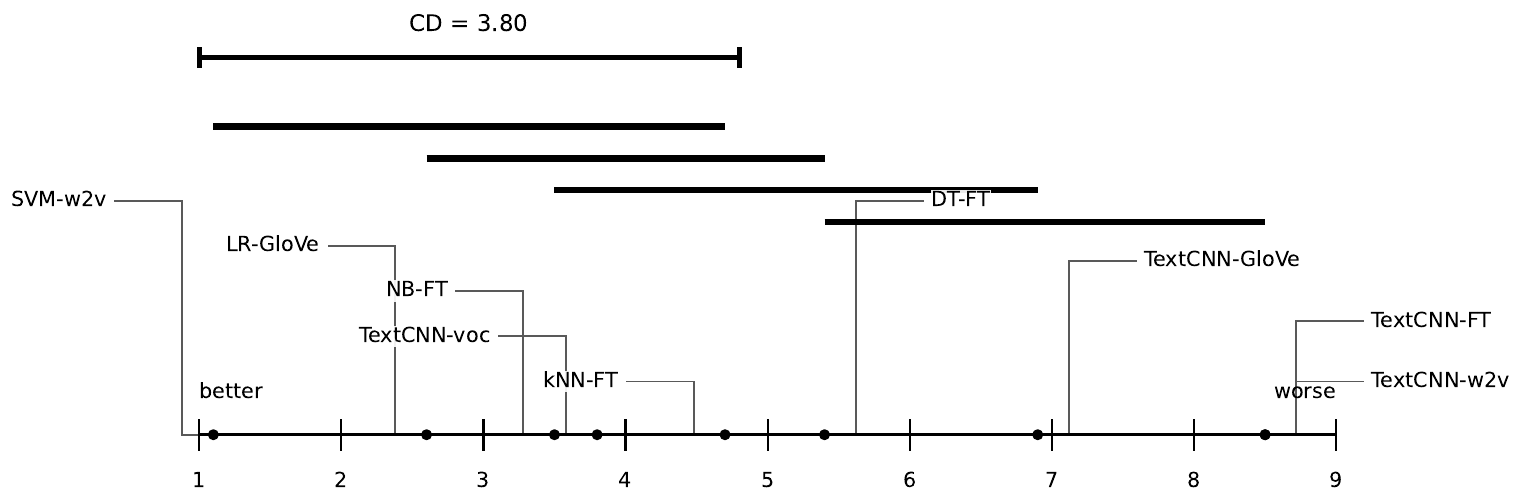}
\caption{Critical Difference (CD) diagram of classifier ranks on Macro-F1 over 10 folds (ML/DL models).}
\label{fig:CD}
\end{figure*}

}

\begin{tcolorbox}
\textbf{RQ1 Summary.} \textcolor{black}{Among the evaluated ML/DL classifiers, SVM combined with \textit{word2vec} achieved the best overall performance on the balanced classification task, while 200-dimensional word embeddings generally produced superior results compared with other dimensionalities. Additional experiments on a naturally imbalanced test set showed that the best-performing classifiers remained effective under realistic low-prevalence conditions, with SVM achieving the highest Macro-F1 among ML/DL approaches. Furthermore, statistical significance analysis confirmed that the observed performance differences among classifiers are systematic rather than random, with SVM-based models consistently ranking among the top-performing approaches.}
\end{tcolorbox}

\subsection{RQ2: Improving Performance with Voting}\label{sec:RQ2_results}

To answer RQ2, we employed an ensemble learning technique, i.e., majority voting strategy, to build an ensemble classifier. We compared the performance of individual classifiers with ensemble classifiers. Specifically, we explored to what extent combining one type of classifier based on three word embedding techniques can achieve better performance, as shown in Table~\ref{T:ensemble classifiers} and Table~\ref{T:voting_embeddings}. 

In Table~\ref{T:ensemble classifiers}, the ``\textit{Mean}'' column presents the \textit{Average} values of each metric of each classifier on three word embeddings from Table~\ref{T: Performance of ML classifiers} and Table~\ref{T: Performance of DL classifiers}. The ``\textit{Best}'' column denotes the corresponding classifier with the best performance from Table~\ref{T: Performance of ML classifiers} and Table~\ref{T: Performance of DL classifiers}. The ``\textit{Voting}'' column refers to the performance of the ensemble classifier (based on classifiers with three word embeddings) on each metric after utilizing a majority voting strategy. For example, the \textit{Voting} values of the SVM classifiers in Table~\ref{T:ensemble classifiers} refer to the voting results among the three individual classifiers SVM\_w2v, SVM\_FT, and SVM\_GloVe. ``\textit{Impro\_M}'' and ``\textit{Impro\_B}'' represent the improved results of the corresponding ensemble classifiers compared with the results in ``\textit{Mean}'' and ``\textit{Best}'' columns, respectively.

According to Table~\ref{T:ensemble classifiers}, when we utilized the voting strategy for each ML classifier based on the three word embeddings, all ensemble classifiers perform better than the mean values of individual classifiers, with F1-score improvement from 0.48\% to 13.10\%. Table~\ref{T:ensemble classifiers} shows that even compared with the best-performing classifiers, most of the ensemble classifiers (4 out of 5) outperform the individual ML/DL classifiers, except for kNN. Moreover, the ensemble TextCNN of the three TextCNN classifiers based on pre-trained word embeddings (see Table~\ref{T: Performance of DL classifiers}) does not perform as well as TextCNN\_FT (i.e., the best one) on Precision, F1-score, and Accuracy. This means that TextCNN\_FT with the \textit{fastText} pre-trained word embedding performs best compared with the other three TextCNN classifiers, and the voting strategy does not have an obvious improvement for the trained DL classifiers.

Additionally, as shown in Table~\ref{T:voting_embeddings}, we utilized the voting strategy for the five ML classifiers based on each of the three word embeddings. For example, ML\_word2vec is the ensemble classifier of the five ML classifiers based on the \textit{word2vec} word embedding (i.e., SVM\_w2v, LR\_w2v, NB\_w2v, DT\_w2v, and kNN\_w2v). Similarly as before, the ``\textit{Mean}'' column refers to the average metric results of ML classifiers (see the last row in Table~\ref{T: Performance of ML classifiers}); the ``\textit{Best}'' column presents the best-performing classifiers in Table~\ref{T: Performance of ML classifiers}; the ``\textit{Voting}'' column denotes the performance results of the ensemble classifiers after voting. It is clear that the three ensemble classifiers outperform the average performance results of the individual classifiers, with improvements from 3.11\% to 7.98\%. Even compared with the best-performing individual classifiers (i.e., the maximum values of the voting samples), the ensemble classifier ML\_fastText can still demonstrate performance improvement on all metrics.

In general, regarding the performance difference of ensemble classifiers with three word embeddings, among the selected five ML models, the \textit{fastText} pre-trained word embedding model achieves relatively better performance improvements on almost all metrics compared with the models using \textit{word2vec} and \textit{GloVe}. This finding affirms a similar observation in the result of RQ1.2.

\begin{tcolorbox}
\textbf{RQ2 Summary.} In comparison to individual ML classifiers, the ensemble classifiers achieve better performances after utilizing the majority voting strategy. However, no significant improvement was observed in the performance of the DL classifiers on our dataset when implementing the ensemble strategy.
\end{tcolorbox}

\subsection{RQ3: Validation in Practice}\label{sec:RQ3_results}
\subsubsection{\textcolor{black}{Qualitative Validation via Survey and Interview}}\label{sec:RQ3_1}
As elaborated in Section~\ref{sec:Survey}, we designed and conducted a survey and semi-structured interviews to solicit the perceptions from the practitioners on the usefulness of our trained models to automatically identify violation symptoms from code review comments. Note that, our survey and interview did not ask participants to ``discover'' additional violation symptoms that were not originally discussed in their code reviews.

Overall, we planned to send customized survey invitation emails to 200 code reviewers who were involved in the discussion of the collected review comments concerning architecture violations. However, we could only derive valid email addresses for 169 of them. We finally obtained 24 valid responses to our survey and interviewed 6 participants after sending 169 survey invitation emails and reminder emails; the survey response rate of 14.2\%, is a reasonable response rate considering that the general response rate is around 5\% in empirical software engineering research \cite{gaese2008CH1}. Table~\ref{T:Demographics} shows the demographic information of the 24 respondents. Twenty two of the respondents have more than 10 years of experience in software development, while two of them have 6-10 years of experience. Among the respondents, 83.3\% (20) are software engineers and 16.7\% (4) are architects. 
Regarding the size of their companies, ten of the respondents reported that their companies have more than 1000 employees, five respondents worked in companies with employees between 501 and 1000, six respondents in companies with employees between 101 and 500, and three respondents in companies with employees between 21 and 100. Besides, the respondents worked in a variety of application domains.

\begin{figure}[htbp]
	\centering
	\includegraphics[width=0.9\linewidth]{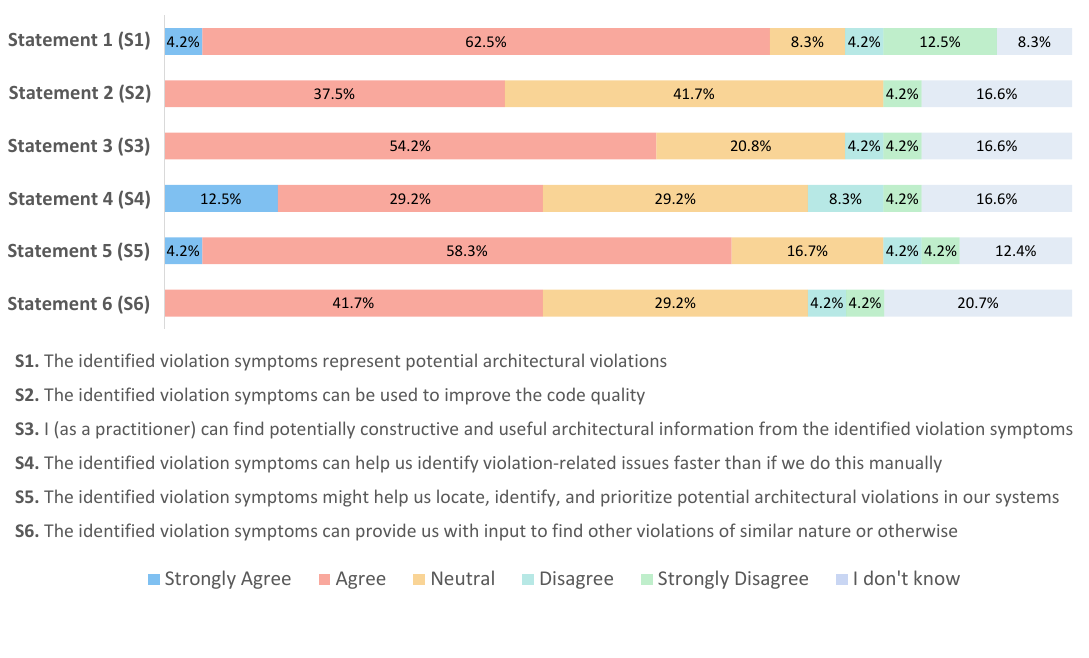}
	\caption{Practitioners' responses to statements regarding the usefulness of the trained models}\label{F:Statement}
\end{figure}

\begin{table}[htbp]
\centering
\footnotesize
\setlength\tabcolsep{1.5pt}
\renewcommand{\arraystretch}{1.1}
\caption{Demographic information of the survey participants and interviewees}\label{T:Demographics}
\begin{threeparttable}
\begin{tabular}{lllllm{40mm}}\toprule
\textbf{\#} & \textbf{Country} & \textbf{Experience} & \textbf{Role} & \textbf{Company size}& \textbf{Company domain}\\\hline
P1 & Ireland & More than 10 years & Software Engineer & employees $\geq$ 1000            & HCM services\\
P2 & Germany & More than 10 years & Software Engineer         & 21 $\leq$ employees $\leq$ 100   & Industrial electronics\\
P3 & Finland & More than 10 years & Software Engineer & 21 $\leq$ employees $\leq$ 100   & Medical software\\
P4 & Norway  & More than 10 years & Architect         & 501 $\leq$ employees $\leq$ 1000 & Industrial DataOps\\
P5 & Israel  & More than 10 years & Software Engineer & employees $\geq$ 1000            & Network\\
P6 & Germany & 6 - 10 years       & Software Engineer         & 101 $\leq$ employees $\leq$ 500  & Development tools\\
P7 & USA     & More than 10 years & Software Engineer         & 501 $\leq$ employees $\leq$ 1000 & Iaas\\
P8 & Norway  & More than 10 years & Software Engineer         & 501 $\leq$ employees $\leq$ 1000 & Cross-platform libraries for application development\\
P9 & Norway  & More than 10 years & Software Engineer         & 101 $\leq$ employees $\leq$ 500  & SW development frameworks\\
P10 & USA     & More than 10 years & Software Engineer & employees $\geq$ 1000            & Software development\\
P11 & Germany & More than 10 years & Software Engineer         & 101 $\leq$ employees $\leq$ 500  & Middleware\\
P12 & USA     & More than 10 years & Software Engineer & employees $\geq$ 1000            & Research\\
P13 (I1) & Norway  & More than 10 years & Software Engineer & 501 $\leq$ employees $\leq$ 1000 & Development framework\\
P14 & Finland & More than 10 years & Architect         & 21 $\leq$ employees $\leq$ 100   & Medical devices\\
P15 & Germany & More than 10 years & Software Engineer         & 501 $\leq$ employees $\leq$ 1000 & Tools and frameworks for software development\\
P16 (I2) & Germany & More than 10 years & Software Engineer         & 101 $\leq$ employees $\leq$ 500  & Middleware Platform \\
P17 & Germany & More than 10 years & Software Engineer & 101 $\leq$ employees $\leq$ 500  & Defence development\\
P18 & Denmark & 6 - 10 years       & Software Engineer & 101 $\leq$ employees $\leq$ 500  & Network\\
P19 (I3) & Canada  & More than 10 years & Software Engineer & employees $\geq$ 1000            & Framework tool\\
P20 & India   & More than 10 years & Architect         & employees $\geq$ 1000            & Development framework\\
P21 (I4) & China   & More than 10 years & Architect         & employees $\geq$ 1000            & Cloud, Data, GenAI\\
P22 & USA     & More than 10 years & Software Engineer & employees $\geq$ 1000            & Tools for software development\\
P23 (I5) & Germany & More than 10 years & Software Engineer & employees $\geq$ 1000            & Game development\\
P24 (I6) & Switzerland & More than 10 years &	Software Engineer & 101 $\leq$ employees $\leq$ 500  & AI-related research and development\\
\bottomrule
\end{tabular}
    \begin{tablenotes}
        \footnotesize
        \item P = Survey Participant, I = Interviewee
      \end{tablenotes}
      \end{threeparttable}
\end{table}

Figure~\ref{F:Statement} presents the feedback from the 24 respondents on the usefulness of the trained models, i.e., the automated identification of violation symptoms from the textual content of code review comments. According to the results, most of the respondents (66.7\%) agreed that the violation symptoms identified by our models represent potential architectural violations (Statement 1). 37.5\% respondents agreed that the identified violation symptoms can be used to improve the code quality (Statement 2), while 41.7\% respondents had a neutral attitude about this statement. 54.2\% of the respondents agreed that the identified violation symptoms contain potentially constructive and useful architectural information (Statement 3).

Statements 3 to 6 investigate whether the trained classifiers can play an auxiliary role in development and code reviews. Approximately 41.7\% respondents strongly agreed (12.5\%) or agreed (29.2\%) that the identified violation symptoms can help identify violation-related issues faster than manual identification (Statement 4), while there are still nearly half of the participants convey their uncertainty; the proportion of positive responses indicates that the classifiers may not fully replace manual efforts, but to some extent they can speed up the identification of violation-related issues during code review. Moreover, the majority (62.5\%) of the respondents strongly agreed (4.2\%) or agreed (58.3\%) that automated identification can help locate and prioritize architectural violation issues (Statement 5). Furthermore, 41.7\% of the respondents agreed that the identified violation symptoms can inspire them to find out other (or similar) violation symptoms (Statement 6). Notably, one respondent strongly disagreed with all six statements, indicating that he/she might not believe that ML/DL-based classifiers can be useful for identifying architectural violations in practice; we did not treat this response as an outlier. Moreover, only a small number of participants (ranging from 2 for Statement 1 to 5 for Statement 6) chose the ``\textit{I Don't Know}'' option. We further analyzed the responses for the six statements and found no significant statistical differences between the identified violation symptoms. Overall, Figure~\ref{F:Statement} shows that positive feedback (strongly agree + agree) surpasses negative (disagree + strongly disagree) and neutral feedback in most of the statements, except for Statement~2. 


In terms of the semi-structured interviews, the six practitioners that accepted our invitation are denoted as I1 to I6 in Table 7. Each interview lasted approximately 30 to 40 minutes, resulting in a total of 189 minutes of audio recordings. Furthermore, the transcripts of these interviews spanned 35 pages with over 22K words. In general, all of them expressed a positive attitude towards such an automated way to identify violation symptoms from code review comments. We present the key common points of the interviewees' opinions below:

\begin{itemize}
    \item \textbf{(IQ1) Potential benefits}: The integration of such an automated identification of violation symptoms of architecture erosion could help to reduce the time and effort of developers, which is recognized by four interviewees. As I3 stated, ``\textit{extra checking could make our work faster and facilitate the code review workflow}''. Additionally, I4 mentioned that ``\textit{Code reviewers prioritize two main categories of issues: those related to functionality and those concerning the architecture, and such an approach is particularly valuable and beneficial for identifying and addressing architectural problems}''.
    \item \textbf{(IQ1) Challenges}: Three out of six interviewees raised concerns regarding data leakage problems when employing or integrating such an automated approach supported by third-party libraries or specific LLMs, especially for inner-source projects. Besides, two out of six interviewees expressed their concerns about ``\textit{Crying Wolf}'' effect of false positives; as I1 mentioned: ``\textit{Any tool that gives you warnings that are false, after a while you start ignoring its warnings}''. If tools give false warnings (or warnings that users consider trivial), it eventually leads users to ignore all warnings, including legitimate ones. 
    This calls for the interpretability of such automated approaches that participants might not only want to know about the identified violation symptoms, but also understand why they are flagged as ``violations''; as I2 mentioned: ``\textit{As a user, I am not satisfied to be told where the issues are located, and I also want to know why they are flagged as violations with explanations}''. Therefore, putting more effort into improving the interpretability could pose another challenge for such automated approaches.
    \item \textbf{(IQ2) Strategies}: I4 thought that the best practices and specific rules (e.g., guidelines on architectural layering, module dependencies, and coupling thresholds) for identifying typical violation symptoms (e.g., layering violations) should be considered and integrated when employing automated techniques during code review. This could help customize such an approach of identifying violation symptoms and then maximize the identification accuracy for inner-source projects inside companies.
    \item \textbf{(IQ3) Visions for improving code quality}: With the advent of LLMs, five out of six interviewees are willing to utilize our approach with customized LLMs for specific tasks in the future, e.g., identifying architecture issues by feeding architecture constraints to the LLMs and providing architecture refactoring suggestions.
\end{itemize}

\subsubsection{\textcolor{black}{Quantitative Validation via a Controlled Experiment}}\label{sec:RQ3_1}
{
\color{black}
To further quantitatively assess the validity of our proposed approaches, we conducted a controlled experiment (see Section~\ref{sec:CtrExp}). We actually received more than 15 answers to questionnaires from the two groups, i.e., 18 from the Control Group and 17 from the Experimental Group, after removing the incomplete answers and very short answers that cannot be used in data analysis. 

Figure~\ref{F:Distribution} presents the distribution of participants' development experience across the control and experimental groups. The final participant pool consisted of developers with 3-15 years of experience from 5 countries. The average experience was 4.9 years for the control group and 5.8 years for the experimental group. Statistical analysis showed no significant difference in experience distribution between the control and experimental groups ($p$ = 0.808 $>>$ 0.05), with a trivial effect size. This confirms that the two groups were statistically balanced in terms of development experience, minimizing potential threats to validity from experience bias.

\begin{figure}[htbp]
    \centering
    \begin{subfigure}[b]{0.48\textwidth}
        \centering
        \includegraphics[width=\linewidth]{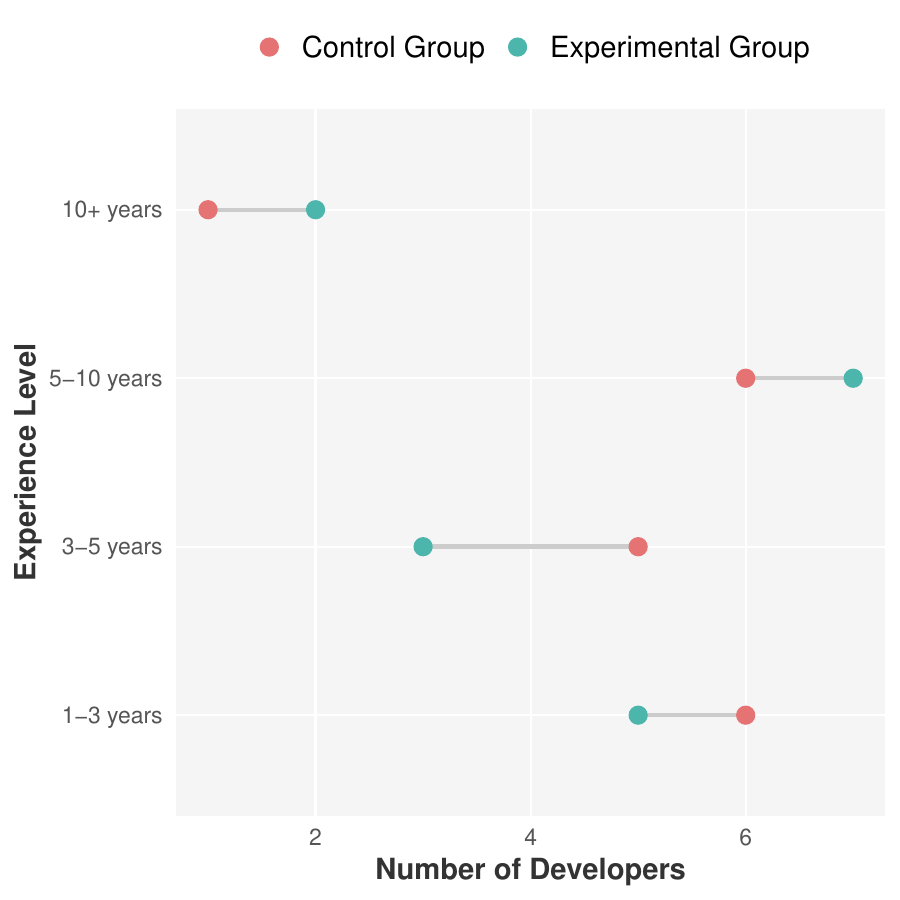}
        \caption{\textcolor{black}{Distribution of participants' years of experience}}
        \label{F:Distribution}
    \end{subfigure}
    \hfill
    \begin{subfigure}[b]{0.48\textwidth}
        \centering
        \includegraphics[width=\linewidth]{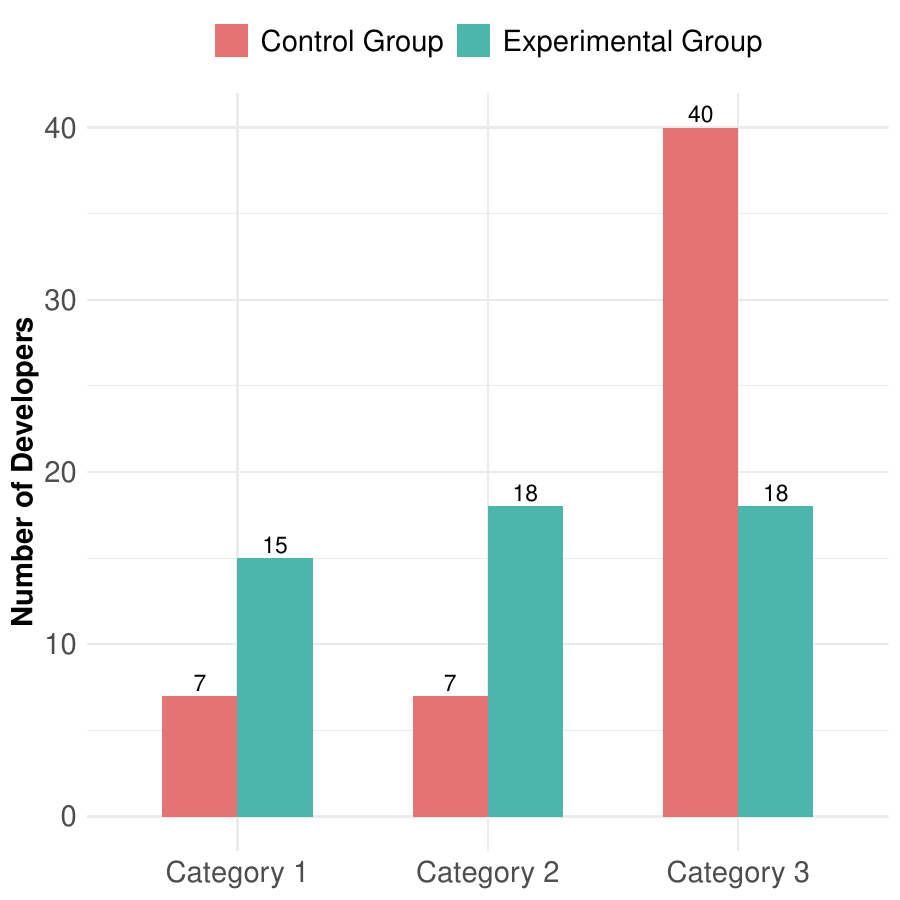}
        \caption{\textcolor{black}{Comparison of review outcomes}}
        \label{F:Group_Bar}
    \end{subfigure}
    \caption{\textcolor{black}{Distribution of participants' experience and comparison of review outcomes between Control and Experimental groups}}
    \label{F:Combined}
\end{figure}

Figure~\ref{F:Group_Bar} illustrates the distribution of review results across the three categories. In the Control Group (54 responses from 18 participants reviewing 3 code snippets each), only 7 cases (13.0\%) fell into Category 1 (identified violation and provided correct fix), 7 cases (13.0\%) into Category 2 (identified but incorrect fix), and the majority 40 cases (74.1\%) into Category 3 (failed to identify). In contrast, the Experimental Group (51 responses from 17 participants reviewing 3 code snippets each) showed substantially better performance: 15 cases (29.4\%) achieved Category 1, 18 cases (35.3\%) reached Category 2, and only 18 cases (35.3\%) remained in Category 3. This indicates that providing violation symptom hints significantly improved developers' ability to identify architecture violations, with the detection rate (Category 1 + Category 2) increasing from 25.9\% (14/54) in the Control Group to 64.7\% (33/51) in the Experimental Group, representing a 150\% relative improvement.

To assess the statistical significance of these differences, we performed Fisher's Exact Test \cite{agresti2013} on the 2×3 contingency table of the three categories between the two groups. The results revealed a statistically significant difference between the two groups ($p < 0.001$), with a medium-to-large effect size (Cramér's $V = 0.42$) \cite{cohen2013}. Specifically, the proportion of correctly identifying and fixing violations (Category 1) increased by 126\% (from 13.0\% to 29.4\%), while the proportion of completely missed violations (Category 3) decreased by 52.4\% (from 74.1\% to 35.3\%) when violation symptom hints were provided. These findings provide strong quantitative evidence that our automated identification of violation symptoms effectively assists developers in recognizing and addressing architecture violations during code review, validating the practical usefulness of our approach.

}

\begin{tcolorbox}
\textbf{RQ3 Summary.} \textcolor{black}{Practitioners' feedback confirms the practical usefulness of automatically identifying violation symptoms: 66.7\% agreed that the identified symptoms accurately represent potential architectural violations. Moreover, a controlled experiment showed that providing symptom hints significantly improves developers' detection rate of architecture violations from 25.9\% to 64.7\%. These results validate that our approach effectively assists practitioners in recognizing and addressing architectural violations during code review.}
\end{tcolorbox}

\subsection{RQ4: Comparison with LLM-based Classifiers}\label{sec:RQ4_results}

To answer RQ4, we conducted additional comparative experiments by constructing LLM-based classifiers. As shown in Table~\ref{tab:LLM_comparison}, we evaluated three LLMs across five shot settings (one zero-shot and four few-shot configurations). GPT-4o demonstrates the highest Precision, achieving a peak of 0.851 in the 4-shot setting and the highest average Precision of 0.759. Qwen-3 achieves its highest Precision of 0.778 in the 10-shot setting and maintains competitive Recall, with a peak of 0.926 at 6-shot and an average Recall of 0.880. DeepSeek-R1 also performs well in Recall, particularly at 4-shot (0.943), resulting in the highest average Recall of 0.898. For F1-scores, GPT-4o achieves the highest peak 0.851 at 2-shot and the highest average 0.811, followed by Qwen-3 (0.817 at 10-shot) and DeepSeek-R1 (0.767 at 4-shot). GPT-4o also leads in Accuracy, reaching a peak of 0.848 at 2-shot and averaging 0.793. Qwen-3 achieves its highest Accuracy (0.807) at 10-shot, while DeepSeek-R1 has the lowest average Accuracy (0.696) despite a 2-shot peak of 0.713. 

Overall, GPT-4o provides the most balanced and robust performance across all metrics and shot settings. Qwen-3 ranks second in average F1-score and Accuracy and shows strong Recall, while DeepSeek-R1 stands out in average Recall but shows lower Precision and overall Accuracy.

\begin{table*}[htbp]
\centering
\setlength\tabcolsep{1.8pt}
\renewcommand{\arraystretch}{1.2}
\caption{Performance comparison of different LLMs across varying shot settings}\label{tab:LLM_comparison}
\scalebox{1.2}{\scriptsize 
\begin{tabular}{cccccccccccccccc}
\toprule
\multirow{2}{*}{\textbf{Model}} & \multicolumn{5}{c}{\textbf{Precision}} & & & & & \multicolumn{5}{c}{\textbf{Recall}} \\ \cline{2-7} \cline{10-15}
 & 0-shot & 2-shot & 4-shot & 6-shot & 10-shot & Average & & &  0-shot & 2-shot & 4-shot  & 6-shot & 10-shot & Average\\ \hline
\textbf{Qwen-3} & 0.688 & 0.692 & 0.691 & 0.711 & \underline{0.778} & 0.712 & & & 0.795 & 0.902 & 0.918 & \underline{0.926} & 0.861 & 0.880 \\
\textbf{GPT-4o} & 0.701 & 0.835 & \underline{0.851} & 0.699 & 0.712 & 0.759 & & & 0.844 & 0.869 & 0.795 & 0.934 & \underline{0.951} & 0.879\\
\textbf{Deepseek-R1} & 0.653 & \underline{0.657} & 0.646 & 0.611 & 0.635 & 0.640 & & & 0.803 & 0.893 & \underline{0.943} & 0.926 & 0.926 & 0.898 \\\hline 
\textbf{Average} & 0.681 & 0.728 & 0.729 & 0.674 & 0.708 & 0.704 & & & 0.814 & 0.888 & 0.885 & 0.929 & 0.913 & 0.886 \\ \bottomrule

\\\toprule
\multirow{2}{*}{\textbf{Model}} & \multicolumn{5}{c}{\textbf{F1-score}} & & & & & \multicolumn{5}{c}{\textbf{Accuracy}} \\ \cline{2-7} \cline{10-15}
 & 0-shot & 2-shot & 4-shot & 6-shot & 10-shot & Average & & &  0-shot & 2-shot & 4-shot  & 6-shot & 10-shot & Average\\ \hline
\textbf{Qwen-3} & 0.738 & 0.783 & 0.789 & 0.804 & \underline{0.817} & 0.786 & & & 0.717 & 0.750 & 0.754 & 0.775 & \underline{0.807} & 0.761\\
\textbf{GPT-4o} & 0.766 & \underline{0.851} & 0.822 & 0.800 & 0.814 & 0.811 & & & 0.742 & \underline{0.848} & 0.828 & 0.766 & 0.783 & 0.793 \\
\textbf{Deepseek-R1} & 0.721 & 0.757 & \underline{0.767} & 0.736 & 0.753 & 0.747 & & & 0.689 & \underline{0.713} & 0.713 & 0.668 & 0.697 & 0.696 \\\hline 
\textbf{Average} & 0.741 & 0.797 & 0.792 & 0.780 & 0.795 & 0.781 & & & 0.716 & 0.770 & 0.765 & 0.736 & 0.762 & 0.750 \\ 
\bottomrule
\end{tabular}
}
\end{table*}

Compared to ML/DL-based classifiers (see Tables~\ref{T: Performance of ML classifiers} and ~\ref{T: Performance of DL classifiers}), the LLM-based classifiers achieve better performance on the same classification task, with an average Precision of 0.704, Recall of 0.886, F1-score of 0.781, and Accuracy of 0.750. This finding indicates that the LLM-based classifiers are more effective in identifying the violation symptoms than traditional ML/DL-based classifiers. Among the LLM-based classifiers, GPT-4o with the 2-shot setting achieves the best F1-score and Accuracy, while GPT-4o with the 4-shot setting achieves the highest Precision and GPT-4o with the 10-shot setting achieves the highest Recall.

We also examined whether ensemble strategy across different shot settings could improve performance (see Tables~\ref{T:LLM-based ensemble classifiers best-model} and~\ref{T:LLM-based ensemble classifiers best-shot}). Table~\ref{T:LLM-based ensemble classifiers best-model} shows the results of ensemble classifiers based on a voting strategy. The ``\textit{Mean}'' column refers to the average performance of each metric across all LLMs and shot settings from Table~\ref{tab:LLM_comparison}, and the ``\textit{Best}'' column captures the best performance per shot setting. The ``\textit{Voting}'' column reports the ensemble performance obtained via majority voting among models with the same shot setting. Likewise, the improvements over the ``\textit{Mean}'' and ``\textit{Best}'' are shown in the ``\textit{Impro\_M}'' and ``\textit{Impro\_B}'' columns, respectively. We observed that the ensemble classifiers often outperform the average results of different LLMs in Table~\ref{T:LLM-based ensemble classifiers best-model}, especially for Recall. However, they generally underperform when compared to the best individual classifiers, with only isolated improvements such as Recall in the 6-shot setting and F1-score in the 4-shot setting.

Similarly, Table~\ref{T:LLM-based ensemble classifiers best-shot} examines ensemble classifiers within a single LLM across different shot settings. The ``\textit{Mean}'' column presents the average metric values across all five shot settings from Table~\ref{tab:LLM_comparison}; the ``\textit{Best}'' column shows the best performance of single LLMs; the ``\textit{Voting}'' column displays the results from the ensemble classifiers across shot settings using a voting strategy. Table~\ref{T:LLM-based ensemble classifiers best-shot} shows that the ensemble classifiers across shot settings generally outperform their respective average performances. However, they usually do not exceed their best individual shot results; DeepSeek-R1 is the only model for which the ensemble classifier slightly surpasses the best individual setting in F1-score and Accuracy, while matching the best Recall.

\begin{sidewaystable}[hp]
\vspace*{0.66\textheight} 
\centering
\small
\setlength\tabcolsep{1.5pt}
\renewcommand{\arraystretch}{1.6}
\caption{Performance comparison of ensemble LLM-based classifiers with different shot settings}\label{T:LLM-based ensemble classifiers best-model}
\begin{minipage}[t]{\textwidth} 
\centering
\resizebox{\textwidth}{!}{
\begin{tabular}{cccccccccccccccccccccccc}
\toprule
\multirow{2}{*}{\textbf{Classifier}} & \multicolumn{5}{c}{\textbf{Precision}}&  & \multicolumn{5}{c}{\textbf{Recall}}&  & \multicolumn{5}{c}{\textbf{F1-score}}&  &\multicolumn{5}{c}{\textbf{Accuracy}}\\ \cline{2-6} \cline{8-12} \cline{14-18} \cline{20-24}
 & Mean & Best & Voting & Imp\_M & Imp\_B & & Mean & Best & Voting & Imp\_M & Imp\_B & & Mean & Best & Voting & Imp\_M & Imp\_B & & Mean & Best & Voting & Imp\_M & Imp\_B\\\hline
\textbf{0-shot}     & 0.681 & 0.701 & 0.699 & 2.64\% & -0.29\% & & 0.814 & 0.844 & 0.836 & 2.68\% & -0.97\% & & 0.741 & 0.766 & 0.761 & 2.68\% & -0.60\% & & 0.716 & 0.742 & 0.738 & 3.05\% & -0.55\% \\
\textbf{2-shot}      & 0.728 & 0.835 & 0.712 & -2.10\% & -14.64\% & & 0.888 & 0.902 & 0.893 & 0.62\% & -0.91\% & & 0.797 & 0.851 & 0.793 & -0.55\% & -6.89\% & & 0.770 & 0.848 & 0.766 & -0.53\% & -9.66\%   \\
\textbf{4-shot}      & 0.729 & 0.851 & 0.743 & 1.92\% & -12.63\% & & 0.885 & 0.943 & 0.926 & 4.63\% & -1.74\% & & 0.792 & 0.822 & 0.825 & 4.08\% & 0.34\% & & 0.765 & 0.828 & 0.803 & 5.00\% & -2.97\%   \\
\textbf{6-shot}      & 0.674 & 0.711 & 0.676 & 0.42\% & -4.82\% & & 0.929 & 0.934 & 0.943 & 1.47\% & 0.88\% & & 0.780 & 0.804 & 0.788 & 0.97\% & -2.06\% & & 0.736 & 0.775 & 0.746 & 1.30\% & -3.70\%   \\
\textbf{10-shot}     & 0.708 & 0.778 & 0.713 & 0.62\% & -8.39\% & & 0.913 & 0.951 & 0.934 & 2.40\% & -1.72\% & & 0.795 & 0.817 & 0.809 & 1.72\% & -1.05\% & & 0.762 & 0.807 & 0.779 & 2.15\% & -3.55\%  \\
\bottomrule
\end{tabular}
}
\end{minipage}

\vspace{2.3cm} 

\begin{minipage}[t]{\textwidth} 
\centering
\small
\setlength\tabcolsep{1.5pt}
\renewcommand{\arraystretch}{1.5}
\caption{Performance comparison among ensemble LLM-based classifiers}\label{T:LLM-based ensemble classifiers best-shot}
\resizebox{\textwidth}{!}{
\begin{tabular}{cccccccccccccccccccccccc}
\toprule
\multirow{2}{*}{\textbf{Classifier}} & \multicolumn{5}{c}{\textbf{Precision}}&  & \multicolumn{5}{c}{\textbf{Recall}}&  & \multicolumn{5}{c}{\textbf{F1-score}}&  &\multicolumn{5}{c}{\textbf{Accuracy}}\\ \cline{2-6} \cline{8-12} \cline{14-18} \cline{20-24}
 & Mean & Best & Voting & Imp\_M & Imp\_B & & Mean & Best & Voting & Imp\_M & Imp\_B & & Mean & Best & Voting & Imp\_M & Imp\_B & & Mean & Best & Voting & Imp\_M & Imp\_B\\\hline
\textbf{Qwen-3}  & 0.712 & 0.778 & 0.718 & 0.85\% & -7.69\% & & 0.880 & 0.926 & 0.918 & 4.28\% & -0.88\% & & 0.786 & 0.817 & 0.806 & 2.50\% & -1.39\% & & 0.761 & 0.807 & 0.779 & 2.37\% & -3.55\% \\
\textbf{GPT-4o} & 0.759 & 0.851 & 0.803 & 5.72\% & -5.64\% & & 0.879 & 0.951 & 0.902 & 2.61\% & -5.17\% & & 0.811 & 0.851 & 0.849 & 4.78\% & -0.23\% & & 0.793 & 0.848 & 0.840 & 5.89\% & -0.97\% \\
\textbf{Deepseek-R1} & 0.640 & 0.657 & 0.650 & 1.47\% & -1.05\% & & 0.898 & 0.943 & 0.943 & 4.93\% & 0.00\% & & 0.747 & 0.767 & 0.769 & 3.01\% & 0.33\% & & 0.696 & 0.713 & 0.717 & 3.06\% & 0.57\%  \\\bottomrule
\end{tabular}
}
\end{minipage}
\end{sidewaystable}

\begin{tcolorbox}
\textbf{RQ4 Summary.} GPT-4o shows the most balanced performance, leading in Precision, F1-score, and Accuracy, while DeepSeek-R1 achieves the highest average Recall and Qwen-3 ranks second in average F1-score and Accuracy. The best LLM-based classifiers outperform traditional ML/DL-based classifiers across all metrics. Ensemble classifiers offer limited improvements in some settings but generally do not surpass the best individual LLM-based classifiers.
\end{tcolorbox}

\section{Discussion}\label{sec:Discussion}
\subsection{Interpretation of Results}\label{sec:Discussion of Results}
\subsubsection{RQ1: Identifying Violation Symptoms}

Our trained ML/DL classifiers can identify violation symptoms in code review comments. The results of RQ1.1 indicate that certain factors such as the type of classifier, the pre-trained word embeddings, and hyperparameters influence the performance of classifiers. Specifically, we found that the SVM classifier with \textit{word2vec} outperforms other ML/DL-based classifiers. We also found that DL classifiers do not always have better performance than ML classifiers, and the outcome may stem from factors such as model complexity, dataset size, and hyperparameter settings; we note that training DL classifiers takes much longer time than training ML classifiers, as DL classifiers have much more complex network architecture and more parameters. In recent years, some studies that leveraged DL-based techniques for SE tasks have achieved better performance that all other state-of-the-art techniques (e.g., ML-based techniques)~\cite{Yang2022pmse}. However, the achieved results imply that ML classifiers might be enough for certain text classification tasks with \textit{small-size datasets}, especially as the trained DL classifiers do not show performance improvement in such a scenario. This finding is also observed in previous studies \cite{Liu2018nmt, Fu2017eoh}: for certain SE tasks, simple ML-based approaches are capable of achieving equal or even better performance than DL-based approaches within less time. Thus, we advise researchers not to blindly apply DL-based techniques and consider simple approaches first.

\begin{framed}
\noindent {\textbf{Finding 1}: \textit{Traditional ML-based approaches have sufficient performance for many SE tasks, such as identification of architecture violations in code reviews}.}
\end{framed}

From the results of RQ1.2, we can see that the TextCNN classifier based on the \textit{fastText} pre-trained word embedding can perform better than other DL models. We thus emphasize that the \textit{fastText} pre-trained technique has a better capacity for training well-performed DL models; this also aligns with the finding in the work of Sesari \textit{et al}. \cite{Sesari2022pte}, that is, the \textit{fastText} pre-trained word embedding model has less bias compared to \textit{word2vec} and \textit{GloVe}.

Besides, the results of RQ1.3 show that 200-dimensional pre-trained word embedding models outperform 100 and 300-dimensional models. The possible reason is that 200-dimensional embeddings strike a balance between expressiveness and robustness for our dataset size: lower-dimensional (100d) embeddings under-represent nuanced lexical patterns, whereas higher-dimensional (300d) embeddings introduce excess parameters, increasing the risk of overfitting. Our results are slightly different from previous studies (e.g., \cite{Ren2019nndSATD, Li2022SATD}) where the 300-dimensional word embedding model has better performance. One possible reason is that the previous studies only measured the average results of classifiers with different dimensional word embedding models, and the average results might be influenced by the individual results. Our results present and compare the classification performance of each ML/DL classifier, which could be more accurate.

\begin{framed}
\noindent {\textbf{Finding 2}: \textit{The 200-dimensional fastText pre-trained word embedding model can be used to train classifiers with better performance when conducting binary classification of violation symptoms from code review comments}.}
\end{framed}

In general, due to a large number of hyperparameters of DL classifiers, we did not specifically focus on the best combination of hyperparameters of the TextCNN classifier, but certain key hyperparameters such as the dimension size of word embedding model. Thus, future work can consider performing hyperparameter tuning (e.g., utilizing the grid search technique) to explore other possible combinations of the hyperparameters to further improve performance.

\textcolor{black}{The additional evaluations under naturally imbalanced conditions further strengthen the robustness of our findings. Although violation symptoms represent only a small fraction of real-world code review comments, the best-performing classifiers maintained competitive Macro-F1 scores, suggesting that the proposed approaches can still effectively distinguish violations from non-violations in practical deployment scenarios. Moreover, the statistical significance analysis confirms that the superiority of top-ranked classifiers (particularly SVM-based models) is not merely an artifact of specific data partitions, increasing confidence in the stability and reproducibility of the results.}

\subsubsection{RQ2: Improving Performance with Voting}

Evaluating the performance of an ensemble classifier offers insight into the performance gains compared with individual classifiers. For voting-based approaches, an accurate output is attained only when a majority of the constituent models generate the correct output \cite{hansen1990nne}. Our results provide further empirical evidence for the efficacy of the majority voting strategy on improving the text classification performance. This technique effectively combines the predictions of multiple classifiers to reach a final decision, leading to improved overall classification performance. Nevertheless, we observed that such performance improvement is not obvious in TextCNN in our case. One possible reason is that our DL classifiers (all TextCNN variants) and the modest dataset size prevented the majority‑vote ensemble from yielding significant gains. We argue that the voting strategy remains valuable to minimize potential prediction errors if there are more outputs for voting.

\begin{framed}
\noindent {\textbf{Finding 3}: \textit{Compared with individual ML/DL-based classifiers for identifying architecture violations from code review comments, ensemble learning is an effective way for improving classification performance. In most cases, ensemble classifiers that utilize the voting strategy can enhance all classification performance metrics}.}
\end{framed}

\subsubsection{RQ3: Validation in Practice}

The existence of architectural violations can be challenging, as they can impede practitioners' understanding of architecture. This particularly influences novice developers who lack practical familiarity with architecture. The practitioners we surveyed and interviewed recognized the identified issues in code snippets, as they had previously noted them during code review. However, they may not have realized these issues were violation symptoms of architecture erosion. Our approach helps practitioners realize the existence of violation symptoms of architecture erosion. The results of RQ3 indicate a predominantly positive feedback, confirming the usefulness of our trained ML/DL classifiers in potentially helping developers during maintenance and evolution. Through analyzing the received responses to the statements (especially Statements 3, 4, and 5), we found that developers convey interest in utilizing automated approaches, tools, or plug-ins to identify violation symptoms during code review, recognizing their potential to enhance the productivity of dealing with architectural violations. Overall, the survey and interview results indicate that developers hold positive attitudes towards such kinds of tools. 
Moreover, automated techniques regarding identifying potential architecture violations could be supplementary means to minimize architecture violations and improve code quality during code review. For example, such techniques could provide code reviewers with warnings that architecture violations need to be refactored, based on the review discussions. Furthermore, according to the participants' feedback, automated techniques could help code reviewers prioritize such refactorings based on various criteria (e.g., the importance of involved components, the number of violation symptoms in those components, and the seniority of code reviewers). Such a prioritization could help to fix violation symptoms that might otherwise have been ignored.

Moreover, according to the respondents' feedback, certain practitioners might hold a conservative attitude toward ML/DL-based classifiers for identifying violation symptoms, which could be observed from the ``\textit{Neutral}'' results in Figure~\ref{F:Statement}, especially for Statement 2. One reason is that they might be skeptical about the performance of the classifiers (false positives or negatives), casting doubt about the reliability of identified violation symptoms. Nevertheless, no classifier can ever achieve 100\% accuracy in classification tasks, and there is always a trade-off between precision and recall. This disclaimer should be made clear when showing classification tasks to practitioners. Another potential reason could be that certain identified review comments may be part of the architectural design \textit{discussions} during the development process, rather than the final architectural decisions. This is particularly common in agile development, where architectures evolve over time and certain decisions are continuously discussed and refined. Such concerns are beyond the target of our research, as the trained classifiers can only provide judgments based on the collected violations, instead of exploring the evolution of architecture violations.

\textcolor{black}{Complementing the qualitative feedback, our controlled experiment provides quantitative evidence that the automated identification of violation symptoms tangibly improves developers' effectiveness in practice. Specifically, when provided with symptom hints, the detection rate of architecture violations increased from 25.9\% to 64.7\%, and the proportion of correctly identified and fixed violations more than doubled. This quantitative gain aligns with practitioners' perceived usefulness reported in the survey: the 66.7\% agreement that identified symptoms represent real violations, and the 62.5\% confidence in their value for prioritizing issues, are empirically corroborated by the observed improvement in detection performance. Together, the qualitative perceptions and quantitative outcomes reinforce each other, suggesting that our approach not only resonates with practitioners' expectations but also delivers measurable assistance in recognizing and addressing architectural violations during code review. Importantly, the controlled experiment complements the survey and interview findings by providing behavioral evidence rather than perception-based evidence alone. While practitioners expressed positive attitudes toward automated identification of violation symptoms, the experiment demonstrates that such support can also translate into measurable improvements in developers' performance during code review.}

\begin{framed}
\noindent {\textbf{Finding 4}: \textit{Despite a few concerns about the accuracy of the identified violation symptoms, \textcolor{black}{the vast majority reacted favorably in surveys and interviews; moreover, a controlled experiment confirmed that symptom hints significantly improve developers' detection performance, providing strong empirical support for automated identification of architecture violation symptoms during code review}}}.
\end{framed}

\subsubsection{RQ4: Comparison: ML/DL-based vs. LLM-based Classifiers}

The results of RQ4 demonstrate that LLMs, when configured as classifiers via prompt engineering, can substantially outperform traditional ML/DL-based approaches on violation symptom identification.The best LLM-based classifiers exceed the best traditional ML/DL-based classifiers on all metrics, although average performance varies across metrics and models. This suggests that LLMs' deep contextual understanding and few-shot adaptability confer distinct advantages for text classification tasks. \textcolor{black}{Although the best LLM-based classifiers achieve better overall performance, the observations obtained from traditional ML/DL models remain valuable. In particular, the comparative analysis of classifiers, embeddings, and ensemble strategies still provides useful insights into the factors that influence automated identification performance and remains relevant for scenarios where computational cost, latency, privacy, or deployment constraints limit the use of LLMs.}

\textcolor{black}{Performance varied among the three LLM-based classifiers for automated identification of violation symptoms. The possible reason could stem from differences in training corpora and pre-training goals, which shape each model's strengths and weaknesses across SE tasks}. Consequently, practitioners should align task requirements (e.g., code understanding, text classification) with each LLM's underlying training focus to select the most suitable LLMs.

Furthermore, ensemble strategy either across models (see Table~\ref{T:LLM-based ensemble classifiers best-model}) or across shot settings for an individual model (see Table~\ref{T:LLM-based ensemble classifiers best-shot}) yields modest improvements over average results in some settings from Table~\ref{tab:LLM_comparison}. However, ensemble classifiers generally do not outperform the best-performing individual LLMs, underscoring that an ensemble's strength lies in balancing errors rather than maximizing peak performance capabilities. Notably, employing a voting strategy cannot significantly improve the performance of ensemble LLM-based classifiers. A plausible explanation could be that the individual LLM-based classifiers already achieve very good performance, leaving limited room for further enhancement through ensembling. In fact, the ensemble technique may even degrade performance, potentially due to the amplification of inherent issues such as LLM-induced \textit{hallucination} \cite{GPT4}. This phenomenon closely resembles the Law of Diminishing Marginal Utility in economics, whereby additional inputs yield progressively smaller benefits beyond a certain threshold~\cite{jevons2013PE}.

\begin{framed}
    \noindent {\textbf{Finding 5}: \textit{\textcolor{black}{Although the best LLM-based classifiers achieve better overall performance, the findings derived from traditional ML/DL approaches remain valuable, especially under constraints on computational cost, latency, privacy, or deployment. While performance varies across different LLMs, the overall comparative trends between LLM-based and traditional ML/DL approaches remain consistent. Moreover, ensembling multiple LLM-based classifiers does not consistently surpass the best individual models, as ensembles balance errors rather than maximize peak capability, with potential amplification of LLM hallucinations.}}}
\end{framed}

\subsection{Implications}\label{sec:Implications}
\indent\textbf{Establish and enrich datasets}. 
Many studies utilized or proposed automated approaches to identify or predict knowledge-related information from textual artifacts. However, no studies focused on the automated identification of violation symptoms from code review comments. This suggests that there is a need for more empirical reports and follow-up tools regarding such issues. Our study provides a starting point for researchers to automatically identify architecture violations in textual artifacts such as review comments. On top of this, we encourage researchers and practitioners to \textbf{establish and share datasets} related to architecture violations to lay a solid foundation for future research.

Moreover, our study is \textbf{scalable} in practice, as practitioners can retrain superior classifiers by adding more code review comments of violation symptoms from different projects, either manually or by using our classifiers. Besides, mining architecture violations from multiple textual artifacts (e.g., issues, source code comments, and pull requests) can provide a more comprehensive examination of architectural violations and also \textbf{enrich the dataset} used for identifying such violations. Furthermore, our findings regarding the experiments in this study might benefit or be adapted by researchers and practitioners to identify other issues, such as security and vulnerability discussions.

\begin{framed}
\noindent {\textbf{Suggestion 1}: \textit{Researchers and practitioners are encouraged to establish and share datasets related to architecture violations. Besides, our study is scalable to retrain superior classifiers by adding more data from different sources, and can also be adapted to identify other issues}.}
\end{framed}

\noindent\textbf{Choose appropriate classifiers}. For traditional ML/DL-based classifiers, an ensemble classifier is a collection of classifiers that work together to classify instances by combining their individual outputs through voting. Researchers can consider exploring different selections of classifiers to further enhance the performance of ensemble classifiers. For example, to some extent, assigning varying weights to individual classifiers might partially mitigate the impact of biased classifiers on voting results. 
While ensemble classifiers based on LLMs require more time and resources, their performance often does not exceed that of certain top-performing individual LLM-based classifiers, particularly when configured with optimal few-shot settings.

\begin{framed}
\noindent {\textbf{Suggestion 2}: \textit{When constructing traditional ML/DL-based classifiers, combining classifiers and assigning different weights to individual classifiers are recommended to further enhance the classification performance of ensemble classifiers for identifying architecture violations. In contrast, for LLM-based classifiers, choosing a suitable few-shot setting is recommended over ensemble techniques.}}
\end{framed}

\textbf{Optimize code review practice.} In the future, researchers and practitioners can explore the development of plug-ins for code review platforms (e.g., Gerrit) that provide warnings of violations to developers or even suggestions to fix those violations. Such plug-ins can be developed utilizing historical code review data or harnessing the existing LLMs through fine-tuning. These plug-ins have the potential to assist developers and maintainers in conducting comprehensive examinations of preexisting architectural violations, and may even serve as reminders of their presence. Additionally, implementing the trained classifiers (e.g., as an extension) into the existing tools (or toolsets) is also worthy of consideration in practice. For example, code review platforms can integrate such classifiers to empower the automated code review process. Such tools may provide more insights and help novices or maintainers troubleshoot architecturally relevant violation issues.

\begin{framed}
\noindent \textbf{Suggestion 3}: \textit{Researchers and practitioners can consider integrating classifiers as plug-ins for code review platforms (e.g., Gerrit) to assist developers in identifying potential architecture violations, or integrating customized classifiers into existing tools to optimize the automated code review process}.
\end{framed}

\textcolor{black}{\textbf{Challenges of Adopting LLMs in Real-World Code Review Workflows}. As demonstrated in our study, LLM-based classifiers, particularly GPT-4o, have shown substantial promise in identifying architecture erosion symptoms during code review. However, the practical adoption of such models in real-world settings is not without challenges, particularly concerning inference cost, latency, and dependency on proprietary APIs. Specifically, LLMs like GPT-4o require considerable computational resources, leading to high inference costs, especially when integrated into continuous development processes. These costs can be further amplified by usage-based pricing models, making operational expenses difficult to predict at scale. The response time of LLMs may introduce delays in time-sensitive code review scenarios. Although individual requests may incur only modest latency, such delays can accumulate and affect the delivery of timely feedback, which is crucial in fast-paced code reviews. Conversely, once trained, traditional ML/DL classifiers can be deployed locally and provide lower and more predictable inference latency, making them more suitable for scenarios requiring real-time feedback. Additionally, relying on proprietary APIs (e.g., OpenAI's offerings) raises concerns over data privacy, security, and long-term operational costs, particularly in large-scale industrial deployments. These challenges highlight the need for carefully designed deployment strategies, such as optimizing models for efficiency or exploring local deployment or hybrid architectures that combine local models with LLM-based APIs, to ensure the viability of LLMs in real-world development environments.}

\begin{framed}
\noindent \textcolor{black}{\textbf{Suggestion 4}: \textit{When selecting between the latest LLMs, traditional ML/DL techniques, or hybrid architectures (combining local models and LLM-based APIs), factors such as performance, data privacy, cost, latency, and stability must be carefully evaluated and traded off to ensure the best fit for real-world code review workflows.}}
\end{framed}

\section{Threats to Validity}\label{sec:Threats to Validity}
In this section, we discuss the potential threats to the validity related to our study and the adopted measures mitigating these threats by following the guidelines proposed by Wohlin \textit{et al}.~\cite{Wohlin2012ESE}. 

{\color{black}
\subsection{Internal Validity}

Internal validity refers to whether the observed effects can be attributed to the presence of classifier-identified hints rather than confounding factors. A primary threat is the variability in participants' experience and expertise, which we mitigate through stratified random assignment and by requiring a minimum level of development and code review experience. Learning and interpretation biases are reduced by providing all participants with the same tutorial and ensuring that tasks are completed independently in a single session. Instrumentation bias is addressed through a pilot study to refine task clarity and evaluation criteria, and by using ground-truth refactoring solutions derived from expert-reviewed histories to objectively assess responses. We also reduce the risk of external interference (e.g., use of AI tools or external resources) by explicitly instructing participants to complete tasks independently under controlled conditions. Finally, to avoid task-related confounding effects, all participants are assigned identical code snippets and tasks. \textcolor{black}{Another threat stems from the relatively small sample size of the controlled experiment and the use of simplified review tasks compared with industrial-scale code review activities. Although statistical tests indicate significant differences between groups, larger-scale replications would further strengthen confidence in the observed effects.}
}

\subsection{Construct Validity}
Construct validity concerns the connection between operational measures and the studied subjects. The first potential threat to construct validity is the evaluation of our selected ML/DL-based and LLM-based classifiers. Regarding the classifiers' performance evaluation, we used the four metrics (recall, precision, F1-score, and accuracy) that are widely used to evaluate the performance of various automated SE techniques (see Section~\ref{sec:Automated Identification}). Therefore, the metrics pose little threat to the construct validity. Another potential threat comes from the convenience sampling method of the survey and interview, as individuals with negative views towards the research might be less inclined to participate in our survey and interview. This is a compromise inherent to the convenience sampling method, and it implies that this threat cannot be completely mitigated. Nevertheless, to improve the response rate and collect as many opinions as possible from practitioners, we sent two rounds of reminder emails to the involved practitioners. It is also possible that some survey respondents did not understand the statements well. To reduce this threat to construct validity, we conducted a pilot survey to refine the survey design. Besides, we provided an ``\textit{I Don't Know}'' option in our survey. Moreover, we also interviewed six participants to collect in-depth opinions besides the six defined statements in the survey. To mitigate this threat, we offered an optional open-ended question that allows respondents to freely share their thoughts with us. 
\textcolor{black}{For the imbalanced evaluation, we additionally employed Macro-F1 to better reflect classifier performance under low-prevalence conditions. Although Macro-F1 is widely recommended for imbalanced classification problems, alternative cost-sensitive evaluation measures or operational utility metrics may lead to different interpretations of classifier effectiveness.}

\subsection{External Validity}
External validity pertains to the generalizability of our experiment results and findings. In certain situations, textual artifacts (e.g., code review, architecture discussions) may be unavailable. In such scenarios, we recognize that our approach may be complemented with traditional code-based architecture conformance checking techniques. Exploring the integration of such an automated approach into existing IDEs or tools remains part of our future work. As noted by Peter \textit{et al}. \cite{peters2017text}, it is not possible to use an exhaustive approach to test all learning models. Therefore, we selected five common ML algorithms, one DL algorithm, three popular pre-trained word embedding models, and three well-performed LLMs, as they are all widely used to conduct text classification. We admit that using different algorithms, pre-trained word embedding models, and LLMs may generate varying results, despite their popularity. We cannot completely mitigate this threat. Selecting different datasets and experimental settings might influence the generalizability of (part of) our findings. We acknowledge that the chosen open-source projects are not representative of all projects from the OpenStack and Qt platforms, and our findings may not be generalized to diverse projects, particularly those from different domains such as closed-source projects or projects on other platforms like Apache. Furthermore, the participants of the survey and interviews are contributors to the selected projects, therefore, the findings of RQ3 might not be generalized to other projects. To support further generalizability, we have made our dataset and scripts \cite{onlinepackage} available to enable other researchers and practitioners to replicate, reuse, and extend our study. \textcolor{black}{Furthermore, the naturally imbalanced evaluation was conducted on a fixed held-out test set derived from four OSS projects. Different violation prevalence rates, project characteristics, or industrial environments may influence classifier performance. Future replications on additional projects and domains are needed to further assess the generalizability of the findings.}

\subsection{Reliability}
Reliability reflects the replicability of a study regarding yielding the same or similar results when other researchers reproduce this study. One potential threat to reliability comes from the experimental settings (e.g., input data and parameters for training classifiers) used in this work. To address this, we used the dataset from our previous study \cite{Li2023wvs} that had publicly shared the labeled data and corresponding scripts for collecting data. \textcolor{black}{Due to the potential non-determinism and variability of LLMs, the performance of LLM-based classifiers may vary across runs because of the stochastic nature of LLM inference and sensitivity to prompting conditions. Therefore, we have made our scripts publicly available~\cite{onlinepackage} to facilitate reproduction and validation.} Another threat might concern the replicability of the survey and interview. In the current study, biases related to survey design were mitigated through careful question formulation and pilot testing. While such biases cannot be entirely eliminated, we provided the survey template, semi-structured interviews, and customized emails in our replication package \cite{onlinepackage} to maximize transparency and reliability, in order to facilitate replication in future studies.

\section{Related Work}\label{sec:Related Work}
In this section, we briefly discuss the current state of research in three associated areas of our study. We first describe related work on code review comments in Section~\ref{sec:Code Review Comments}; we then introduce previous studies on architecture violations in Section~\ref{sec:Architecture violations}; we finally present the benefits of utilizing ML/DL and LLM-related techniques to analyze artifacts in software development in Section~\ref{sec:Automated Identification}.

\subsection{Code Review Comments}\label{sec:Code Review Comments}
Code review comments contain massive knowledge related to software development, and a variety of studies analyze software defects and evolution through mining review comments and commit records. Han \textit{et al}. \cite{Han2022cmd} conducted an empirical study by manually checking 25,415 code review comments to investigate code smells identified in modern code review from four OSS projects. Their results show that most reviewers provide constructive suggestions to help developers fix code, and developers are willing to repair the smells through suggested refactoring operations. El Asri \textit{et al}. \cite{ElAsri2019ess} empirically investigated the impact of expressed sentiments on the code review duration and its outcome. They found that reviews with negative comments on average took longer time to complete than the reviews with positive/neutral comments. Kashiwa \textit{et al}. \cite{Kashiwa2022SATDmcr} reported Self-Admitted Technical Debt (SATD) during code review by checking 156,372 review records from OpenStack and Qt. They found that 28-48\% SATD are introduced during code review, and 20\% SATD comments are created because of reviewers' requests.

Besides, Uchôa \textit{et al}. \cite{Uchoa2020hdm} investigated the impact of code review on the evolution of design degradation through mining and analyzing a plethora of code reviews from seven OSS projects. They found that there is a wide fluctuation of design degradation during the revisions of certain code reviews. Paixão \textit{et al}. \cite{Paixao2020bti} explored how developers perform refactorings in code review, and they found that refactoring operations are most often used in code reviews that implement new features. Besides, they observed that the refactoring operations were rarely refined or undone along the code review, and such refactorings often contribute to new code smells and bugs.



\subsection{Architecture Violations}\label{sec:Architecture violations}
Over the past decades, there have been various investigations on architecture violations. Brunet \textit{et al}. \cite{Brunet2012av} performed a longitudinal study to explore the evolution of architectural violations in 19 bi-weekly versions of four open source systems. They investigated the life cycle and location of architecture violations over time by comparing the intended and recovered architectures of a system. They found that architectural violations tend to intensify as software evolves and a few design entities are responsible for the majority of violations. More interestingly, some violations seem to be recurring after being eliminated. Mendoza \textit{et al}. \cite{Mendoza2021avd} proposed a tool called ArchVID based on model-driven engineering techniques for identifying architecture violations from source code; this tool supports recovering and visualizing the implemented architecture. 

Besides, Terra \textit{et al}.~\cite{Terra2015rsr} reported their experience in fixing architectural violations. They proposed a recommendation system ArchFix that provides refactoring guidelines for developers and maintainers to repair architectural violations in the module architecture view of object-oriented systems. Their results show that their approach can provide correct recommendations for 75\% and 79\% of the detected architecture violations on two selected systems, respectively. Maffort \textit{et al}.~\cite{Maffort2016mav} proposed an approach to check architecture conformance for detecting architecture violations based on defined heuristics. They claimed that their approach relies on the defined heuristic rules and can rapidly raise architecture violation warnings. 



\subsection{Analyzing Software Repositories with ML/DL and LLM Techniques}\label{sec:Automated Identification}

The application of ML/DL and LLM techniques to extract informative knowledge from various artifacts during software development is gaining popularity in the SE community. In recent years, a plethora of empirical studies have been conducted to investigate automated techniques for supporting SE tasks. Khalajzadeh \textit{et al}. \cite{Khalajzadeh2022sda} conducted an empirical study by extracting and manually analyzing app reviews from 12 OSS projects and developed ML/DL models to automatically detect and classify human-centric issues from app reviews and discussions. Their results show that automated techniques can help developers to recognize and appreciate human-centric issues of end-users more easily. Nasab \textit{et al}. \cite{Nasab2021ais} developed 15 ML/DL models to automatically identify security discussions, and then they collected practitioners' feedback through a validation survey and confirmed the promising applications of the models in practice. 

Hou \textit{et al.} \cite{hou2024large} conducted a systematic literature review to explore the applications of LLMs across various phases of software development, from requirements gathering to program repair. Their study provides a comprehensive overview of recent advancements in the applications of LLMs in SE tasks, highlighting significant opportunities and critical areas that require further research and development to fully leverage LLMs in software engineering. 
Ahmed \textit{et al.} \cite{Ahmed2025can} investigated the potential of LLMs to substitute costly human annotation efforts in evaluating SE artifacts. They found that state-of-the-art LLMs can achieve inter-rater agreements comparable to human annotators across various code and code-related tasks. 
De Martino \cite{Martino2025LLM4MSR} \textit{et al.} presented a mixed-method empirical study combining a rapid literature review and survey to examine the use of LLMs in Mining Software Repositories (MSR). They uncovered 14 approaches and actions to mitigate 9 identified threats in applying LLMs to MSR, and proposed PRIMES 2.0, an enriched six-stage empirical framework with 23 substeps to enhance transparency, reproducibility, and methodological robustness in LLM-based MSR studies.


\subsection{Conclusive Summary}\label{sec:Conclusive Summary}

\textcolor{black}{While previous studies addressed relevant topics pertaining to code review, architecture violations, and the use of ML/DL/LLM techniques in mining software repositories, none of them have specifically utilized these automated approaches to classify architecture violations during code review. Building on developers' perspectives, we further collected feedback from practitioners involved in the code review discussions to evaluate the usefulness of the classifiers.} 

\textcolor{black}{Regarding code review comments analysis, previous studies \cite{Han2022cmd, ElAsri2019ess, Kashiwa2022SATDmcr, Uchoa2020hdm, Paixao2020bti} have made contributions to understanding various aspects of the code review process. However, these studies primarily focused on code-level concerns (e.g., code smells \cite{Han2022cmd}, sentiment analysis \cite{ElAsri2019ess}, technical debt~\cite{Kashiwa2022SATDmcr}, and design degradation \cite{Paixao2020bti}) typically through manual analysis approaches. While invaluable for understanding code review dynamics, they did not specifically target architecture-level issues or employ automated identification techniques for such concerns. Our work complements these studies by shifting the focus from code-level to architecture-level violations and introducing automation to identify these more subtle, yet critical issues. This represents an important evolution in code review research, moving from descriptive analysis toward prescriptive tool support.} 

\textcolor{black}{In the domain of architecture violations, prior research \cite{Mendoza2021avd, Brunet2012av, Terra2015rsr, Maffort2016mav} established important foundations for understanding and detecting architectural problems. These studies typically employed approaches (e.g., visualization tools \cite{Mendoza2021avd}, longitudinal analysis \cite{Brunet2012av}, heuristic rules and architecture conformance checking \cite{Maffort2016mav}) often examining source code directly during development or maintenance phases. While effective for their intended purposes, these approaches often require significant upfront investment in architecture documentation and mapping between architectural elements and source code. Our work offers a complementary perspective by examining architecture violations through the lens of developer discussions during code review. This approach requires minimal setup while capturing violations as they are actually perceived and discussed by development teams. By focusing on the code review context, we bridge the gap between theoretical architecture conformance and practical architectural decision-making in routine development workflows.}

\textcolor{black}{Concerning automated techniques for software repository mining, prior studies \cite{hou2024large, Nasab2021ais, Khalajzadeh2022sda, Ahmed2025can, Martino2025LLM4MSR} have demonstrated the potential of ML/DL and LLM techniques across various SE tasks. These studies have successfully applied automated approaches to identify security issues \cite{Nasab2021ais}, classify human-centric app reviews \cite{Khalajzadeh2022sda}, and address many other SE challenges \cite{hou2024large}. Our work extends this research thread by specifically targeting the identification of architecture violation symptoms—a nuanced and context-dependent problem that has not been adequately addressed through automation. Furthermore, unlike many previous automation studies, we combined direct feedback from practitioners involved in the original code review discussions with a controlled experiment that quantitatively evaluated the usefulness of classifier-identified hints. This mixed-method validation provides stronger evidence of our approach's practical utility than purely technical evaluations commonly used in this research area. This practitioner-centered validation represents a significant methodological advancement beyond the purely technical evaluations commonly used in this research area.}

\textcolor{black}{The synthesis of these three research directions reveals a gap that our work addresses: the lack of automated approaches for identifying architecture violations specifically within code review discussions. Previous studies either focused on manual analysis of code reviews, examined architecture violations through code-based techniques, or applied automation to other SE problems. Our approach uniquely bridges these areas by leveraging advanced ML/DL and LLM techniques to automatically identify architecture violation symptoms in code review comments, providing early warnings about potential architecture erosion directly within developers' existing workflows. This integration of architectural concern detection into the code review process represents a practical step toward maintaining architectural integrity throughout software evolution. Moreover, by combining practitioner feedback with a controlled experiment, our study provides both qualitative and quantitative evidence of the practical usefulness of the proposed approach. By positioning our work within these established research areas while highlighting its novel contributions, we demonstrate how automated identification of architecture violation symptoms can serve as a valuable complement to both traditional architecture conformance checking and modern code review practices, ultimately supporting better architectural decision-making and system sustainability.}

\section{Conclusions and Future Work}\label{sec:Conclusion}
During code review, reviewers typically spend a substantial amount of effort in comprehending code changes, as significant information (e.g., architecturally relevant information) for inspecting code changes may be dispersed across several files that the reviewers are not acquainted with. Automated identification of architecture violations from the discussions between reviewers and developers can save valuable time and effort to locate and check potential issues. 

In this work, we attempted to address the issue of automated identification of violation symptoms in code review comments. To this end, we performed a series of experiments on a dataset of architecture violations from code review comments. Subsequently, to validate the usefulness of our trained ML/DL-based classifiers, we conducted a survey and semi-structured interviews that acquired feedback from the involved developers who discussed architecture violations in code reviews. Furthermore, to compare LLM-based classifiers with traditional ML/DL-based classifiers, we performed the same classification task using three state-of-the-art LLMs (i.e., \textit{GPT-4o}, \textit{Qwen-3}, and \textit{DeepSeek-R1}).

Specifically, we developed 15 ML-based and 4 DL-based classifiers to identify violation symptoms from developer discussions in code reviews (i.e., code review comments from four OSS projects in Gerrit). The results show that the SVM classifier based on \textit{word2vec} pre-trained word embedding performs the best with an F1-score of 0.779. In most cases, classifiers with the \textit{fastText} pre-trained word embedding model can achieve relatively good performance. Furthermore, 200-dimensional pre-trained word embedding models outperform classifiers that use 100 and 300-dimensional models. In addition, an ensemble classifier based on the majority voting strategy can further enhance the classifier and outperforms the individual classifiers. Moreover, practitioners' perception of the usefulness of the classifiers' results validates the promising applications of automated identification of violation symptoms in practice. Especially, such classifiers can enable and inspire practitioners to find architecturally-related violation issues, prioritize, and eventually handle them. Additionally, the best LLM-based classifiers outperform traditional ML/DL-based classifiers on identifying violation symptoms, and GPT-4o achieves comparatively better performance than Qwen-3 and DeepSeek-R1. Interestingly, we found that the ensemble technique cannot consistently improve the performance of LLM-based classifiers. \textcolor{black}{Overall, we evaluated the proposed approaches using both balanced and naturally imbalanced settings. The results show that SVM-based classifiers achieve the best performance among traditional ML/DL approaches, while GPT-4o achieves the best overall performance among the evaluated LLMs. Additional statistical analyses confirm that the observed differences among classifiers are statistically significant, and the imbalanced evaluation demonstrates that the approaches remain effective under realistic low-prevalence conditions.}


\textcolor{black}{Future work will focus on extending the evaluation to larger and more diverse industrial datasets, investigating domain adaptation techniques for cross-project generalization, and improving robustness under extreme class imbalance. We also plan to explore explainable AI techniques and architecture-aware LLMs that can provide interpretable rationales and actionable refactoring suggestions, thereby further increasing practitioners' trust and adoption in real-world code review environments.}

\section*{Data Availability}
The replication package of this work has been made available at \cite{onlinepackage}.

\section*{Acknowledgements}
We would like to thank all the practitioners who participated in our survey, interviews, and controlled experiments. This work has been partially supported by the National Natural Science Foundation of China (NSFC) with Grant No. 92582203 and 62402348.



\bibliographystyle{ACM-Reference-Format}
\bibliography{ref}

@misc{onlinepackage,
   title = {{Replication Package for the Paper: Towards Automated Identification of Violation Symptoms of Architecture Erosion}},
   author={Li, Ruiyin and Liang, Peng and Avgeriou, Paris and Wang, Yifei},
   url = {https://github.com/RuiyinL/AEr_Identification},
   year = {2026}
}

@article{Li2023wvs,
   author = {Li, Ruiyin and Liang, Peng and Avgeriou, Paris},
   title = {{Warnings: Violation symptoms indicating architecture erosion}},
   journal = {Information and Software Technology},
   volume = {164},
   pages = {107319},
   year = {2023}
}

@inproceedings{Li2022sae,
   author = {Li, Ruiyin and Soliman, Mohamed and Liang, Peng and Avgeriou, Paris},
   title = {{Symptoms of architecture erosion in code reviews: A study of two OpenStack projects}},
   bookTitle = {Proceedings of the 19th IEEE International Conference on Software Architecture (ICSA)},
   publisher = {IEEE},
   pages = {24--35},
   year = {2022}
}

@article{Li2022SMS,
   title = {{Understanding software architecture erosion: A systematic mapping study}},
   author = {Li, Ruiyin and Liang, Peng and Soliman, Mohamed and Avgeriou, Paris},
   journal = {Journal of Software: Evolution and Process},
   volume = {34},
   number = {3},
   pages = {e2423},
   year = {2022}
}

@inproceedings{Li2021uae,
   author = {Li, Ruiyin and Liang, Peng and Soliman, Mohamed and Avgeriou, Paris},
   title = {{Understanding architecture erosion: The practitioners' perceptive}},
   bookTitle = {Proceedings of the 29th IEEE/ACM International Conference on Program Comprehension (ICPC)},
   publisher = {IEEE},
   pages = {311--322},
   year = {2021}
}

@misc{SnowballStemmer,
   author={GeeksforGeeks},
   title = {{GeeksforGeeks, Snowball Stemmer – NLP}},
   url = {https://www.geeksforgeeks.org/snowball-stemmer-nlp/},
   year={2026}
}

@article{PerryWolf1992AEr,
   author = {Perry, Dewayne E and Wolf, Alexander L},
   title = {Foundations for the study of software architecture},
   journal = {ACM SIGSOFT Software Engineering Notes},
   volume = {17},
   number = {4},
   pages = {40--52},
   year = {1992}
}

@article{Basili1994gmq,
   author = {Basili, Victor R and Caldiera, Gianluigi and Rombach, H Dieter},
   title = {The goal question metric approach},
   journal = {Encyclopedia of Software Engineering},
   pages = {528--532},
   year = {1994}
}

@inproceedings{Brunet2012av,
   author = {Brunet, Joao and Bittencourt, Roberto Almeida and Serey, Dalton and Figueiredo, Jorge},
   title = {On the evolutionary nature of architectural violations},
   bookTitle = {Proceedings of the 19th Working Conference on Reverse Engineering (WCRE)},
   publisher = {IEEE},
   pages = {257--266},
   year = {2012}
}

@article{Mendoza2021avd,
   author = {Mendoza, Camilo and Bocanegra, José and Garcés, Kelly and Casallas, Rubby},
   title = {Architecture violations detection and visualization in the continuous integration pipeline},
   journal = {Software: Practice and Experience},
   volume = {51},
   number = {8},
   pages = {1822--1845},
   year = {2021}
}

@article{Terra2015rsr,
   author = {Terra, Ricardo and Valente, Marco Tulio and Czarnecki, Krzysztof and Bigonha, Roberto S},
   title = {A recommendation system for repairing violations detected by static architecture conformance checking},
   journal = {Software: Practice and Experience},
   volume = {45},
   number = {3},
   pages = {315--342},
   year = {2015}
}

@article{Maffort2016mav,
   author = {Maffort, Cristiano and Valente, Marco Tulio and Terra, Ricardo and Bigonha, Mariza and Anquetil, Nicolas and Hora, André},
   title = {Mining architectural violations from version history},
   journal = {Empirical Software Engineering},
   volume = {21},
   number = {3},
   pages = {854--895},
   year = {2016}
}

@article{Miranda2016acc,
   author = {Miranda, Sergio and Rodrigues Jr, Elder and Valente, Marco Tulio and Terra, Ricardo},
   title = {Architecture conformance checking in dynamically typed languages},
   journal = {Journal of Object Technology},
   volume = {15},
   number = {3},
   pages = {1--34},
   year = {2016}
}

@inproceedings{Efstathiou2018weso,
   author = {Efstathiou, Vasiliki and Chatzilenas, Christos and Spinellis, Diomidis},
   title = {Word embeddings for the software engineering domain},
   bookTitle = {Proceedings of the 15th International Conference on Mining Software Repositories (MSR)},
   publisher = {ACM},
   pages = {38--41},
   year = {2018}
}

@inproceedings{Li2021mpl,
   author = {Li, Zengyang and Qi, Xiaoxiao and Yu, Qinyi and Liang, Peng and Mo, Ran and Yang, Chen},
   title = {Multi-programming-language commits in OSS: An empirical study on Apache projects},
   bookTitle = {Proceedings of the 29th IEEE/ACM International Conference on Program Comprehension (ICPC)},
   publisher = {IEEE},
   pages = {219--229},
   year = {2021}
}

@inproceedings{Uchoa2020hdm,
   author = {Uchôa, Anderson and Barbosa, Caio and Oizumi, Willian and Blenilio, Publio and Lima, Rafael and Garcia, Alessandro and Bezerra, Carla},
   title = {How does modern code review impact software design degradation? An in-depth empirical study},
   bookTitle = {Proceedings of the 36th IEEE International Conference on Software Maintenance and Evolution (ICSME)},
   publisher = {IEEE},
   pages = {511--522},
   year = {2020}
}

@inproceedings{Paixao2020bti,
   author = {Paixão, Matheus and Uchôa, Anderson and Bibiano, Ana Carla and Oliveira, Daniel and Garcia, Alessandro and Krinke, Jens and Arvonio, Emilio},
   title = {Behind the intents: An in-depth empirical study on software refactoring in modern code review},
   bookTitle = {Proceedings of the 17th International Conference on Mining Software Repositories (MSR)},
   publisher = {ACM},
   pages = {125--136},
   year = {2020}
}

@article{Bird2010nlp,
   author = {Bird, Steven and Klein, Ewan and Loper, Edward},
   title = {Natural language processing with Python: Analyzing text with the natural language toolkit},
   journal = {Language Resources and Evaluation},
   volume = {44},
   pages = {421--424},
   year = {2010}
}

@incollection{gaese2008CH3,
  title={Personal Opinion Surveys},
  author={Kitchenham, Barbara A. and Pfleeger, Shari L.},
  booktitle={Guide to Advanced Empirical Software Engineering},
  chapter = 3,
  pages={63--92},
  year={2008},
  publisher={Springer}
}

@incollection{gaese2008CH1,
  title={Software Engineering Data Collection for Field Studies},
  author={Singer, Janice and Sim, Susan E. and Lethbridge, Timothy C.},
  booktitle={Guide to Advanced Empirical Software Engineering},
  chapter = 1,
  pages={9--34},
  year={2008},
  publisher={Springer}
}

@article{Lethbridge2005sse,
   author = {Lethbridge, Timothy C and Sim, Susan Elliott and Singer, Janice},
   title = {Studying software engineers: Data collection techniques for software field studies},
   journal = {Empirical Software Engineering},
   volume = {10},
   number = {3},
   pages = {311--341},
   year = {2005}
}

@article{Khalajzadeh2022sda,
   author = {Khalajzadeh, Hourieh and Shahin, Mojtaba and Obie, Humphrey O and Agrawal, Pragya and Grundy, John},
   title = {Supporting Developers in Addressing Human-centric Issues in Mobile Apps},
   journal = {IEEE Transactions on Software Engineering},
   volume = {49},
   number = {4},
   pages = {2149--2168},
   year = {2022}
}

@inproceedings{Le2018emad,
   author = {Le, Duc Minh and Link, Daniel and Shahbazian, Arman and Medvidovic, Nenad},
   title = {An empirical study of architectural decay in open-source software},
   bookTitle = {Proceedings of the 15th IEEE International Conference on Software Architecture (ICSA)},
   publisher = {IEEE},
   pages = {176--185},
   year = {2018}
}

@article{DeSilva2012AErsurvey,
   author = {De Silva, Lakshitha and Balasubramaniam, Dharini},
   title = {{Controlling software architecture erosion: A survey}},
   journal = {Journal of Systems and Software},
   volume = {85},
   number = {1},
   pages = {132--151},
   year = {2012}
}

@article{Nasab2021ais,
  title={Automated identification of security discussions in microservices systems: Industrial surveys and experiments},
  author={Nasab, Ali Rezaei and Shahin, Mojtaba and Liang, Peng and Basiri, Mohammad Ehsan and Raviz, Seyed Ali Hoseyni and Khalajzadeh, Hourieh and Waseem, Muhammad and Naseri, Amineh},
  journal={Journal of Systems and Software},
  volume={181},
  pages={111046},
  year={2021},
  publisher={Elsevier}
}

@inproceedings{Hassaine2012ADvISE,
   author = {Hassaine, Salima and Guéhéneuc, Yann-Gaël and Hamel, Sylvie and Antoniol, Giuliano},
   title = {ADvISE: Architectural decay in software evolution},
   bookTitle = {Proceedings of the 16th European Conference on Software Maintenance and Reengineering (CSMR)},
   publisher = {IEEE},
   pages = {267--276},
   year = {2012}
}

@article{Hochstein2005cad,
   author = {Hochstein, Lorin and Lindvall, Mikael},
   title = {Combating architectural degeneration: A survey},
   journal = {Information and Software Technology},
   volume = {47},
   number = {10},
   pages = {643--656},
   year = {2005}
}

@inproceedings{Mair2013tesa,
   author = {Mair, Matthias and Herold, Sebastian},
   title = {Towards extensive software architecture erosion repairs},
   bookTitle = {Proceedings of the 7th European Conference on Software Architecture (ECSA)},
   publisher = {Springer},
   pages = {299--306},
   year = {2013}
}

@inproceedings{Le2016rad,
   author = {Le, Duc Minh and Carrillo, Carlos and Capilla, Rafael and Medvidovic, Nenad},
   title = {Relating architectural decay and sustainability of software systems},
   bookTitle = {Proceedings of the 13th Working IEEE/IFIP Conference on Software Architecture (WICSA)},
   publisher = {IEEE},
   pages = {178--181},
   year = {2016}
}

@article{Martin2000dpdp,
   author = {Martin, Robert C},
   title = {Design principles and design patterns},
   journal = {Object Mentor},
   volume = {1},
   number = {34},
   pages = {1--34},
   year = {2000}
}

@book{Wohlin2012ESE,
   author = {Wohlin, Claes and Runeson, Per and Höst, Martin and Ohlsson, Magnus C and Regnell, Björn and Wesslén, Anders},
   title = {Experimentation in Software Engineering},
   publisher = {Springer Science \& Business Media},
   year = {2012}
}

@inproceedings{Bacchelli2013eoc,
   author = {Bacchelli, Alberto and Bird, Christian},
   title = {Expectations, outcomes, and challenges of modern code review},
   bookTitle = {Proceedings of the 35th International Conference on Software Engineering (ICSE)},
   publisher = {IEEE},
   pages = {712--721},
   year = {2013}
}

@article{Adolph2011CC,
   author = {Adolph, Steve and Hall, Wendy and Kruchten, Philippe},
   title = {Using grounded theory to study the experience of software development},
   journal = {Empirical Software Engineering},
   volume = {16},
   number = {4},
   pages = {487--513},
   year = {2011}
}

@inproceedings{Mikolov2013word2vec,
   author = {Mikolov, Tomas and Sutskever, Ilya and Chen, Kai and Corrado, Greg S and Dean, Jeff},
   title = {Distributed representations of words and phrases and their compositionality},
   bookTitle = {Proceedings of the 27th Annual Conference on Neural Information Processing Systems (NeurIPS)},
   publisher = {Curran Associates},
   pages = {3111--3119},
   year = {2013}
}

@article{Bojanowski2017fastText,
   author = {Bojanowski, Piotr and Grave, Edouard and Joulin, Armand and Mikolov, Tomas},
   title = {Enriching word vectors with subword information},
   journal = {Transactions of the Association for Computational Linguistics},
   volume = {5},
   pages = {135--146},
   year = {2017}
}

@inproceedings{Pennington2014GloVe,
   author = {Pennington, Jeffrey and Socher, Richard and Manning, Christopher D},
   title = {Glove: Global vectors for word representation},
   bookTitle = {Proceedings of the 19th Conference on Empirical Methods in Natural Language Processing (EMNLP)},
   publisher = {ACL},
   pages = {1532--1543},
   year = {2014}
}

@inproceedings{Herold2015dvc,
   author = {Herold, Sebastian and English, Michael and Buckley, Jim and Counsell, Steve and Cinn{\'{e}}ide, Mel {\'{O}}},
   title = {Detection of violation causes in reflexion models},
   bookTitle = {Proceedings of the 22nd IEEE International Conference on Software Analysis, Evolution, and Reengineering (SANER)},
   publisher = {IEEE},
   pages = {565--569},
   year = {2015}
}

@inproceedings{Fontana2016aer,
   author = {Fontana, Francesca Arcelli and Roveda, Riccardo and Zanoni, Marco and Raibulet, Claudia and Capilla, Rafael},
   title = {An experience report on detecting and repairing software architecture erosion},
   bookTitle = {Proceedings of the 13th Working IEEE/IFIP Conference on Software Architecture (WICSA)},
   publisher = {IEEE},
   pages = {21--30},
   year = {2016}
}

@inproceedings{Macia2012anomalies,
   author = {Macia, Isela and Arcoverde, Roberta and Garcia, Alessandro and Chavez, Christina and von Staa, Arndt},
   title = {On the relevance of code anomalies for identifying architecture degradation symptoms},
   bookTitle = {Proceedings of the 16th European Conference on Software Maintenance and Reengineering (CSMR)},
   publisher = {IEEE},
   pages = {277--286},
   year = {2012}
}

@inproceedings{Kim2014TextCNN,
   author = {Kim, Yoon},
   title = {Convolutional neural network for sentence classification},
   bookTitle = {Proceedings of the 19th Conference on Empirical Methods in Natural Language Processing (EMNLP)},
   publisher = {ACL},
   pages = {1746--1751},
   year = {2014}
}

@article{he2009imbalance,
  title={Learning from imbalanced data},
  author={He, Haibo and Garcia, Edwardo A},
  journal={IEEE Transactions on Knowledge and Data Engineering},
  volume={21},
  number={9},
  pages={1263--1284},
  year={2009},
  publisher={IEEE}
}

@article{Ren2019nndSATD,
   author = {Ren, Xiaoxue and Xing, Zhenchang and Xia, Xin and Lo, David and Wang, Xinyu and Grundy, John},
   title = {Neural network-based detection of self-admitted technical debt: From performance to explainability},
   journal = {ACM Transactions on Software Engineering and Methodology},
   volume = {28},
   number = {3},
   pages = {1--45},
   year = {2019}
}

@inproceedings{Sesari2022pte,
   author = {Sesari, Emeralda and Hort, Max and Sarro, Federica},
   title = {An Empirical Study on the Fairness of Pre-trained Word Embeddings},
   bookTitle = {Proceedings of the 60th Annual Meeting of the Association for Computational Linguistics (ACL)},
   publisher = {ACL},
   pages = {129--144},
   year = {2022}
}

@inproceedings{Dietterich2000emml,
   author = {Dietterich, Thomas G},
   title = {Ensemble methods in machine learning},
   bookTitle = {Proceedings of the 1st International Workshop of Multiple Classifier Systems (MCS)},
   publisher = {Springer},
   pages = {1--15},
   year = {2000}
}

@inproceedings{Qi2018pwe,
  author={Qi, Ye and Sachan, Devendra and Felix, Matthieu and Padmanabhan, Sarguna and Neubig, Graham},
  title={When and why are pre-trained word embeddings useful for neural machine translation?},
  booktitle={Proceedings of the 32nd Conference of the North American Chapter of the Association for Computational Linguistics (NAACL)},
  pages={529--535},
  publisher = {ACL},
  year={2018}
}

@article{Radford2018ilu,
  title={Improving language understanding by generative pre-training},
  author={Radford, Alec and Narasimhan, Karthik and Salimans, Tim and Sutskever, Ilya and others},
  journal={In: Preprint},
  pages={1--12},
  publisher = {OpenAI},
  year={2018}
}

@article{peters2017text,
  title={Text filtering and ranking for security bug report prediction},
  author={Peters, Fayola and Tun, Thein Than and Yu, Yijun and Nuseibeh, Bashar},
  journal={IEEE Transactions on Software Engineering},
  volume={45},
  number={6},
  pages={615--631},
  year={2017},
  publisher={IEEE}
}

@article{Venters2018ss,
   author = {Venters, Colin C and Capilla, Rafael and Betz, Stefanie and Penzenstadler, Birgit and Crick, Tom and Crouch, Steve and Nakagawa, Elisa Yumi and Becker, Christoph and Carrillo, Carlos},
   title = {Software sustainability: Research and practice from a software architecture viewpoint},
   journal = {Journal of Systems and Software},
   volume = {138},
   pages = {174--188},
   year = {2018}
}

@book{hutter2019aml,
  title={Automated Machine Learning: Methods, Systems, Challenges},
  author={Hutter, Frank and Kotthoff, Lars and Vanschoren, Joaquin},
  year={2019},
  publisher={Springer Nature}
}

@article{Li2022SATD,
  title={Identifying self-admitted technical debt in issue tracking systems using machine learning},
  author={Li, Yikun and Soliman, Mohamed and Avgeriou, Paris},
  journal={Empirical Software Engineering},
  volume={27},
  number={6},
  pages={1--37},
  year={2022},
  publisher={Springer}
}

@article{hansen1990nne,
  title={Neural network ensembles},
  author={Hansen, Lars Kai and Salamon, Peter},
  journal={IEEE Transactions on Pattern Analysis and Machine Intelligence},
  volume={12},
  number={10},
  pages={993--1001},
  year={1990},
  publisher={IEEE}
}

@book{Bass2021sap,
   author = {Bass, Len and Clements, Paul and Kazman, Rick},
   title = {Software Architecture in Practice (4th Edition)},
   publisher = {Addison-Wesley Professional},
   edition = {4th},
   year = {2021}
}

@inproceedings{Herold2020asd,
   author = {Herold, Sebastian},
   title = {An initial study on the association between architectural smells and degradation},
   bookTitle = {Proceedings of the 14th European Conference on Software Architecture (ECSA)},
   publisher = {Springer},
   pages = {193--201},
   year = {2020}
}

@inproceedings{Liu2018nmt,
   author = {Liu, Zhongxin and Xia, Xin and Hassan, Ahmed E and Lo, David and Xing, Zhenchang and Wang, Xinyu},
   title = {{Neural-machine-translation-based commit message generation: How far are we?}},
   bookTitle = {Proceedings of the 33rd ACM/IEEE International Conference on Automated Software Engineering (ASE)},
   publisher = {ACM},
   pages = {373--384},
   year = {2018}
}

@inproceedings{Fu2017eoh,
   author = {Fu, Wei and Menzies, Tim},
   title = {Easy over hard: A case study on deep learning},
   bookTitle = {Proceedings of the 11th Joint Meeting on the European Software Engineering Conference and the ACM SIGSOFT Symposium on the Foundations of Software Engineering (ESEC/FSE)},
   publisher = {ACM},
   pages = {49--60},
   year = {2017}
}

@article{Yang2022pmse,
   author = {Yang, Yanming and Xia, Xin and Lo, David and Bi, Tingting and Grundy, John and Yang, Xiaohu},
   title = {Predictive models in software engineering: Challenges and opportunities},
   journal = {ACM Transactions on Software Engineering and Methodology},
   volume = {31},
   number = {3},
   pages = {1--72},
   year = {2022}
}

@book{Han2022dm,
   author = {Han, Jiawei and Pei, Jian and Tong, Hanghang},
   title = {Data Mining: Concepts and Techniques},
   publisher = {Morgan kaufmann},
   year = {2022}
}

@article{Minaee2021dltc,
   author = {Minaee, Shervin and Kalchbrenner, Nal and Cambria, Erik and Nikzad, Narjes and Chenaghlu, Meysam and Gao, Jianfeng},
   title = {Deep learning-based text classification: A comprehensive review},
   journal = {ACM Computing Surveys},
   volume = {54},
   number = {3},
   pages = {1--40},
   year = {2021}
}

@article{Han2022cmd,
   author = {Han, Xiaofeng and Tahir, Amjed and Liang, Peng and Counsell, Steve and Blincoe, Kelly and Li, Bing and Luo, Yajing},
   title = {Code smells detection via modern code review: A study of the OpenStack and Qt communities},
   journal = {Empirical Software Engineering},
   volume = {27},
   number = {6},
   pages = {1--42},
   year = {2022}
}

@article{ElAsri2019ess,
   author = {El Asri, Ikram and Kerzazi, Noureddine and Uddin, Gias and Khomh, Foutse and Idrissi, MA Janati},
   title = {An empirical study of sentiments in code reviews},
   journal = {Information and Software Technology},
   volume = {114},
   pages = {37-54},
   year = {2019}
}

@article{Watson2022SLRDL,
   author = {Watson, Cody and Cooper, Nathan and Palacio, David Nader and Moran, Kevin and Poshyvanyk, Denys},
   title = {A systematic literature review on the use of deep learning in software engineering research},
   journal = {ACM Transactions on Software Engineering and Methodology},
   volume = {31},
   number = {2},
   pages = {1-58},
   year = {2022}
}

@article{Kashiwa2022SATDmcr,
   author = {Kashiwa, Yutaro and Nishikawa, Ryoma and Kamei, Yasutaka and Kondo, Masanari and Shihab, Emad and Sato, Ryosuke and Ubayashi, Naoyasu},
   title = {An empirical study on self-admitted technical debt in modern code review},
   journal = {Information and Software Technology},
   volume = {146},
   pages = {106855},
   year = {2022}
}

@article{Davila2021SLRmcr,
   author = {Davila, Nicole and Nunes, Ingrid},
   title = {A systematic literature review and taxonomy of modern code review},
   journal = {Journal of Systems and Software},
   volume = {177},
   pages = {110951},
   year = {2021}
}

@article{Paixao2021crac,
   author = {Paixao, Matheus and Krinke, Jens and Han, DongGyun and Ragkhitwetsagul, Chaiyong and Harman, Mark},
   title = {The impact of code review on architectural changes},
   journal = {IEEE Transactions on Software Engineering},
   volume = {47},
   number = {5},
   pages = {1041--1059},
   year = {2021}
}

@article{badampudi2023modern,
  title={Modern Code Reviews—Survey of Literature and Practice},
  author={Badampudi, Deepika and Unterkalmsteiner, Michael and Britto, Ricardo},
  journal={ACM Transactions on Software Engineering and Methodology},
  volume={32},
  number={4},
  pages={1--61},
  year={2023},
  publisher={ACM}
}

@article{Cohen1960cans,
   author = {Cohen, J.},
   title = {A coefficient of agreement for nominal scales},
   journal = {Educational and Psychological Measurement},
   volume = {20},
   number = {1},
   pages = {37-46},
   year = {1960}
}

@article{Sharma2018sss,
   author = {Sharma, Tushar and Spinellis, Diomidis},
   title = {A survey on software smells},
   journal = {Journal of Systems and Software},
   volume = {138},
   pages = {158-173},
   year = {2018}
}

@article{li2025ChatGPT,
  title={{Unveiling the Role of ChatGPT in Software Development: Insights from Developer-ChatGPT Interactions on GitHub}},
  author={Li, Ruiyin and Liang, Peng and Wang, Yifei and Cai, Yangxiao and Sun, Weisong and Li, Zengyang},
  journal={ACM Transactions on Software Engineering and Methodology},
  year={2026}
}

@article{hou2024large,
  title={Large language models for software engineering: A systematic literature review},
  author={Hou, Xinyi and Zhao, Yanjie and Liu, Yue and Yang, Zhou and Wang, Kailong and Li, Li and Luo, Xiapu and Lo, David and Grundy, John and Wang, Haoyu},
  journal={ACM Transactions on Software Engineering and Methodology},
  volume={33},
  number={8},
  pages={1--79},
  year={2024},
  publisher={ACM}
}

@techreport{GPT4,
   author = {Achiam, Josh and Adler, Steven and Agarwal, Sandhini and Ahmad, Lama and Akkaya, Ilge and Aleman, Florencia Leoni and Almeida, Diogo and Altenschmidt, Janko and Altman, Sam and Anadkat, Shyamal},
   title = {GPT-4 Technical Report},
   year = {2023} 
}

@misc{GPT-4o,
    title = {{GPT-4o}},
    url = {https://github.com/marketplace/models/azure-openai/gpt-4o},
    author = {OpenAI},
    month = {November},
    year = {2024}
}

@article{deepseekai2025,
      title={DeepSeek-R1: Incentivizing Reasoning Capability in LLMs via Reinforcement Learning}, 
      author={DeepSeek-AI},
      year={2025},
      journal={arXiv preprint arXiv:2501.12948}
}

@inproceedings{Ahmed2025can,
  title={Can LLMs replace manual annotation of software engineering artifacts?},
  author={Ahmed, Toufique and Devanbu, Premkumar and Treude, Christoph and Pradel, Michael},
  booktitle={Proceedings of the 22nd IEEE/ACM International Conference on Mining Software Repositories (MSR)},
  pages={526--538},
  publisher = {IEEE},
  year={2025}
}

@book{jevons2013PE,
  title={The Theory of Political Economy},
  author={Jevons, William},
  year={2013},
  publisher={Springer}
}

@article{Martino2025LLM4MSR,
   author = {De Martino, Vincenzo and Castaño, Joel and Palomba, Fabio and Franch, Xavier and Martínez-Fernández, Silverio},
   title = {A Methodological Framework for LLM-Based Mining of Software Repositories},
   journal = {arXiv preprint arXiv:2508.02233},
   year = {2025}
}

@article{Buda2018SMSCNN,
   author = {Buda, Mateusz and Maki, Atsuto and Mazurowski, Maciej A},
   title = {A systematic study of the class imbalance problem in convolutional neural networks},
   journal = {Neural Networks},
   volume = {106},
   pages = {249--259},
   year = {2018}
}

@inproceedings{elkan2001foundations,
   author = {Elkan, Charles},
   title = {The Foundations of Cost-Sensitive Learning},
   bookTitle = {Proceedings of the 17th International Joint Conference on Artificial Intelligence (IJCAI)},
   pages={973--978},
   publisher={AAAI Press},
   year = {2001}
}

@inproceedings{zadrozny2001costs,
   author = {Zadrozny, Bianca and Elkan, Charles},
   title = {Learning and Making Decisions When Costs and Probabilities are Both Unknown},
   bookTitle = {Proceedings of the 7th ACM SIGKDD International Conference on Knowledge Discovery and Data Mining (KDD)},
   publisher={ACM},
   pages={204--213},
   year = {2001}
}

@article{chawla2002smote,
   author = {Chawla, Nitesh V. and Bowyer, Kevin W. and Hall, Lawrence O. and Kegelmeyer, W. Philip},
   title = {{SMOTE: Synthetic Minority Over-sampling Technique}},
   journal = {Journal of Artificial Intelligence Research},
   volume = {16},
   pages = {321--357},
   year = {2002}
}

@inproceedings{lin2017focal,
   author = {Lin, Tsung-Yi and Goyal, Priya and Girshick, Ross and He, Kaiming and Doll{\'a}r, Piotr},
   title = {Focal Loss for Dense Object Detection},
   bookTitle = {Proceedings of the 16th IEEE International Conference on Computer Vision (ICCV)},
   publisher={IEEE},
   pages={2999--3007},
   year = {2017}
}

@inproceedings{cui2019classbalanced,
   author = {Cui, Yin and Jia, Menglin and Lin, Tsung-Yi and Song, Yang and Belongie, Serge},
   title = {Class-Balanced Loss Based on Effective Number of Samples},
   bookTitle = {Proceedings of the 32nd IEEE/CVF Conference on Computer Vision and Pattern Recognition (CVPR)},
   publisher={IEEE},
   pages={1--11},
   year = {2019}
}

@inproceedings{guo2017calibration,
   author = {Guo, Chuan and Pleiss, Geoff and Sun, Yu and Weinberger, Kilian Q.},
   title = {On Calibration of Modern Neural Networks},
   bookTitle = {Proceedings of the 34th International Conference on Machine Learning (ICML)},
   publisher={PMLR},
   pages={1321--1330},
   year = {2017}
}

@article{platt1999probabilistic,
  title={Probabilistic outputs for support vector machines and comparisons to regularized likelihood methods},
  author={Platt, John and others},
  journal={Advances in Large Margin Classifiers},
  volume={10},
  number={3},
  pages={61--74},
  year={1999},
  publisher={Cambridge}
}

@article{demsar2006statistical,
   author = {Dem{\v{s}}ar, Janez},
   title = {{Statistical Comparisons of Classifiers over Multiple Data Sets}},
   journal = {Journal of Machine Learning Research},
   volume = {7},
   pages = {1--30},
   year = {2006}
}

@article{dudani1976distance,
   author = {Dudani, Sahibsingh A.},
   title = {{The Distance-Weighted k-Nearest-Neighbor Rule}},
   journal = {IEEE Transactions on Systems, Man, and Cybernetics},
   volume = {SMC-6},
   number = {4},
   pages = {325--327},
   year = {1976}
}

@article{sokolova2009systematic,
   author = {Sokolova, Marina and Lapalme, Guy},
   title = {{A Systematic Analysis of Performance Measures for Classification Tasks}},
   journal = {Information Processing \& Management},
   volume = {45},
   number = {4},
   pages = {427--437},
   year = {2009}
}

@article{powers2011evaluation,
   author = {Powers, David M. W.},
   title = {{Evaluation: From Precision, Recall and F-Measure to ROC, Informedness, Markedness \& Correlation}},
   journal = {Journal of Machine Learning Technologies},
   volume = {2},
   number = {1},
   pages = {37--63},
   year = {2011}
}

@article{brown2020language,
   author = {Brown, Tom and Mann, Benjamin and Ryder, Nick and Subbiah, Melanie and Kaplan, Jared and Dhariwal, Prafulla and others},
   title = {Language Models are Few-Shot Learners},
   journal = {arXiv preprint arXiv:2005.14165},
   year = {2020}
}

@article{liu2023pretrainpromptpredict,
   author = {Liu, Pengfei and Yuan, Weizhe and Fu, Jinlan and Jiang, Zhengbao and Hayashi, Hiroaki and Neubig, Graham},
   title = {Pre-train, Prompt, and Predict: A Systematic Survey of Prompting Methods in Natural Language Processing},
   journal = {arXiv preprint arXiv:2107.13586},
   year = {2021}
}

@misc{qwen3techreport,
  title = {{Qwen3: Think Deeper, Act Faster}},
  author = {{Qwen Team}},
  year = {2025},
  month = {April},
  url = {https://qwenlm.github.io/blog/qwen3/}
}

@book{agresti2013,
  title={Categorical Data Analysis},
  author={Agresti, Alan},
  year={2013},
  publisher={John Wiley \& Sons}
}

@book{cohen2013,
  title={Statistical Power Analysis for the Behavioral Sciences},
  author={Cohen, Jacob},
  year={2013},
  publisher={Routledge}
}

\end{document}